\documentclass[a4paper,11pt,aps,preprintnumbers,amsmath,amssymb,superscriptaddress,nofootinbib]{article}
\pdfoutput=1
\usepackage{jcappub}

\usepackage{multirow}
\usepackage{amssymb}
\usepackage{enumerate}
\usepackage{amssymb}
\usepackage{amsmath}
\usepackage{tikz}
\usepackage{feynmf}
\usepackage{makecell}
\usepackage{slashed}
\usepackage{slashbox}
\usepackage{natbib}
\usepackage{graphicx}
\hypersetup{colorlinks=true}

\usepackage{changes,color}

\usepackage{mathrsfs,amssymb} 
\usepackage{tensor} 
\usepackage{cancel}
\usepackage[normalem]{ulem}
\usepackage{slashbox}

\definecolor{darkgreen}{rgb}{0.01, 0.75, 0.24}
\definecolor{darkorange}{rgb}{1.0, 0.55, 0.0}
\usepackage{pifont}% http://ctan.org/pkg/pifont
\newcommand{\beq}{\begin{equation}}
\newcommand{\eeq}{\end{equation}}
\newcommand\bea{\begin{eqnarray}}
\newcommand\eea{\end{eqnarray}}

\newcommand{\eq}[1]{Eq.~(\ref{#1})}
\newcommand{\eqs}[2]{Eqs.~(\ref{#1}) and (\ref{#2})}
\newcommand{\eqss}[3]{Eqs.~(\ref{#1}), (\ref{#2}) and (\ref{#3})}

\newcommand{\ov}{\overline}
\renewcommand{\l}{\left}
\renewcommand{\r}{\right}
\newcommand{\la}{\langle}
\newcommand{\ra}{\rangle}

\title{Axi-Majoron : One-shot solution to most of the big puzzles of particle
cosmology}

\author{Gabriela Barenboim$^{1,2}$}
\author{Pyungwon Ko$^3$}
\author{and Wan-il Park$^{4,1,2}$}
\affiliation{
$^1$Instituto de Física Corpuscular, CSIC-Universitat de València, Paterna 46980, Spain\\
$^2$Departament de Física Teòrica, Universitat de València, Burjassot 46100, Spain	\\
$^3$Korea Institute for Advanced Study, Seoul 02455, Republic of Korea\\
$^4$Division of Science Education and Institute of Fusion Science, Jeonbuk National University, Jeonju 54896, Republic of Korea\\
}

\abstract{
The details of the minimal cosmological standard model (MCSM) proposed in [The minimal cosmological standard model, arXiv:2403.05390.] are discussed.
The model is based on the scale-symmetry and the global Peccei-Quinn(PQ) symmetry with a key assumption that the latter is broken only in the gravity sector in a scale-invariant manner.
We show that the model provides a quite simple unified framework for the unknown history of the Universe from inflation to the epoch of big-bang nucleosynthesis, simultaneously addressing key puzzles of high energy theory and cosmology: (i) the origin of scales, (ii) primordial inflation, (iii) matter-antimatter asymmetry, (iv) tiny neutrino masses, (v) dark matter, and (vi) the strong \textit{CP}-problem.
Scale symmetry can be exact, and the Planck scale is dynamically generated.
The presence of Gauss-Bonnet term may safely retain dangerous nonperturbative symmetry-breaking effects negligible, allowing a large-field trans-Planckian inflation along the PQ-field. 
Isocurvature perturbations of axi-Majorons are suppressed.
A sizable amount of PQ-number asymmetry is generated at the end of inflation, and conserved afterwards.
Domain wall problem is absent due to the nonrestoration of the symmetry and the nonzero PQ-number asymmetry.
Baryogenesis can be realized by either the transfer of the PQ-number asymmetry through the seesaw sector, or by resonant leptogenesis.
Dark matter is purely cold axi-Majorons from the mis-alignment contribution with the symmetry-breaking scale of $\mathcal{O}(10^{12}) {\rm GeV}$.
Hot axi-Majorons from the decay of the inflaton become a natural source for a sizable amount of dark radiation.
Inflationary gravitational waves have information about the mass parameters of the lightest left-handed and right-handed neutrinos, thanks to the presence of an early matter-domination era driven by the long-lived lightest right-handed neutrino species.
}

\emailAdd{gabriela.barenboim@uv.es}
\emailAdd{pko@kias.re.kr}
\emailAdd{wipark@jbnu.ac.kr}

\begin{document}

\maketitle

\section{Introduction}
\label{sec:intro}

In modern physics, the subatomic world is described by the standard model(SM) of particle physics \cite{Glashow:1961tr,Weinberg:1967tq}, and the large scale associated with our observable Universe is governed by Einstein gravity. 
They are associated with two vastly different scales, the electroweak scale for the former and the Planck scale for the latter, as set by various experiments.
These scales appear as dimensionful constants in the corresponding theories.
The presence of such constants may mean the existence of a defining higher level theory in which dimensionful parameters are absent\footnote{Scale-invariance has been discussed in various contexts including cosmological applications (see \cite{Wetterich:2020cxq} for example and references therein).}.
Then, all scales should be generated dynamically, and differences among scales would be controlled by dimensionless parameters.
Question still remains though for the origin of the dimensionless parameters.
They might require extra-structure of spacetime including our 4-$D$ spacetime.

Recently, there have been several works discussing dynamical generation of scales in a scale invariant version of gravity and extensions of the SM \cite{Garcia-Bellido:2011kqb,Bezrukov:2012hx,Salvio:2014soa,Ferreira:2016vsc,Ferreira:2016wem}.
An interesting finding is that the Planck scale can be dynamically generated even without a Coleman-Weinberg(CW)-type symmetry-breaking \cite{Coleman:1973jx}, thanks to the conservation of scale current which has a kernel settling down to a constant \cite{Garcia-Bellido:2011kqb,Ferreira:2016wem}.
Low energy scales, such as the electroweak scale, can be generated in connection with the field responsible for the generation of the Planck scale, or separately via CW-mechanism in the presence of matter fields other than SM fields.
Scale-invariance can be maintained even at quantum level, provided that the renormalization scale (which usually appears as an explicit dimensionful parameter) is replaced by a fundamental scalar field.
Preservation of scale-invariance decouples the massless dilaton from the other fields, hence makes the theory free from fifth-force constraints \cite{Blas:2011ac,Ferreira:2016kxi}.
This implies that a fully scale-invariant theory can be consistent with low energy phenomenology.

Meanwhile, any theory trying to explain the current state of the Universe should be able to accommodate a phase of primordial inflation \footnote{It was found that in the scenario of an emergent gravity with scale-invariance a slow-roll inflation can arise quite naturally \cite{Garcia-Bellido:2011kqb,Kannike:2015apa,Ferreira:2016vsc,Ferreira:2018qss}.}, generation of matter-antimatter asymmetry, proper candidate(s) of dark matter, neutrino masses inferred by neutrino flavor oscillation data \cite{Gonzalez-Garcia:2007dlo}, and probably a solution to the strong \textit{CP}-problem such as invisible axion \cite{Kim:1979if,Shifman:1979if,Dine:1981rt,Zhitnitsky:1980tq}.
All of these big puzzles of cosmology and phenomenology cannot be solved within the SM only, and definitely demand a theory beyond the standard model(BSM). 
%Also, any BSM theories should include a sector(or interactions) responsible for the tiny mass of left-handed neutrinos. 
In this regard, the idea of unifying axion and Majoron \cite{Shin:1987xc} is quite interesting, since it can be realized in a scale-invariant manner, simultaneously address the tiny neutrino masses, and provide an axion solution to the strong \textit{CP}-problem in a very simple structure.
Then, it is tempting to see if the idea can be combined with inflation and generation of matter-antimatter asymmetry while keeping the scale-invariance.

In Ref.~\cite{Barenboim:2024akt}, we proposed a minimal scale-invariant $U(1)_{\rm PQ}$-symmetric extension of the SM \cite{Peccei:1977hh,Peccei:1977ur}, dubbed as `\textit{the minimal cosmological standard model(MCSM)}', to address the aforementioned issues altogether other than the notorious cosmological constant (CC) problem.
In this work, we show the details of the MCSM.
It is a DFSZ-type axion model \cite{Dine:1981rt,Zhitnitsky:1980tq} unified with the singlet Majoron model, in which Peccei-Quinn(PQ) field not only provides the axion solution, but it is also responsible for the generation of the masses of right-handed neutrinos.
Hence, we call the associated pseudo-Goldstone boson as `\textit{axi-Majoron}'. 
In the sense of unifying axion and Majoron, our scenario is similar to the earlier works such as Refs.~\cite{Shin:1987xc} and \cite{Volkas:1988cm,Clarke:2015bea}, based on KSVZ \cite{Kim:1979if,Shifman:1979if} and DFSZ \cite{Dine:1981rt,Zhitnitsky:1980tq} axion models, respectively.
Also, in regard of providing a unified framework addressing aforementioned puzzles, one can find works such as Refs.~\cite{Ballesteros:2016xej,Sopov:2022bog}.
Our scenario is different from those works in various aspects including dynamical generation of the Planck scale, inflationary physics, generation of matter-antimatter asymmetry and physics of axions.
In particular, assuming $U(1)_{\rm PQ}$-symmetry to be broken in the gravity sector only\footnote{There are some earlier works such as Refs.~\cite{Takahashi:2015waa} and \cite{Hashimoto:2021xgu} where 
the PQ symmetry-breaking nonminimal gravitational interaction was considered.
In Ref.~\cite{Takahashi:2015waa}, the authors payed attention only to the impact of such an interaction on the axion-isocurvature perturbations which can be suppressed by a temporary axion mass much larger than the expansion rate during inflation thanks to the interaction.
In Ref.~\cite{Hashimoto:2021xgu}, initial position of the field was set at the valley of the angular potential for simplicity, removing the impact of angular motion.
In both works, the PQ-field are located at the angular minimum position during inflation.
In our case, the angular motion and the nonzero angular displacement of the PQ-field at the end of inflation play the critical role for the post inflation cosmology.}, we show that the dynamics of the inflaton along axi-Majoron direction can generate a large amount of PQ-number asymmetry which is eventually converted to the baryon number asymmetry through the seesaw sector.
It is a kind of Affleck-Dine(AD) mechanism \cite{Affleck:1984fy} realized by Peccei-Quinn field.
Interestingly, such an asymmetry corresponding to a motion of axi-Majoron allows not only AD-mechanism but also spontaneous leptogenesis \cite{} although the latter turns out to be subdominant.
Even in the case of PQ-number asymmetry not being enough, our scenario allows resonant leptogenesis \cite{Pilaftsis:2003gt} to work.
Also, very intriguingly, in regard to the axion solution to the strong \textit{CP}-problem, the symmetry-breaking terms in the gravity sector in our scenario are harmless at least at perturbative level because of the field-dependent nature, and may be so even at nonperturbative level thanks to topological curvature-terms such as Gauss-Bonnet \cite{Kallosh:1995hi}.
Isocurvature perturbations are suppressed, and the domain-wall problem is absent due to the angular motion of PQ-field with nonrestoration of the symmetry.
Also, a sizable amount of dark radiation from hot axi-Majorons can be obtained naturally, and may ameliorate the Hubble tension \cite{DiValentino:2021izs} \cite{Vagnozzi:2019ezj}.
The standard thermal bath is recovered by decays of the long-lived lightest right-handed neutrino species.
Because of the presence of an early matter domination era, primordial gravitational waves from inflation are expected to have a characteristic spectral feature, containing information of the masses of the lightest left-handed and the lightest right-handed neutrinos.
In combinations with dark radiation and inflationary observables, it may provide a unique way to probe such physical quantities.
Our scenario provides thus a very minimal, unique unified framework for the unknown history of the Universe from inflation to the BBN epoch.

This paper is organized as follows.
In Sec.~\ref{sec:model}, the model is described.
In Sec.~\ref{sec:inf}, inflation with regard to dynamics of the homogeneous mode, evolutions of perturbations, and reheating are discussed. 
In Sec.~\ref{sec:lepto}, we show how leptogenesis works in our scenario via various mechanisms thanks to the dynamics of PQ-field.
In Sec.~\ref{sec:DM-DR}, dark matter and dark radiation from axi-Majorons are briefly discussed.
In Sec.~\ref{sec:conc}, conclusions are drawn.

\section{The Model}
\label{sec:model}

We assume that scale-invariance is a fundamental symmetry and preserved in both of the matter and the gravity sectors, and the Planck scale is dynamically generated by spontaneous breaking of scale symmetry.
We also assume that the global Peccei-Quinn symmetry, $U(1)_{\rm PQ}$, is preserved but only in the matter sector, since gravity does not seem to respect global symmetries.
Based on this symmetry, in order to realize the axion solution to the strong \textit{CP}-problem and the seesaw mechanism for neutrino masses and mixings in a minimal unified setup, we consider a scale-invariant type II two Higgs doublet extension of the SM \cite{Branco:2011iw}\footnote{KSVZ-type realization such as Ref.~\cite{Ballesteros:2016xej} is also possible, and heavy extra quarks can contribute to dark matter relic density in this case.} with the seesaw sector 
in the form of the singlet Majoron model \cite{Chikashige:1980ui}. 
%%%
\begin{table}[htp]
\begin{center}
\begin{tabular}{|c|c|c|c|c|c|c|c|c|c|}
\hline
\backslashbox{Charge}{Field} & $Q_L$ & $u_R$ & $d_R$ & $\ell_L$ & $e_R$ & $\nu_R$ & $H_1$ & $H_2$ & $\Phi$ \\ 
\hline
$q_{\rm PQ}$ & 3/2 & 1/2 & 1/2 & 3/2 & 1/2 & 1/2 & 1 & -1 & 1 \\
%\hline
%$q_{BL}$ & 1/3 & 1/3 & 1/3 & -1 & -1 & -1 & 0 & 0 & -2 \\
\hline\end{tabular}
\end{center}
\caption{Charge assignment of fields under $U(1)_{\rm PQ}$ and $U(1)_{B-L}$ symmetries.
$\nu_R$ is the right-handed neutrino(RHN) field, $H_1=\l( H_1^+, \ H_1^0 \r)^T$ and $H_2=\l( H_2^+, \ H_2^0 \r)^T$ are down- and up-type Higgs doublets, respectively.
$\Phi$ is the Peccei-Quinn field. 
Flavor indices of matter fields were suppressed.
}
\label{tab:PQ-charges}
\end{table}%
%%%
There are four scalar fields as the minimal set: a real scalar $\chi$ which is responsible for generating Planck scale via spontaneous breaking of scale symmetry, type II two Higgs doublets, and one complex scalar which plays the role of the Peccei-Quinn field, denoted here as $\Phi$.
The assignment of PQ-charges to fields is shown in Table~\ref{tab:PQ-charges}.

Our model is minimal in the sense of the number of fields introduced to accommodate various phenomenological/cosmological/theoretical necessities.
It might be compared to the model discussed in Ref.~\cite{Ballesteros:2016xej} for example.
The difference is how the axion solution is realized, either DFSZ-type or KSVZ-type.
While our scenario introduces only one extra Higgs doublet as a minimal DFSZ axion model, the compared model introduces two extra heavy quarks as a minimal KSVZ axion model.
Aesthetically, we think our model is much nicer in regard of its symmetrical structure and the very attractive efficiency achieved in such a simple structure with a minimal set of fields.

We characterize our model by the following action:
\bea \label{eq:S-G}
S_{\rm G} &=& - \frac{1}{2} \int d^4 x \sqrt{-{\tilde g}} \mathcal{F} {\tilde R} 
\ ,
\\ \label{eq:S-M}
S_{\rm M} &=& \int d^4 x \sqrt{-{\tilde g}} \l[ {\tilde g}^{\mu \nu} \l( \frac{1}{2} \partial_\mu \chi \partial_\nu \chi + \partial_\mu \Phi^\dagger \partial_\nu \Phi + \sum_{i=1}^2 \partial_\mu H_i^\dagger \partial_\nu H_i \r) +  \mathcal{L}_{\rm ss} - {\tilde V} \r] \ ,
\eea
where the SM Lagrangian for fermions and gauge bosons were omitted for simplicity,  $\mathcal{F} \equiv \mathcal{F}_{\rm s} + \mathcal{F}_{\rm sb}$ with
\bea \label{eq:F-s}
\mathcal{F}_{\rm s} &\equiv& \xi_\chi \chi^2 + 2 \xi_\phi  \l| \Phi \r|^2 + 2 \xi_{H_1} \l| H_1 \r|^2 + 2 \xi_{H_2} \l| H_2 \r|^2  \ ,
\\ \label{eq:F-sb}
\mathcal{F}_{\rm sb} &\equiv& \xi_+ \l( \Phi^2 + {\rm c.c.} \r) - i \xi_- \l( \Phi^2 - {\rm c.c.} \r)  +  \xi_{H_+} \l( H_1^\dag H_2 + {\rm c.c.} \r) - i \xi_{H_-} \l( H_1^\dag H_2 - {\rm c.c.} \r) \ ,
\eea
and the seesaw sector Lagrangian and the scalar potential are given by
\bea \label{eq:L-ss}
- \mathcal{L}_{\rm ss} &=& \frac{1}{2} y_N \Phi^*\ov{\nu_R^c} \nu_R + y_\nu \ov{\ell}_L {\tilde H_2} \nu_R + {\rm H.c.} \  ,
\\
\label{eq:V-scalar}
{\tilde V} &=& \frac{\lambda_\chi}{4} \chi^4 + \lambda_\phi |\Phi|^4 + \lambda_{h_1} |H_1|^4 + \lambda_{h_2} |H_2|^4
\nonumber \\
&-& \frac{1}{2} \lambda_{\chi \phi} \chi^2 |\Phi|^2 - \frac{1}{2} \sum_{I=1}^2  \l( \lambda_{\chi h_i} \chi^2 + 2 \lambda_{\phi h_i} |\Phi|^2 \r) |H_i|^2  - \frac{1}{2} \l( \lambda_{\phi h} \Phi^2 H_1^\dag H_2 + {\rm c.c.} \r) 
\nonumber \\
&-& \lambda_{12} |H_1|^2 |H_2|^2 - \lambda_{12,\times} |H_1^\dag H_2|^2 \ .
\eea
In \eq{eq:L-ss}, ${\tilde H_2} \equiv i \tau_2 H_2^*$ with $\tau_2$ being the second Pauli matrix, and the flavor indices for $\nu_R$ and $\ell_L$ were omitted.
We assume that the nonminimal gravitational couplings in \eqs{eq:F-s}{eq:F-sb} are positive definite and field-dependent quantities such as\footnote{Such an exponential-field-dependence might be hinting a nonperturbative origin for the couplings.}
\beq \label{eq:xi-alpha}
\xi_\alpha(\chi, \varphi_\alpha) \equiv \xi_{\alpha,0} \times \exp \l( - \frac{c_\alpha \chi}{|\varphi_\alpha|} \r) \ ,
\eeq
where the subscript `$_\alpha$' stands for all different $\xi$s, and $|\varphi_\alpha| = \l( \chi,  \l| \Phi \r|, \l| H_1 \r|, \l| H_2 \r|, \l| H_1^\dag H_2 \r|^{1/2} \r)$ depending on $\xi_\alpha$.
For the numerical coefficient $c_\alpha$, we assume $c_\chi=0$ for $\varphi_\alpha = \chi$, otherwise 
\beq
\sqrt{\frac{\xi_{\chi,0} \l| \varphi_{\alpha, 0} \r|}{M_{\rm P}}} \ll c_\alpha \ll \sqrt{\frac{\xi_{\chi,0}}{\xi_{\alpha,0}}} \ ,
\eeq
where $|\varphi_{\alpha,0}|$ is the low energy vacuum expectation value of the corresponding field.
Then, $\xi_\alpha(\chi, \varphi_\alpha)$ is relevant only for $|\varphi_\alpha| \gg c_\alpha \chi$ with $\xi_\alpha(\chi, \varphi_\alpha) \simeq \xi_{\alpha, 0}$, while turned off for $|\varphi_\alpha| \ll c_\alpha \chi$ at low energy.

The gravity part of Lagrangian can have additional scale-invariant curvature-terms such as ${\tilde R}^2, \  {\tilde R}_{\mu \nu} {\tilde R}^{\mu \nu}, \  {\tilde R}_{\mu \nu \rho \sigma} {\tilde R}^{\mu \nu \rho \sigma}$, and ${\tilde R}_{\mu \nu \rho \sigma} {}^*{\tilde R}^{\mu \nu \rho \sigma}$ with ${}^*{\tilde R}_{\mu \nu \rho \sigma} \equiv \epsilon_{\mu\nu\mu'\nu'}{\tilde R}\indices{^{\mu'}^{\nu'}_{\rho\sigma}}/2$.
They might be originated simply from two topological terms, ${\tilde R}_{\mu \nu \rho \sigma} {}^*{\tilde R}^{\mu \nu \rho \sigma}$ and ${}^*{\tilde R}_{\mu \nu \rho \sigma} {}^*{\tilde R}^{\mu \nu \rho \sigma}$.
Gauss-Bonnet term (i.e., the latter) may resolve the axion quality problem if the numerical coefficient is at least of order unity \cite{Kallosh:1995hi}.
Combined with scale-symmetry, the term allows trans-Planckian large-field inflation along Peccei-Quinn field direction, as discussed in the next section.
Those topological terms do not affect our discussion for all the other physics.
So, we assume the presence of the term, but omit them in the subsequent discussion.

In the theory characterized by the action in \eqs{eq:S-G}{eq:S-M}, Planck scale at low energy is determined dominantly by the vacuum expectation value (VEV) of $\chi$, since  $\xi_\alpha |\varphi_{\alpha,0}|^2 \ll M_{\rm P}^2$ for $\varphi_\alpha \neq \chi$ is expected.
It is then given by
\beq
M_{\rm P} \equiv \mathcal{F}_0^{1/2} \simeq \xi_\chi^{1/2} \chi_0 \simeq 2.44 \times 10^{18} {\rm GeV} \ ,
\eeq
where the subscript `$_0$' stands for the low energy VEV of the associated quantity.
The determination of Planck scale implies breaking of scale-invariance.
Since there are no explicit symmetry-breaking terms, this should be achieved by spontaneous symmetry-breaking.
In Ref.~\cite{Ferreira:2016wem}, it was shown that Planck scale can emerge naturally in scale-invariant theories even without CW-type dimensional transmutation.
This is due to the fact that the conserved scale-current has a kernel which eventually settles down to a constant after some general expansion in a very early epoch, at least well before the last about $60$ inflationary $e$-folds.
Scale-anomaly can compromise this picture, but it was shown that, once the renormalization scale, which usually appears as a dimensionful parameter, is replaced by a field, scale symmetry becomes exact, and the existence of a constant kernel remains valid 
\cite{Ferreira:2018itt}.
The constant value of the kernel defines an ellipse in the multidimensional scalar-field space \footnote{
Quantum effects on $\xi_i$ can change the elliptic configuration \cite{Ferreira:2016wem}, but the impact on our discussion is minor.
Hence, we ignore the effects in this work.}, and eventually determines the Planck scale at low energy, once all fields settle down to their true vacuum positions. 
This whole picture is still valid even in the case nonminimal couplings depend on fields in such a way of \eq{eq:xi-alpha}.

Without loss of generality, decomposing the scalar fields as 
\beq
\Phi = \frac{1}{\sqrt{2}} \phi e^{i \theta} = \frac{1}{\sqrt{2}}\l( \phi_r (x) + i \phi_i (x) \r)
\eeq
and $H_i = h_i e^{i \theta_i}/\sqrt{2} \ (i = 1, 2)$ for the two electrically neutral Higgs fields, one can express the scalar potential ${\tilde V}$ as
\bea \label{eq:V-scalar-lowE}
{\tilde V}(\chi, h, \phi) 
%&=& \frac{\lambda_\chi'}{4}  \chi^4 + \frac{\lambda_\phi'}{4} \l( \phi^2 - \zeta_{\phi \chi} \chi^2 \r)^2 + \frac{\lambda_h}{4} \l( \zeta_1 h_1^2 + \zeta_2 h_2^2 + \zeta_{h \phi} \phi^2 - \zeta_{h \chi} \chi^2 \r)^2 
%\nonumber \\
%&-& \frac{1}{4} \lambda_{\phi h} \phi^2 h_1 h_2 \cos \l( 2 \theta - \theta_1 + \theta_2 \r) \ ,
%\nonumber \\
&=& \frac{1}{4} \chi^4 \l[ \lambda_\chi' + \lambda_\phi' \l( {\tilde \phi}^2 - \zeta_{\phi \chi} \r)^2 + \lambda_h \l( \zeta_1 {\tilde h}_1^2 + \zeta_2 {\tilde h}_2^2 - \zeta_{h \phi} {\tilde \phi}^2 - \zeta_{h \chi} \r)^2 
\r.
\nonumber \\
&-& \l. \lambda_{\phi h} {\tilde \phi}^2 {\tilde h}_1 {\tilde h}_2 \cos \l( 2 \theta - \theta_1 + \theta_2 \r) \r] \ ,
\eea
where ${\tilde \phi} \equiv \phi / \chi$, ${\tilde h}_i \equiv h_i/\chi$, and all $\lambda$s and $\zeta$s in \eq{eq:V-scalar-lowE} are assumed to be positive real free parameters.
When the $\xi_\chi$-term in \eq{eq:F-s} dominates the others in $\mathcal{F}$ at low energy, the potential in Einstein frame, $U \equiv {\tilde V} / \Omega^2$ with $\Omega \equiv \mathcal{F}/M_{\rm P}^2$, can be expressed as
\bea \label{eq:VE-scalar-lowE}
U({\tilde h}_i, {\tilde \phi}) 
&\simeq & \frac{M_{\rm P}^4}{4 \xi_\chi^2} \l[ \lambda_\chi' + \lambda_\phi' \l( {\tilde \phi}^2 - \zeta_{\phi \chi} \r)^2 + \lambda_h \l( \zeta_1 {\tilde h}_1^2 + \zeta_2 {\tilde h}_2^2 - \zeta_{h \phi} {\tilde \phi}^2 - \zeta_{h \chi} \r)^2 
\r.
\nonumber \\
&-& \l. \lambda_{\phi h} {\tilde \phi}^2 {\tilde h}_1 {\tilde h}_2 \cos \l( 2 \theta - \theta_1 + \theta_2 \r) \r] \ .
\eea
At the minimum of $U$ satisfying $\partial U/\partial {\tilde h}_i = 0$ for $\tilde{h}_{1,2}$, one finds
\bea \label{eq:hi-vev}
\zeta_1 \tilde{h}_1^2 &=& \zeta_2 \tilde{h}_2^2
= \frac{1}{2} \l[ \zeta_{h \chi} + \l( \zeta_{h \phi} + \frac{\lambda_{\phi h}}{4 \lambda_h \sqrt{\zeta_1 \zeta_2}} \r) \tilde{\phi}^2 \r] \ .
\eea
The VEV of $\tilde{\phi}$ is then found as
\beq
\tilde{\phi}_0^2 \simeq \zeta_{\phi \chi} + \frac{\lambda_{\phi h}}{2 \lambda_\phi' \sqrt{\zeta_1 \zeta_2}} \l( \zeta_{h \phi} + \frac{\lambda_{\phi h}}{8 \lambda_h \sqrt{\zeta_1 \zeta_2}} \r) \simeq \zeta_{\phi \chi} \ ,
\eeq
and the tree-level cosmological constant(CC) term is given by
\bea
U_0 &\simeq& \frac{M_{\rm P}^4}{4 \xi_\chi^2} \l\{ \lambda_\chi' + \lambda_\phi' \zeta_{\phi \chi}^2 \l[ 1 - \frac{1 + \frac{\lambda_{\phi h} \zeta_{h \chi}}{4 \lambda_\phi' \sqrt{\zeta_1 \zeta_2} \zeta_{\phi \chi}}}{1 - \frac{\lambda_{\phi h}}{2 \lambda_\phi' \sqrt{\zeta_1 \zeta_2}} \l( \zeta_{h \phi} + \frac{\lambda_{\phi h}}{8 \lambda_h \sqrt{\zeta_1 \zeta_2}} \r) } \r] \r\}
\nonumber \\
&\simeq& \frac{M_{\rm P}^4}{4 \xi_\chi^2} \l[ \lambda_\chi' - \frac{\lambda_{\phi h}}{2 \lambda_\phi' \sqrt{\zeta_1 \zeta_2}} \l( \zeta_{h \phi} + \frac{\lambda_{\phi h}}{8 \lambda_h \sqrt{\zeta_1 \zeta_2}} \r) \lambda_\phi' \zeta_{\phi \chi}^2 \r]
\eea
There would be quantum corrections to the CC-term, but we will not discuss them, since it can be canceled by a proper choice of $\lambda_\chi'$ and we are not trying to solve the CC-problem in this work.

We may express $U$ at low energy in a form,
\bea \label{eq:V-low-energy}
U(h, \phi) 
&=& U_0 + \frac{\lambda_\phi'}{4} \l( \phi^2 - \phi_{0, \chi}^2 \r)^2 + \frac{\lambda_h}{4} \l( \zeta_1 h_1^2 + \zeta_2 h_2^2 - \zeta_{h \phi} \phi^2 - h_{0,\chi}^2 \r)^2 
\nonumber \\
&-& \frac{1}{4} \lambda_{\phi h} \phi^2 h_1 h_2 \cos \l( 2 \theta - \theta_1 + \theta_2 \r) \ ,
\eea
where  
\bea
\phi_{0, \chi} &\equiv& \sqrt{\zeta_{\phi\chi} \xi_\chi^{-1}} M_{\rm P} \ ,
\\
h_{0,\chi} &\equiv& \sqrt{\zeta_{h\chi} \xi_\chi^{-1}} M_{\rm P} \ .
%\\
%h_0 &=& \sqrt{\zeta_{h\phi} \phi_0^2 + \zeta_{h\chi} \xi_\chi^{-1} M_{\rm P}^2 } \ .
\eea
%Note that, thanks to the Yukawa interaction $y_N$ in \eq{eq:L-ss}, the spontaneous breaking of $U(1)_{\rm PQ}$ caused by a nonzero $\phi_0$ is responsible for Majorana masses of RHN fields.
%For this reason, we will call the associated pseudo-Goldstone boson as \textit{axi-Majoron}.
%
%We will be interested in $\la \phi \ra \simeq \phi_0 = \mathcal{O}(10^{12}) \ {\rm GeV}$ for axion dark matter which will be discussed in Sec. \ref{sec:DM-DR}.
%
Also, \eq{eq:hi-vev} with $\tilde{\phi}$ replaced to $\tilde{\phi}_0$ gives the VEV of $h_i$ as
\beq \label{eq:hi-0}
h_{i,0}^2 = \frac{M_{\rm P}^2}{2 \zeta_i \xi_\chi} \l[ \zeta_{h \chi} + \l( \zeta_{h \phi} + \frac{\lambda_{\phi h}}{4 \lambda_h \sqrt{\zeta_1 \zeta_2}} \r) \tilde{\phi}_0^2 \r] \lesssim \mathcal{O}(10^2) {\rm GeV}
\eeq 
which should satisfy $v_h \equiv \sqrt{h_{1,0}^2 + h_{2,0}^2} \simeq 246 {\rm GeV}$ for the correct electroweak physics.
This implies that the quartic couplings between $\Phi$ and $H_i$ are constrained as $\lambda_{\phi H_i} \lesssim \l( h_{i,0} / \phi_0 \r)^2$.
The radiative corrections to the coupling between $\Phi$ and $H_2$, caused by the couplings $y_N$ and $y_\nu$, are also constrained so as to satisfy
\beq \label{eq:mixing-radiative-corrections}
\mathcal{O}(10^{-2}) \times \l( y_\nu \r)_{\alpha i}^2 y_{N_i}^2 \lesssim \l( \frac{v_h}{\phi_0} \r)^2 \sin^2 \beta
\eeq 
where the subscripts `$_{\alpha i}$' are flavor indices of left-handed and right-handed neutrino species, respectively, $\tan \beta \equiv h_{2,0}/h_{1,0}$, and the mass-matrix of right-handed neutrinos was assumed to be diagonal.
Combined with the seesaw formula in \eq{eq:seesaw-relation}, the condition constrains $y_{N_i}$ as
\beq \label{eq:yN-EWPT-bnd}
y_{N_i} \lesssim \mathcal{O}(10^{-5}) \times \l( \frac{0.05 {\rm eV}}{m_{\nu_\alpha}} \r)^{1/3} \l( \frac{10^{12} {\rm GeV}}{\phi_0} \r) \sin^{\frac{4}{3}} \beta
\eeq
which is relevant for right-handed neutrino flavors associated with the largest $\l( y_\nu \r)_{\alpha i}$.
It can be satisfied either trivially or marginally, depending on other parameters, as will become clear in the subsequent discussion.
Also, if the contributions of $\phi_0$ and $\chi_0$ to $h_{i,0}$ (see \eq{eq:hi-0}) have different signs, the constraint can become weaker.
Hence, we ignore it in the discussion from now on.

In this work, we are interested in the case of $|\Phi| \gtrsim M_{\rm P}$ while $|H_i| \lll M_{\rm P}$ initially some time well before the last about $60$ $e$-folds of the primordial inflation responsible for our Universe\footnote{
We will discuss else where the case where initially $H_1$ and $H_2$ have very large field values during inflation.
}.
In such a circumstance, the two Higgs doublets do not play a critical role in our discussion\footnote{For $\la |H_{1,2}| \ra \ll M_{\rm P}$, the nonminimal gravitational couplings of the two Higgs-doublets can be essentially turned off(see \eq{eq:xi-alpha}).
Then, the electroweak symmetry-breaking is expected to take place as the case in the standard cosmology as long as the final reheating temperature after inflation is higher than the electroweak scale.
}.
They are relevant only for low energy phenomenology which is separable from our discussion in this work.
Also, the details of the vacuum structure of the electroweak sector does not affect our argument.
Hence, we omit Higgs fields in the subsequent discussion except in some relevant places.

In regard of relevant scalar fields and their interactions only, the simplified model is given by the following action:
\bea \label{eq:S-G-sim}
S_{\rm G} &=& - \frac{1}{2} \int d^4 x \sqrt{-{\tilde g}} {\tilde R} \l[ \xi_\chi \chi^2 + 2 \xi_\phi |\Phi| ^2 + \xi_+ \l( \Phi^2 + {\rm c.c.} \r) - i \xi_- \l( \Phi^2 - {\rm c.c.} \r) \r]
\ ,
\\ \label{eq:S-M-sim}
S_{\rm M} &=& \int d^4 x \sqrt{-{\tilde g}} \l[ {\tilde g}^{\mu \nu} \l( \frac{1}{2} \partial_\mu \chi \partial_\nu \chi + \partial_\mu \Phi^\dagger \partial_\nu \Phi \r) + \mathcal{L}_{\rm ss} - {\tilde V} \r] \ ,
\eea
where
\beq \label{eq:V-scalar-sim}
{\tilde V} = \frac{\lambda_\chi}{4} \chi^4 + \lambda_\phi |\Phi|^4  - \frac{1}{2} \lambda_{\chi \phi} \chi^2 |\Phi|^2 \ .
\eeq
Redefining $\Phi$ as
\beq
\Phi \to e^{i \theta_0} \Phi
\eeq
so that 
\beq
\l( \xi_+ - i\xi_- \r) e^{i 2\theta_0} = \sqrt{\xi_+^2 + \xi_-^2} \equiv \xi_a \ ,
\eeq
we can express the action as
\bea \label{eq:S-G-field-redef}
S_{\rm G} &=& - \frac{1}{2} \int d^4 x \sqrt{-{\tilde g}} {\tilde R} M_{\rm P}^2 \Omega \ ,
\\ \label{eq:S-M-can-field}
S_{\rm M} &=& \int d^4 x \sqrt{-{\tilde g}} \l[ \frac{1}{2} {\tilde g}^{\mu \nu} \l( \partial_\mu \chi \partial_\nu \chi +  \sum_a \partial_\mu \phi^a \partial_\nu \phi^a \r) + \mathcal{L}_{\rm ss} - {\tilde V} \r] \ ,
\eea
where in \eq{eq:S-G-field-redef}  the frame-function $\Omega$ is now given by
\beq \label{eq:Omega-org}
\Omega 
\equiv \frac{\xi_\chi \chi^2}{M_{\rm P}^2} + \frac{\xi_\phi \phi^2}{M_{\rm P}^2} \l( 1 + \alpha \cos (2 \theta) \r)
= \frac{\xi_\chi \chi^2}{M_{\rm P}^2} + \frac{\xi_r \phi_r^2}{M_{\rm P}^2} + \frac{\xi_i \phi_i^2}{M_{\rm P}^2}
\eeq
with $\alpha \equiv \xi_a/\xi_\phi$, $\xi_r \equiv \xi_\phi + \xi_a$, $\xi_i \equiv \xi_\phi - \xi_a$, and $\phi^a \equiv \l( \phi_r, \phi_i \r)^T$ in \eq{eq:S-M-can-field}.
We assume
\beq
\xi_r > \xi_i \quad \Leftrightarrow \quad \xi_\phi > \xi_a  \quad \Leftrightarrow \quad \alpha < 1
\eeq
in order for $\Omega$ to be positive definite even in the large field region, $\xi_\phi^{1/2} \phi > M_{\rm P}$, which will be of our interest for a primordial inflation.

Applying the analysis of Ref.~\cite{Ferreira:2016wem} to our scenario, when 
$h_i \lll \chi, \phi_r, \phi_i$ in the very early epoch including inflation era, we are left with a three-dimensional field space $\varphi_\alpha = \l( \chi, \phi_r, \phi_i \r)$. 
%In terms of new field variables defined as \footnote{The index $\alpha$ should not be confused with $\alpha$ defined in \eq{eq:Omega-org}.}
%\beq
%X_\alpha \equiv \xi_\alpha \varphi_\alpha^2 \ ,
%\eeq
%with $\alpha = (1,2,3)$, $\xi_\alpha = \l( \xi_\chi, \xi_r, \xi_i \r)$, and $\varphi_\alpha = \l( \chi, \phi_r, \phi_i \r)$, 
The constant kernel is given by
\beq \label{eq:scale-current-kernel}
K = \frac{1}{2} \sum_\alpha \kappa_\alpha \varphi_\alpha^2 \ ,
\eeq
where $\kappa_\alpha \equiv 1 + 6 \xi_\alpha$ with $\xi_\alpha = \l( \xi_1, \xi_2, \xi_3 \r) =  \l( \xi_\chi, \xi_r, \xi_i \r)$.
It is of the same form as the case when $\xi_\alpha$s are pure numerical constants.
If the initial condition and the dynamics of fields are such that 
\beq \label{eq:cond-for-chi}
\kappa_\chi \chi^2 \gg \kappa_r \phi_r^2 + \kappa_i \phi_i^2
% \ \xrightarrow[]{\xi_\chi \ll 1/6} \ \chi^2 \gg \kappa_r \phi_r^2 + \kappa_i \phi_i^2
\eeq
during the whole evolution of fields, $K$ would always be dominated by the contribution of $\chi$. 
In terms of new field variables defined as
\beq
X_\alpha \equiv \xi_\alpha \varphi_\alpha^2 \ ,
\eeq
the condition \eq{eq:cond-for-chi} corresponds to
\beq \label{eq:cond-for-fixed-planck-scale}
X_1 \gg \frac{\xi_1}{\kappa_1} \l[ \l( \frac{\kappa_2}{\xi_2} \r) X_2 + \l( \frac{\kappa_3}{\xi_3} \r) X_3 \r] \ .
\eeq 
For $\xi_a \ll \xi_\phi$ which will be of our interest, $\xi_2 \approx \xi_3$ and we can take $\xi_\alpha$s such as
\beq \label{eq:xi-cond}
\frac{\xi_1}{\kappa_1} \lll \frac{\xi_2}{\kappa_2} \approx \frac{\xi_3}{\kappa_3} \ \l( < \frac{1}{6} \r) \ ,
\eeq
while \eq{eq:cond-for-fixed-planck-scale} is always satisfied.
Then, one can have 
\beq \label{eq:X-ini}
X_1(0) \ll X_2(0) + X_3(0)
\eeq
as the initial condition for the primordial inflation responsible for our Universe.
Also, as long as $c_\phi \chi \ll \phi$ is satisfied at least during inflation, one can treat $\xi_\phi$ as a constant. 
We assume this is the case with $c_\phi$ chosen properly.
In this case, if ${\tilde V}$ in \eq{eq:V-scalar-sim} is dominated by the $\lambda_\phi \phi^4$-term during inflation, in the slow-roll limit of the primordial inflation, the evolution equation in the Jordan frame is approximated as \footnote{Actually, since the dynamics is constrained to be on the ellipse defined in \eq{eq:scale-current-kernel}, one of the equations in \eq{eq:eom-of-X} is redundant.
Also, in the field-equations there are terms from the field-dependence of $\xi_\alpha$. 
However, those terms are subdominant and have negligible impacts on the field-dynamics.
So, we ignored them.}~\cite{Ferreira:2016wem}
\bea \label{eq:eom-of-X}
\l( \begin{array}{c}
X_1' \\ X_2' \\ X_3'
\end{array} \r)
&=& 8 \l( \frac{\sum_i X_i}{\sum_i \kappa_i X_i} \r) \frac{1}{{\tilde V}_S}
\nonumber \\
&\times&
\l( \begin{array}{cc}
\xi_1 X_1 \l[ \kappa_2 \l( {\tilde V}_{22} + {\tilde V}_{23} \r) + \kappa_3 \l( {\tilde V}_{23} + {\tilde V}_{33} \r) \r]
\\
-\xi_2 \l[ \kappa_1 X_1 \l( {\tilde V}_{22} + \frac{{\tilde V}_{23}}{2} \r) - \kappa_3 \l( X_2 \l( \frac{{\tilde V}_{23}}{2} + {\tilde V}_{33} \r) -  X_3 \l( {\tilde V}_{22} + \frac{{\tilde V}_{23}}{2}  \r)  \r) \r]
\\
-\xi_3 \l[ \kappa_1 X_1 \l( {\tilde V}_{33} + \frac{{\tilde V}_{23}}{2} \r) + \kappa_2 \l( X_2 \l( \frac{{\tilde V}_{23}}{2} + {\tilde V}_{33} \r) -  X_3 \l( {\tilde V}_{22} + \frac{{\tilde V}_{23}}{2}  \r)  \r) \r]
\end{array} \r) \ ,
\eea
where ``$\prime$'' denotes the derivative with respect to $N = \ln a(t)$, and
\beq
{\tilde V}_{22} \equiv \frac{\lambda_\phi}{4} \phi_r^4, \ {\tilde V}_{23} \equiv \frac{\lambda_\phi}{2} \phi_r^2 \phi_i^2, \ {\tilde V}_{33} \equiv \frac{\lambda_\phi}{4} \phi_i^4 \ .
\eeq
Note that we have ${\tilde V}_{ij} / {\tilde V} \lesssim 1$ for $\phi_r \gtrsim \phi_i$.
Hence, from \eq{eq:eom-of-X}, it is clear that, for $\xi_1 \lll 1/8$ with the initial condition satisfying \eqs{eq:cond-for-fixed-planck-scale}{eq:X-ini}, $X_1$ barely changes during the dynamics of $X_{2,3}$ moving toward minima along each direction.
In such a circumstance, if $X_3=0$ as a simple possibility, the EOM of $X_2$ can be trivially integrated as 
\beq \label{eq:X2-sol}
X_2(N) - X_2(0) + \frac{\kappa_1-\kappa_2}{\kappa_2} X_1(0) \ln \l[ \frac{X_1(0)+X_2(N)}{X_1(0)+X_2(0)} \r] = - 8 \xi_2 \l( \frac{\kappa_1}{\kappa_2} \r) X_1(0) N \ .
\eeq
It is valid for $X_2(N) \gtrsim X_1(0)$ with the change of $X_1$ ignored, and gives $e$-folds as
\beq \label{eq:N-efolding-for-X2}
N \approx \frac{\kappa_2}{8 \xi_2 \kappa_1} \frac{X_2(0) - X_2(N)}{X_1(0)} 
\ \xrightarrow[]{X_2(0) \gg X_2(N)} \ \frac{\kappa_2}{8 \xi_2 \kappa_1} \frac{X_2(0)}{X_1(0)} 
> \frac{3}{4} \frac{X_2(0)}{X_1(0)} \ . 
\eeq
Hence, one can obtain sufficiently large amount of $e$-folds if $X_2(0) \gg X_1(0)$.
Note that, since $X_1$ barely changes for a sufficiently small $\xi_1$ while $X_{2,3}$ becomes negligible in the end, it is possible to assume that Planck scale is essentially fixed with a value given by
\beq
M_{\rm P} \simeq \sqrt{\xi_\chi} \chi_0 \simeq \sqrt{ \frac{2 \xi_\chi K}{1 + 6 \xi_\chi} } \ .
\eeq
Then, the physics becomes the same as that of nonminimal gravitational couplings of scalar fields other than $\chi$ with the fixed Planck scale. 
We take this approximation in the subsequent discussion to describe the dynamics $X_{2,3}$ in the more transparent and clear, conventional way.

\begin{figure}[h] 
\begin{center}
\includegraphics[width=0.48\textwidth]{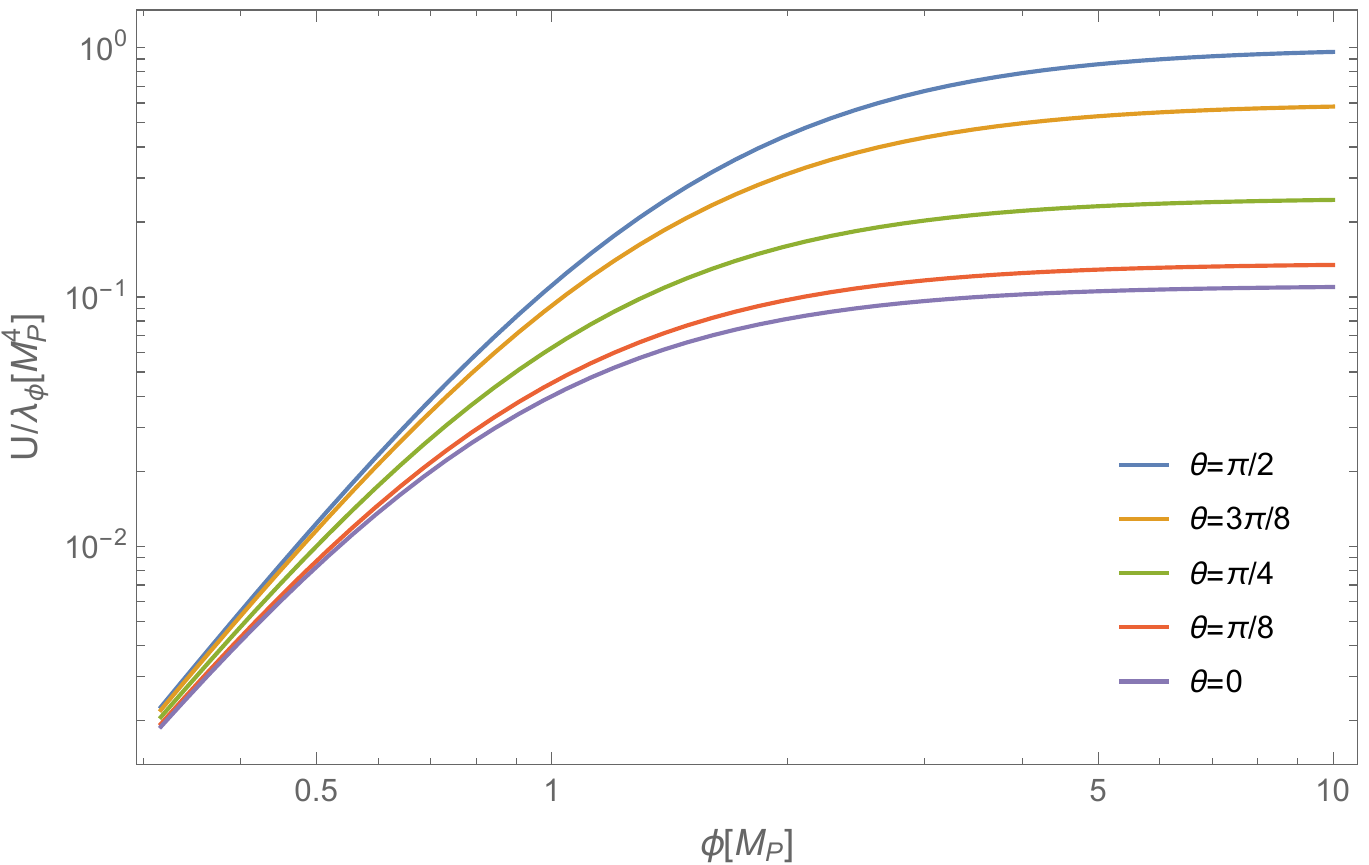}
\hspace{1em}
\includegraphics[width=0.48\textwidth]{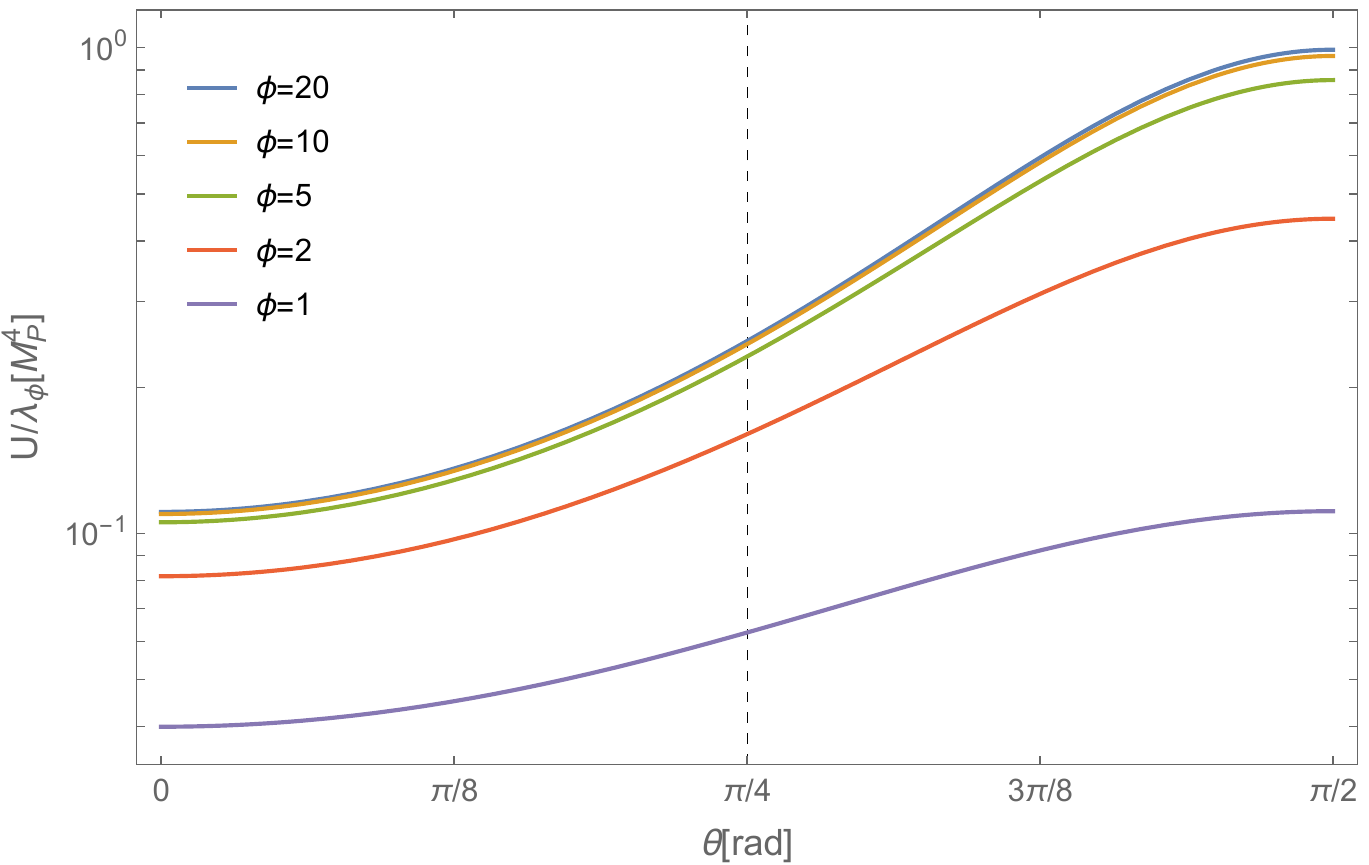}
\caption{
Potential energy density $U$ with $\xi_\phi=1$, $\lambda_\phi = 6 \times 10^{-10}$, and $\alpha = 0.5$ as a function of $\phi$ for various $\theta$ (left), and as a function of $\theta$ for various $\phi$ (right).
}
\label{fig:U-plot}
\end{center}
\end{figure}

Once the Planck scale is (nearly) fixed in this way, we can follow the conventional way of converting the Jordan frame to the Einstein frame.
The frame-function in \eq{eq:Omega-org} is now expressed as
\beq
\Omega 
= 1 + \frac{\xi_\phi \phi^2}{M_{\rm P}^2} \l( 1 + \alpha \cos (2 \theta) \r)
= 1 + \frac{\xi_r \phi_r^2}{M_{\rm P}^2} + \frac{\xi_i \phi_i^2}{M_{\rm P}^2} \ .
\eeq
Under a Weyl rescaling, 
\beq
g_{\mu \nu} = \Omega {\tilde g}_{\mu \nu} 
\eeq
with $g_{\mu \nu}$ the metric in the Einstein frame, the action of the gravity part obtains a form in the Einstein gravity \cite{Wald:1984rg},
\beq
S_{\rm G} = - \frac{1}{2} \int d^4 x \sqrt{-g} M_{\rm P}^2 \l[ R - \frac{3}{2} g^{\mu \nu} \l( \partial_\mu \ln \Omega \r) \l( \partial_\nu \ln \Omega \r) \r] \ , 
\eeq
while the action of the matter sector with scalar fields only is given by
\beq
S_{\rm M} = \int d^4 x \sqrt{-g} \l\{ \frac{1}{2 \Omega} \l[ \sum_a \l( \partial \phi^a \r)^2 \r] - \frac{{\tilde V}}{\Omega^2} \r\}
\eeq 
The action for the scalar fields can now be written as 
\beq
S_{\rm scalar} = \int d^4 x \sqrt{-g} \l[ \frac{1}{2} \mathcal{K}_{ab} g^{\mu \nu} \partial_\mu \phi^a \partial_\nu \phi^b - U \r] \ ,
\eeq
where the symmetric field-space metric $\mathcal{K}_{ab}$ in the basis of $\phi^a = \l( \phi_r, \phi_i \r)^T$ is given by
\beq
\mathcal{K}_{ab} = \Omega^{-2} \l[ \Omega \delta_{ab} + \frac{3M_{\rm P}^2}{2} \frac{\partial \Omega}{\partial \phi^a} \frac{\partial \Omega}{\partial \phi^b} \r] \equiv \Omega^{-2} K_{ab} \ .
\eeq
At low energy with $\phi \lll M_{\rm P}$, $\Omega \to 1$ and the kinetic terms return to the canonical form.
The shape of the potential $U$ is shown in Fig.~\ref{fig:U-plot}.
%%%%%%%%%%%%%%
%%%%%%%%%%%%%%%

The nonminimal symmetry-breaking gravitational interaction, $\xi_a$-term, contributes to the mass of the axi-Majoron, $a_\phi \equiv \theta \phi$, the angular mode of $\Phi$ field for a given nonzero $\phi$.
If $\xi_\phi$ were a constant, in the Einstein frame, the mass-squared of axi-Majoron at low energy obtains a tree-level contribution given by\footnote{The low-energy definition of axi-Majoron differs due to the symmetry-breaking nonzero VEVs of the PQ-charged two Higgses, but such a change does not affect our argument.
}
\beq \label{eq:m-a}
\Delta m_{a, \rm tree}^2 \equiv \frac{\partial^2 U}{\phi^2 \partial \theta^2} = 8 \alpha \xi_\phi \frac{U}{M_{\rm P}^2} \ .
\eeq
Although the contribution could be sizable relative to the genuine QCD-contribution, it does not disturb the axion solution, since $U \to 0$ at low energy.
However, the symmetry-breaking terms play the key role for baryogenesis after inflation, as discussed in Sec.~\ref{sec:lepto}.

Meanwhile, the exchange of massless graviton causes interactions such as $\mathcal{F} \partial^2 \mathcal{F}$ and $\mathcal{F} \sum\limits_a \l( \partial \varphi^a \r)^2$ \cite{Hill:2020oaj}.
Those interactions result in a loop-induced additional axi-Majoron potential other than the QCD-contribution\footnote{We thank to Hyun Min Lee for pointing out this effect.}.
The axi-Majoron mass from the contribution is  
\bea \label{eq:Delta-ma-loop}
\Delta m_{a, \rm loop} 
&=& \mathcal{O}(10^{-1}) \times \alpha^{1/2} \xi_\phi \lambda_\phi \l( \frac{\phi_0^2}{M_{\rm P}} \r) \ln^{1/2} \l( \frac{\chi^2}{m_\phi^2 } \r)
\nonumber \\
&=& \mathcal{O}(1) \times \xi_\phi  \l( \frac{\alpha}{10^{-4}} \r)^{1/2} \l( \frac{\lambda_\phi}{6 \times 10^{-10}} \r) \l( \frac{\phi_0}{10^{12} {\rm GeV}} \r)^2 \ {\rm keV} \ ,
\eea 
where we took $\chi = 100 M_{\rm P}$ for the numerical estimation in the second line.
We will be interested in $\xi_\phi \gtrsim \mathcal{O}(1)$ during inflation (see Sec.~\ref{sec:inf}).
\eq{eq:Delta-ma-loop} implies that, if $\xi_\phi \gtrsim \mathcal{O}(1)$ even at low energy, the axion solution to the strong \textit{CP}-problem can work only if $\alpha \ll \mathcal{O}(10^{-20})$ which is too small for baryogenesis in our scenario. 
This danger is remedy thanks to the field-dependence of $\xi_\phi$ as assumed in \eq{eq:xi-alpha}.
The coupling is essentially turned off at low energy due to the exponential field-dependence.
Hence, the symmetry-breaking nonminimal couplings do not cause the axion-quality problem in our scenario.

\section{Inflation along $\Phi$ direction}
\label{sec:inf}

We consider an initial condition, $h \ll M_{\rm p} \lesssim \phi$.
Consistency with our assumption discussed in the previous section, i.e., 
\beq
\kappa_r \phi_{r, \rm ini}^2 + \kappa_i \phi_{i, \rm ini}^2 \ll \kappa_\chi \chi_{\rm ini}^2
\eeq
with $M_{\rm P} \simeq \sqrt{\xi_\chi} \chi_{\rm ini}$ in this analysis constrains the initial position of $\phi$ as 
\beq
\sqrt{\xi_\phi} \phi_{\rm ini} \ll \sqrt{
\frac{1+6 \xi_\chi}{\xi_\chi} \times \frac{\xi_\phi}{1 + 6 \xi_\phi \l( 1 + \alpha \cos \l( 2 \theta \r) \r)}
}
M_{\rm P}
\ \xrightarrow[]{\xi_\chi \lll 1 \ \& \ \xi_\phi \gg 1} \ \frac{M_{\rm P}}{\sqrt{6 \xi_\chi}} \ .
\eeq 
Therefore, for a sufficiently small $\xi_\chi$ with $\xi_\phi \gtrsim \mathcal{O}(1)$, it is possible to have $\sqrt{\xi_\phi} \phi_{\rm ini} \gg M_{\rm P}$ while Planck scale is nearly fixed, and we assume this is the case.

In this work, we will be interested in $\xi_{\phi,0} = \mathcal{O}(1-10^2)$, concerning various cosmological aspects of this model.
Hence, if $c_\phi \chi_0 \ll M_{\rm P}$ in \eq{eq:xi-alpha}, the unitarity-cutoff might be slightly lower than the Planck scale (e.g., $\Lambda_{\rm cut} \sim M_{\rm P}/\sqrt{\xi_{\phi,0}}$ \cite{Burgess:2009ea,Burgess:2010zq,Barbon:2009ya}), but it is well above any energy and mass scales involved 
in the field dynamics.
Also, even though the field excursion during inflation spans over the Planck scale as will be discussed subsequently, it may present no problem as long as scale-invariance is preserved.

In the large field region around or over the Planck scale, the kinetic terms of the scalar fields $\phi_r$ and $\phi_i$ in Einstein frame are neither canonical nor diagonal.
In order to analyze the dynamics of fields in such a curved field-space, we resort to the conventional method for multifield inflation in a curved field-space \cite{Sasaki:1995aw,GrootNibbelink:2001qt,Gordon:2000hv}.
From now on we work in the basis of $\phi^{a=1,2} \equiv (\phi_r, \phi_i)^T$ unless otherwise stated.
We define the Levi-Civita connection, covariant field-derivative $\nabla_a$, and covariant time-derive in the field space as
\bea
\Gamma_{ab}^c &=& \frac{1}{2} \mathcal{K}^{cd} \l( \mathcal{K}_{da, b} + \mathcal{K}_{db, a} - \mathcal{K}_{ab, d} \r) \ ,
\\ 
\nabla_a \phi^c &=& \partial_a \phi^c + \Gamma_{ab}^c \phi^b \ ,
\\
\nabla_t &\equiv& \dot{\phi}^a \nabla_a \ ,
\eea
where $\partial_a \equiv \partial / \partial \phi^a$, $\mathcal{K}^{ab}$ is the inverse of $\mathcal{K}_{ab}$, and $\mathcal{K}_{ab, c} \equiv \partial \mathcal{K}_{ab} / \partial \phi^c$.
Each component of the metric can be found in a straightforward manner and given in Appendix~\ref{appen:metric-compo} in terms of $K_{ab}$ for convenience.
The speed of the inflaton is given by $\dot{\phi}_I \equiv \sqrt{ \mathcal{K}_{ab} \dot{\phi}^a \dot{\phi}^b}$.
Then, the vectors tangent and normal to the trajectory of the field dynamics are respectively given by
\bea
T^a &=& \frac{\dot{\phi}^a}{\dot{\phi}_I}  
\\
N_a &=& \sqrt{\mathcal{K}} \epsilon_{ab} T^b
\eea
where $\mathcal{K}$ is the determinant of the metric $\mathcal{K}_{ab}$, and $\epsilon_{ab}$ is the antisymmetric Levi-Civita tensor.

\subsection{Dynamics of the homogeneous mode}
For the flat FLRW metric, $g_{\mu \nu} = {\rm diag}\l( 1, -a^2(t), -a^2(t), -a^2(t) \r)$, the equations of motions of the background fields are given by
\beq \label{eq:phi-vec-eom}
0 = \nabla_t \dot{\phi}^a + 3H \dot{\phi}^a + \mathcal{K}^{ab} U_b \ ,
\eeq
where $U_a \equiv \partial U / \partial \phi^a$, and Friedman equation is given by
\beq
3 H^2 M_{\rm P}^2 = \frac{1}{2} \mathcal{K}_{ab} \dot{\phi}^a \dot{\phi}^b + U \ .
\eeq
Contracting the equation of motion of the field vector $\phi^a$ in \eq{eq:phi-vec-eom} with either $T_a$ or $N_a$, one obtains the equations of motions along the inflaton trajectory and orthogonal to that as
\bea
0 &=&\ddot{\phi}_I + 3H \dot{\phi}_I + U_T
\\
\nabla_t T^a &=& - \frac{U_N N^a}{\dot{\phi}_I} \equiv - \dot{\theta}_T N^a \ ,
\eea
where $U_T \equiv T^a U_a$, $U_N \equiv N^a U_a$, and $\dot{\theta}_T \equiv U_N/\dot{\phi}_I$ is the turning rate of the inflaton trajectory.

One can solve numerically these equations of motion of fields straightforwardly.
However, it is instructive to see a limiting case which can be proven simpler to understand.
For that, one can take a look at the case of $\alpha \lll 1$ for which approximate analytic expressions can be obtained as follows.
The frame function in $\l( \phi, \theta \r)$-basis is approximated as
\beq \label{eq:Omega-small-alpha}
\Omega \approx 1 + \frac{\xi_\phi \phi^2}{M_{\rm P}^2} \ .
\eeq
The field-space metric is approximated as
\bea
K_{11} &\approx& 1 + \frac{\xi_\phi \l( 1 + 6 \xi_\phi \r) \phi^2}{M_{\rm P}^2} \ ,
\nonumber \\
K_{22} &\approx& \phi^2 + \frac{\l[ 1 + 6 \xi_\phi \l( \alpha \cos (2 \theta) \r)^2 \r] \xi_\phi \phi^4}{M_{\rm P}^2} \ ,
\nonumber \\
K_{12} &\approx& 6 \l( \frac{\xi_\phi \phi}{M_{\rm P}} \r)^2  \phi \alpha \cos \l( 2 \theta \r) \ll 1 \ .
\eea
In order to have a clear picture, let us ignore the mixing term $K_{12}$ for the moment\footnote{This term, however, is included in numerical analysis.}.
Then, the canonically normalized fields can be defined as
\bea
d \varphi_1 &\equiv& \Omega^{-1} \sqrt{1 + \frac{\xi_\phi \l( 1 + 6 \xi_\phi \r) \phi^2}{M_{\rm P}^2}} d\phi \ ,
\\
d \varphi_2 &\equiv& \Omega^{-1/2} \phi d \theta \ .
\eea
Note that, as expected, the field $\varphi_1$ plays the same role as the canonically normalized inflaton of the SM Higgs inflation \cite{Bezrukov:2007ep}.
In the large field region, $\xi_\phi^{1/2} \phi \gg M_{\rm P}$, we have
\bea
\phi &\simeq& \frac{M_{\rm P}}{\sqrt{\xi_\phi}} \exp \l[ \sqrt{\frac{\xi_\phi}{1+6\xi_\phi}} \frac{\varphi_1}{M_{\rm P}} \r] \ ,
\\
\theta &\simeq& \frac{\varphi_2}{M_{\rm P}/\sqrt{\xi_\phi}} \ ,
\eea
where $\varphi_2$ should be regarded as a periodic field, i.e.,
\beq
\varphi_2 = \varphi_2 + 2\pi \l( \frac{M_{\rm P}}{\sqrt{\xi_\phi}} \r) \ .
\eeq 
In terms of these approximate canonical fields, the potential is given by
\bea \label{eq:U-approx}
U(\varphi_1,\varphi_2) 
&=& \frac{\lambda_\Phi M_{\rm P}^4}{4 \xi_\phi^2}  \l[ 1  +  \alpha \cos \l( \frac{2 \sqrt{\xi_\phi} \varphi_2}{M_{\rm P}} \r) + \exp\l[ - 2 \sqrt{\frac{\xi_\phi}{1+6\xi_\phi}} \frac{\varphi_1}{M_{\rm P}} \r] \r]^{-2} \ .
\eea
For $\alpha \ll\ 1$, it is essentially the same as the potential of the SM Higgs-inflation \cite{Bezrukov:2007ep} with the expansion rate of inflation given by
\beq \label{eq:H-inflation}
H_I \simeq \frac{\lambda_\Phi^{1/2} M_{\rm P}}{2 \sqrt{3} \xi_{\phi,0}}
\eeq
in the large field region.

\begin{figure}[h] 
\begin{center}
\includegraphics[width=0.32\textwidth]{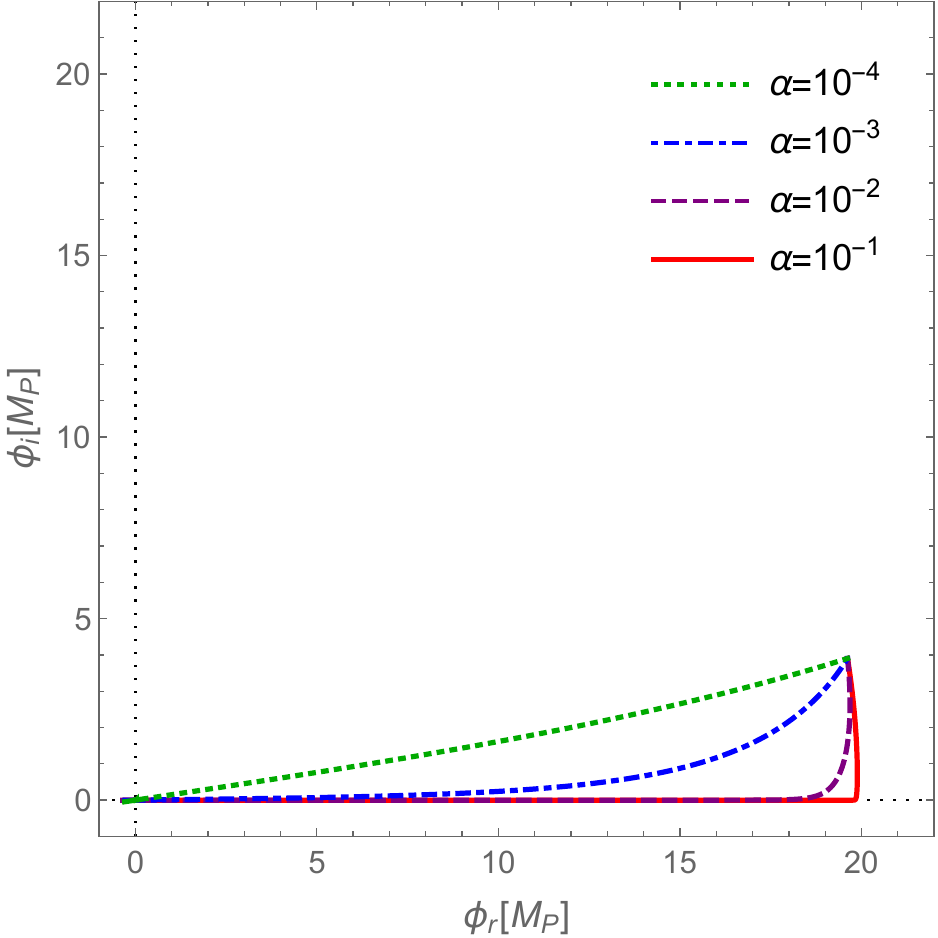}
\includegraphics[width=0.32\textwidth]{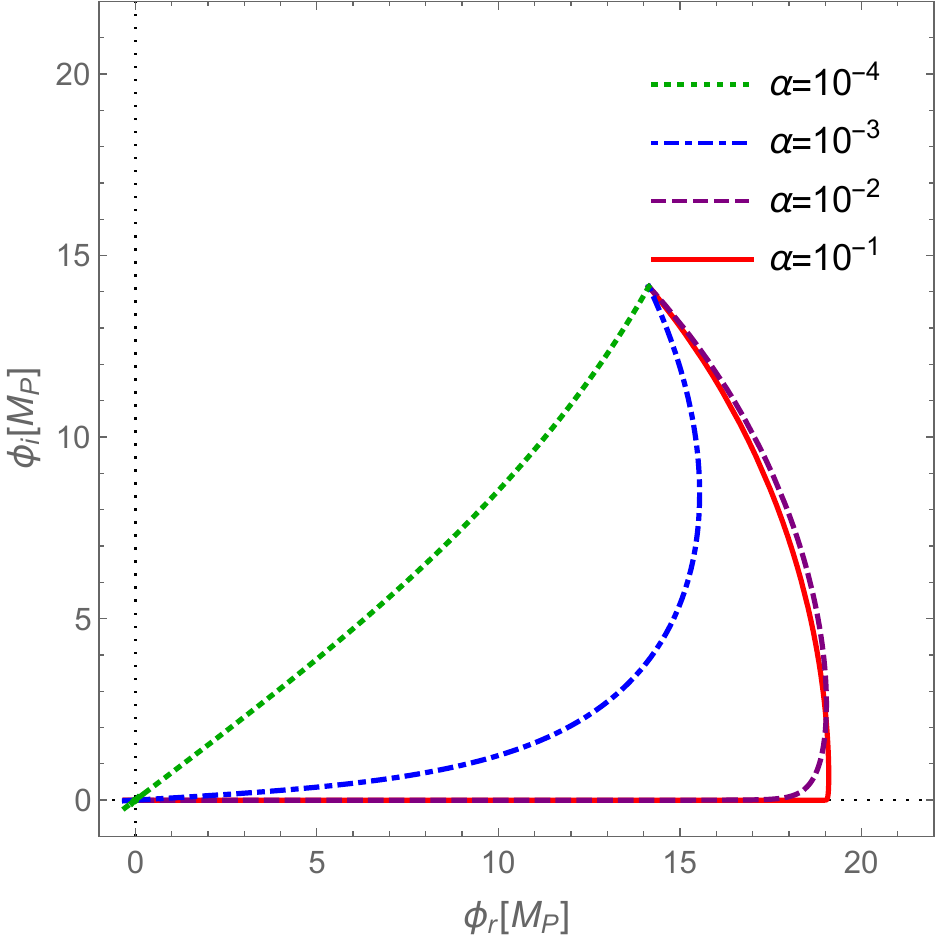}
\includegraphics[width=0.32\textwidth]{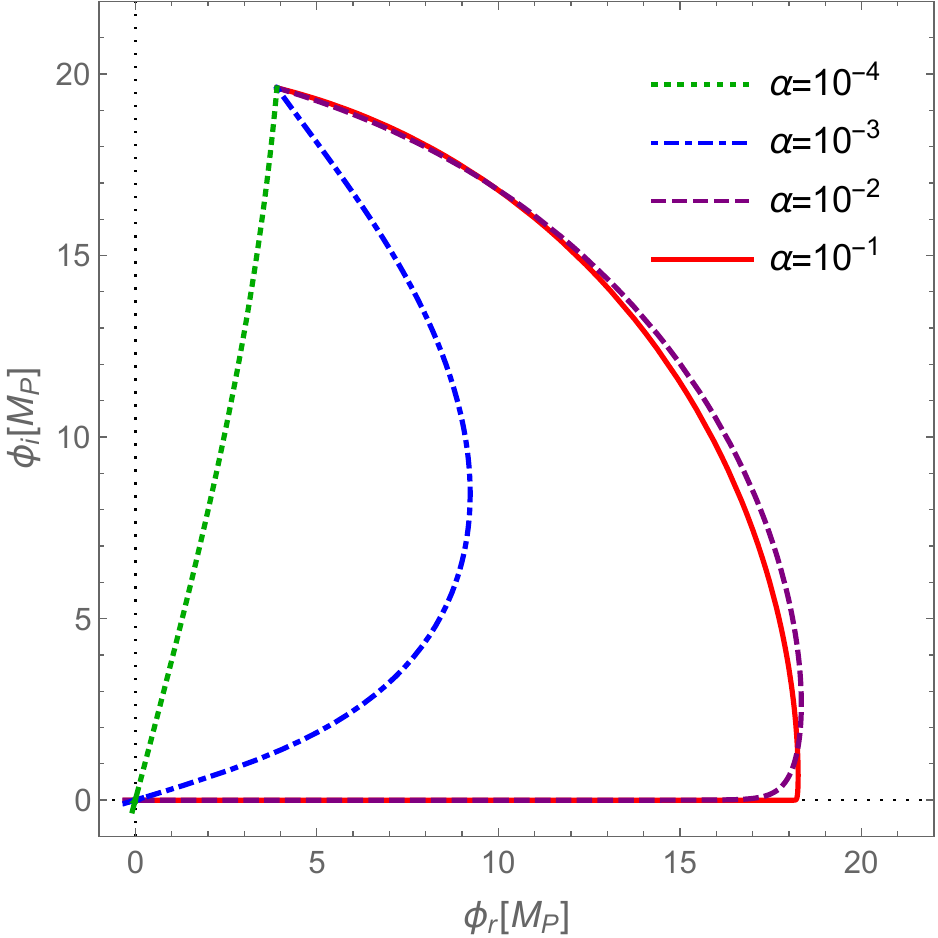}
\caption{
Trajectories of the field dynamics for $\alpha = 10^{-4},10^{-3}, 10^{-2}, 10^{-1}$ from left to right with $\xi_{\phi,0}=1$, $\lambda_\phi=6\times10^{-10}$, $\phi_{\rm ini}=20 M_{\rm P}$, and $\theta_{\rm ini}=\pi/16, \pi/4, 7\pi/16$(from left to right).
}
\label{fig:Path-vs-alpha}
\end{center}
\end{figure}

Meanwhile, from the derivatives along $\varphi_2$ direction, one finds (see Appendix~\ref{appen:U-deriv})
\bea
\frac{M_{\rm P}}{U} \frac{\partial U}{\partial \varphi_2} &=& 4 \sqrt{\xi_\phi} \alpha \sin \l( \frac{2 \sqrt{\xi_\phi} \varphi_2}{M_{\rm P}} \r) \ ,
\\ \label{eq:angular-curvature}
\frac{M_{\rm P}^2}{U} \frac{\partial^2 U}{\partial \varphi_2^2} &=& 8 \xi_\phi \alpha \cos \l( \frac{2 \sqrt{\xi_\phi} \varphi_2}{M_{\rm P}} \r) \ ,
\\ 
\label{eq:radial-angular-mixing}
\frac{M_{\rm P}^2 }{U} \frac{\partial^2 U}{\partial \varphi_1\partial \varphi_2} &=& \frac{\frac{12 \xi_\phi}{\sqrt{1+6\xi_\phi}} \alpha \exp\l[ 2 \sqrt{\frac{\xi_\phi}{1+6\xi_\phi}} \frac{\varphi_1}{M_{\rm P}} \r] \sin \l( \frac{2 \sqrt{\xi_\phi} \varphi_2}{M_{\rm P}} \r) }{\l[ 1 + \l( 1  +  \alpha \cos \l( \frac{2 \sqrt{\xi_\phi} \varphi_2}{M_{\rm P}} \r) \r) \exp\l[ 2 \sqrt{\frac{\xi_\phi}{1+6\xi_\phi}} \frac{\varphi_1}{M_{\rm P}} \r] \r]^2} \ .
\eea
From \eq{eq:angular-curvature}, the mass-squared along $\varphi_2$-direction is found to be 
\beq \label{eq:m-sq-varphi2}
m_{\varphi_2}^2 \approx 24 \alpha \xi_\phi H_I^2 \cos \l( \frac{2 \sqrt{\xi_\phi} \varphi_2}{M_{\rm P}} \r)
\eeq
Hence, if $\alpha \sim \alpha_{\rm c}$ with
\beq
\alpha_{\rm c} \equiv 1/(24 \xi_{\phi,0}),
\eeq
the mass along $\varphi_2$ direction (i.e., angular direction in the complex field space of $\Phi$) is comparable to the expansion rate given in \eq{eq:H-inflation}.
This implies that, in the large field region where the potential along the radial direction (i.e., along $\varphi_1$ or $\phi$) becomes quite flat, if $\alpha \gtrsim \alpha_{\rm c}$, the dynamics of field would be initially along $\varphi_2$ toward the vicinity of $\varphi_2 \sim 0$ and afterward moves nearly along $\varphi_1$ toward the origin, as will be shown by numerical analysis.
On the other hand, if $\alpha \lll \alpha_{\rm c}$, the trajectory will be nearly radial toward the origin.
That is, the trajectory of the field would look like one of the following three cases in $(\phi_r, \phi_i)$ space:
\begin{itemize}
\item{C-I} : $\sqrt{|m_{\varphi_2}^2|} \gtrsim H_I$ ($\Leftrightarrow \alpha \gtrsim \alpha_{\rm c} $) - $\varphi_2 \to 0$ (angular motion) and then $\varphi_1 \to 0$ (radial motion).
\item{C-II} : $|m_{\varphi_2}^2| \sim |m_{\varphi_1}^2|$ ($ \Leftrightarrow \alpha \ll \alpha_{\rm c}$) - a curvilinear trajectory (mixture of angular and radial motion).
\item{C-III} : $|m_{\varphi_2}^2| \ll |m_{\varphi_1}^2|$ ($\Leftrightarrow \alpha \lll \alpha_{\rm c}$) - a nearly radial trajectory.
\end{itemize}
In Fig.~\ref{fig:Path-vs-alpha}, trajectories of field dynamics are shown for different values of $\alpha$ with a fixed initial position as an example.
%%%%%%%%%%%%%%
%%%%%%%%%%%%%%%
Note that the angular motion in the case of $\alpha = 0.1$ is closer to the origin relative to the case of $\alpha=10^{-2}$. 
This is because, as $\alpha$ becomes sufficiently large, the effect of the mixing between $\varphi_1$ and $\varphi_2$ becomes significant, if $\theta_{\rm ini} = \mathcal{O}(1)$ as inferred in \eq{eq:radial-angular-mixing}

\begin{figure}[h] 
\begin{center}
\includegraphics[width=0.48\textwidth]{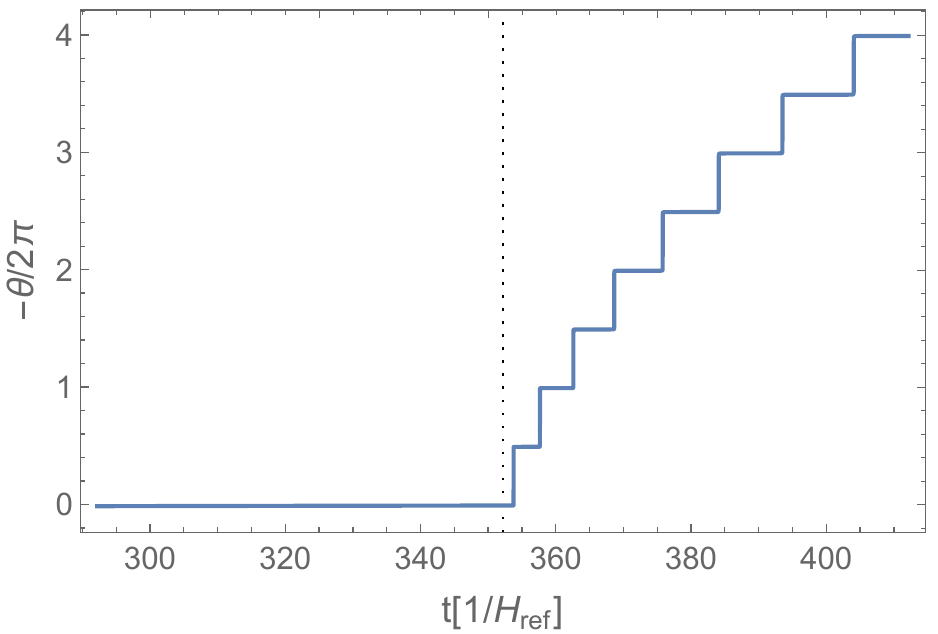}
\hspace{1em}
\includegraphics[width=0.48\textwidth]{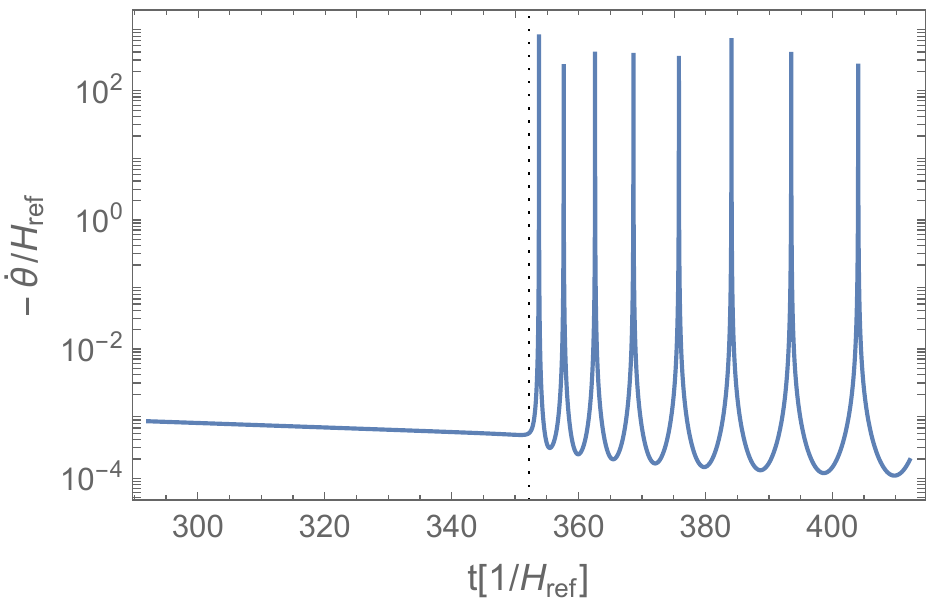}
\caption{
The time dependence of the phase $\theta$ (left) and its time-derivative (right) for $\alpha = 10^{-3}$ with $\xi_{\phi,0}=1$, $\lambda_\phi=6 \times 10^{-10}$, $\phi_{\rm ini}=20 M_{\rm P}$, and $\theta_{\rm ini}=\pi/4$.
$H_{\rm ref} \equiv \sqrt{U(\phi_{\rm ini}, \theta_{\rm ini})}/\sqrt{3} M_{\rm P} \approx H_I$.
The vertical dotted line indicates the end of inflation.}
\label{fig:angular-motion}
\end{center}
\end{figure}

As it can be seen from the dynamics of the inflaton with $\theta_{\rm ini} \neq 0$, a salient and interesting feature of our hybrid axi-Majoron inflation is that the PQ field $\Phi$ can have angular motion even after inflation, if $\alpha$ is neither too large nor too small.
Figure \ref{fig:angular-motion} shows the time-dependence of $\theta$ and its time-derivative.
%%%%%%%%%%%%%%
%%%%%%%%%%%%%%%
The angular momentum in a comoving volume is conserved after inflation, since the symmetry-breaking effect becomes negligible as $\phi$ becomes smaller than the Planck scale.
This corresponds to the asymmetry of $U(1)_{\rm PQ}$-charged particle $\Phi$ as in the case of Affleck-Dine mechanism \cite{Affleck:1984fy} \footnote{In Ref.~\cite{Co:2019wyp}, generation of Peccei-Quinn number asymmetry and subsequent baryogenesis was discussed, but in a different realization.}.
The asymmetry can be transferred to the visible sector, as the baryon number asymmetry, via the usual way of Affleck-Dine mechanism and/or the mechanism of spontaneous baryogenesis \cite{Cohen:1987vi}.
More details will be discussed in Sec.~\ref{sec:lepto}.

\subsection{Perturbations}
For the estimations of perturbations\footnote{There are perturbations of $\chi$ field too. However, for $\xi_\chi \ll \xi_{r,i}$ the background dynamics is nearly along $\Phi$ with barely changing $\chi$.
In addition, in a simplified two field case with negligible kinetic mixing which is the case of $\xi_\chi \lll 1$, one can see that the growing rate of isocurvature perturbation is negative (i.e., suppression instead of growth) in the slow-roll and slow-turn limit \cite{Peterson:2010np}.
Hence, we can ignore the perturbations of $\chi$ in our discussion. 
}, we take the Newtonian gauge in an isotropic universe for the metric perturbations, i.e., the perturbed metric is given by
\beq
ds^2 = \l( 1 + 2 \psi \r) dt^2 - a(t)^2 \l( 1 - 2 \psi \r) \delta_{ij} dx^i dx^j \ .
\eeq
In terms of the Mukhanov-Sasaki variable for gauge invariant perturbations \cite{Sasaki:1986hm,Mukhanov:1988jd}, 
\beq
Q^a \equiv \delta \phi^a + \frac{\dot{\phi}^a}{H} \psi
\eeq
with $Q_T \equiv T_aQ^a$ and $Q_N \equiv N_aQ^a$, adiabatic and isocurvature perturbations can be defined as
\bea
\mathcal{R} &\equiv& \frac{H}{\dot{\phi}_I} Q_T \ ,
\\
\mathcal{S} &\equiv& \frac{H}{\dot{\phi}_I} Q_N \ .
\eea
Also, one can define slow-roll functions as \cite{GrootNibbelink:2000vx,GrootNibbelink:2001qt}
\bea
\epsilon &\equiv& - \frac{\dot{H}}{H^2} = \frac{\dot{\phi}_I^2}{2 H^2 M_{\rm P}^2}
\\
\eta^a &\equiv& - \frac{\nabla_t \dot{\phi}^a}{H \dot{\phi}_I} = \eta_T T^a + \eta_N N^a
\eea
with
\bea \label{eq:etaT}
\eta_T &\equiv& T_a \eta^a = - \frac{\ddot{\phi}_I}{H \dot{\phi}_I} \ ,
\\ \label{eq:etaN}
\eta_N &\equiv& N_a \eta^a = \frac{U_N}{H \dot{\phi}_I} = \frac{\dot{\theta}_T}{H} \ .
\eea
They are related to the conventional slow-roll parameters defined in terms of the potential in single field scenarios as
\bea
\epsilon &\simeq& \epsilon_{U_T}
\\
\eta_T &\simeq& \eta_{U_T} - \epsilon_{U_T}
\eea
where $\epsilon_{U_T} \equiv \l( M_{\rm P} U' /U \r)^2/2$ and $\eta_{U_T} \equiv M_{\rm P}^2 U'' / U$ with the derivatives taken with respect to the inflaton field.

The quadratic order action for the perturbations is given by \cite{Achucarro:2012yr}
\bea
S_2 &=& \frac{1}{2} \int d^4 x \sqrt{-g} \l[ \frac{\dot{\phi}_I^2}{H^2} g^{\mu\nu} \l( \partial_\mu \mathcal{R} \r) \l(\partial_\nu \mathcal{R} \r) + g^{\mu \nu} \l( \partial_\mu Q_N \r) \l( \partial_\nu Q_N \r) \r.
\nonumber \\
&& \l. - m_{\rm eff}^2 Q_N^2 + 4 \dot{\phi}_I \eta_N \dot{\mathcal{R}} Q_N \r] \ .
\eea
It gives the EOMs of the perturbations as
\bea \label{eq:R-eom}
\ddot{\mathcal{R}} + \l( 3 + 2 \epsilon - 2 \eta_T \r) H \dot{\mathcal{R}} + \frac{k^2}{a^2} \mathcal{R}
&=&
-2 \dot{\theta}_T \frac{H}{\dot{\phi}_I} \l[ \dot{Q}_N + \l( 3 - \eta_T - 2\epsilon + \frac{\dot{\eta}_N}{H \eta_N} \r)HQ_N \r]
\\ \label{eq:QN-eom}
\ddot{Q}_N + 3H\dot{Q}_N + \frac{k^2}{a^2} Q_N + m_{\rm eff}^2 Q_N &=& 2 \dot{\theta}_T \frac{\dot{\phi}_I}{H} \dot{\mathcal{R}}
\eea
where $k$ is the comoving wave number associated with cosmological scales.
The effective mass of the fluctuations orthogonal to the inflaton direction is given by 
\beq
m_{\rm eff}^2 = m^2 - \dot{\theta}_T^2 \ ,
\eeq
where
\beq \label{eq:m-sq-orthogonal}
m^2 = U_{NN} + \epsilon H^2 M_{\rm P}^2 \mathbb{R}
\eeq
with $U_{NN} \equiv N^a N^b \nabla_a \nabla_b U$ and $\mathbb{R}$ being the Ricci scalar associated with the field-space metric $\mathcal{K}_{ab}$.

For numerical analysis, we took $\xi_{\phi,0}=1$, $\alpha = 10^{-3}$, $\lambda_\phi=6 \times 10^{-10}$, $\phi_{\rm ini}=20 M_{\rm P}$, and $\theta_{\rm ini}=\pi/4$ as an illustrative set of parameters.
%%%%%%%%%%%%%%
 \begin{figure}[h] 
\begin{center}
\includegraphics[width=0.48\textwidth]{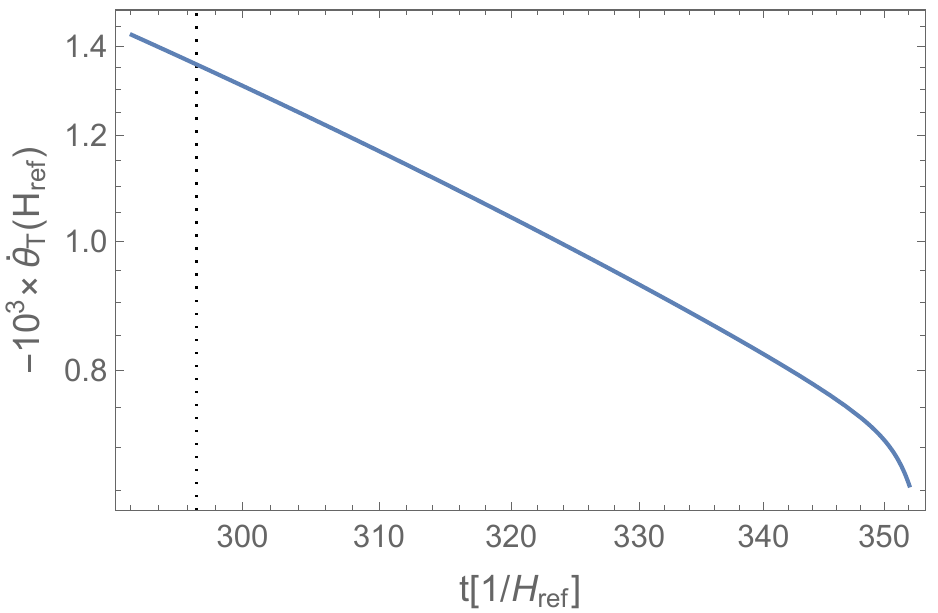}
\hspace{1em}
\includegraphics[width=0.48\textwidth]{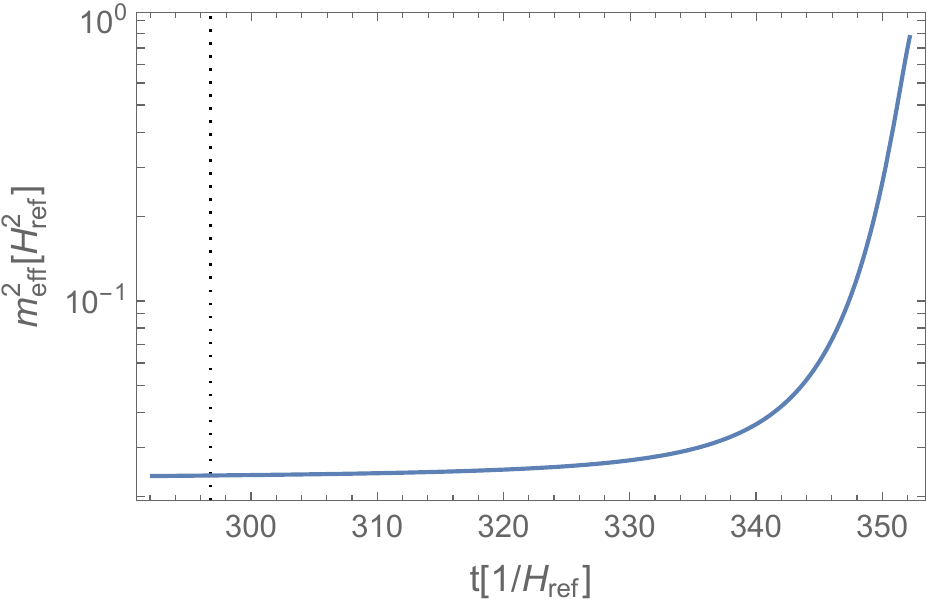}
\caption{
Left: the turning rate, $\dot{\theta}_T(t)$, for $t \leq t_e$.
Right: the effective mass-squared $m_{\rm eff}^2$ associated with isocurvature perturbations for the same parameter set as the one used in Fig.~\ref{fig:angular-motion}.
The vertical dotted line corresponds to $52$ $e$-folds to the end of inflation at $t=t_e \simeq 352 H_{\rm ref}^{-1}$.
Roughly, it corresponds to the time when the comoving scale associated with the pivot scale of the Planck mission exits the horizon during inflation. 
}
\label{fig:m-eff-for-QN}
\end{center}
\end{figure}
%%%%%%%%%%%%%%%
In Fig.~\ref{fig:m-eff-for-QN}, $\dot{\theta}_T$ and $m_{\rm eff}^2$ are depicted.
From the left panel of the figure, one can see that, $|\dot{\theta}_T| = \mathcal{O}(10^{-3}) \times H_I$, i.e., the turning rate of the inflaton trajectory is quite small.
Also, the right panel shows that, $m_{\rm eff}^2/H_I^2 \lesssim \mathcal{O}(10^{-2})$ when large cosmological scales of our interest exit horizon, but it increases rapidly and becomes of $\mathcal{O}(0.1-1)$ within the last about 10 $e$-folds before the end of inflation.
%
%It can be understood as follows in $\l( \phi, \theta \r)$-basis.
%For $\alpha \lll 1$, the main contribution of $m^2$ in \eq{eq:m-sq-orthogonal} is from the second term in the right-hand side of the equation.
%The field-space Ricci scalar is approximated as 
%\bea
%M_{\rm P}^2 \mathbb{R} &\approx&
%\frac{M_{\rm P}^2}{\mathcal{K}_{\phi \phi} } \l[ \frac{1}{2} \l( \frac{\partial \ln \mathcal{K}_{\phi \phi}}{\partial \phi} + \frac{\partial \ln \mathcal{K}_{\theta \theta}}{\partial \phi} \r) \times \frac{\partial \ln \mathcal{K}_{\theta \theta}}{\partial \phi} 
%- \frac{1}{\mathcal{K}_{\theta \theta}} \frac{\partial^2 \mathcal{K}_{\theta \theta}}{\partial \phi^2} \r]
%\\
%&\approx& 4 \xi_\phi \l[ \l( 1+ 3 \xi_\phi \r) + \l( 1+ 6 \xi_\phi \r) \frac{\xi_\phi \phi^2}{M_{\rm P}^2} \r] \l[ 1 + \frac{\l( 1 + 6 \xi_\phi \r) \xi_\phi \phi^2}{M_{\rm P}^2} \r]^{-2}
%\\
%&\approx& \l\{
%\begin{array}{ll}
%\frac{4 M_{\rm P}^2}{\l( 1 + 6 \xi_\phi \r) \phi^2} & \quad \textrm{for} \ \xi_\phi \phi^2 \gg M_{\rm P}^2
%\\
%4 \xi_\phi \l( 1+ 3 \xi_\phi \r) & \quad \textrm{for} \ \xi_\phi \phi^2 \ll M_{\rm P}^2 \ .
%\end{array}
%\r.
%\eea   
%Hence, starting from a value much smaller than unity, $\mathbb{R}$ becomes of $\mathcal{O}(10) \times \xi_\phi^2$ and $\epsilon \to 1$ toward the end of inflation.
%As a result, there appears a mild suppression of the perturbation $Q_N$ around the end of inflation.

When $\dot{\theta} /H \lll 1$ during inflation that is the case in our scenario, the source terms in \eqs{eq:R-eom}{eq:QN-eom} (i.e., the right-hand sides of the equations) are negligible, and both $\mathcal{R}$ and $Q_N$ are nearly frozen soon after crossing the horizon although there appears a mild suppression of $Q_N$ due to the increasing $m_{\rm eff}^2$ toward the end of inflation.
Thus, there is no sizable impact of a two-dimensional curvilinear trajectory of the inflaton on the evolution of adiabatic and isocurvature perturbations associated with cosmological scales of the present universe, and non-Gaussianities are not enhanced \cite{Peterson:2010mv} either.
Also, even though the inflaton trajectory involves axi-Majoron direction which has anomalous couplings to non-Abelian gauge fields of the SM, the perturbations of gauge fields are not enhanced because the velocity of axi-Majoron during inflation is maintained to be smaller than the expansion rate by at least a couple of orders of magnitude during the whole relevant $e$-folds, as shown in Fig.~\ref{fig:angular-motion}.

\begin{figure}[h] 
\begin{center}
\includegraphics[width=0.48\textwidth]{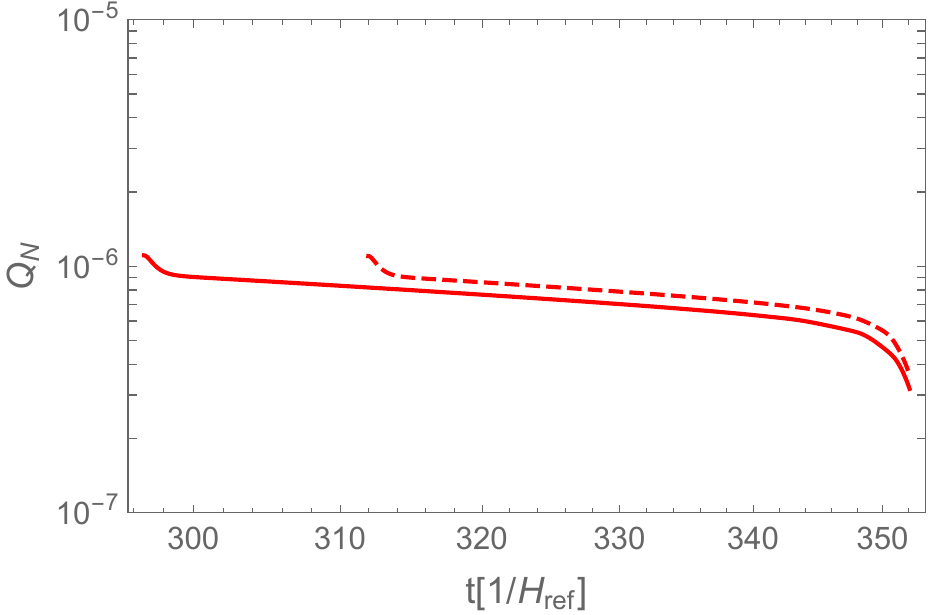}
\hspace{1em}
\includegraphics[width=0.48\textwidth]{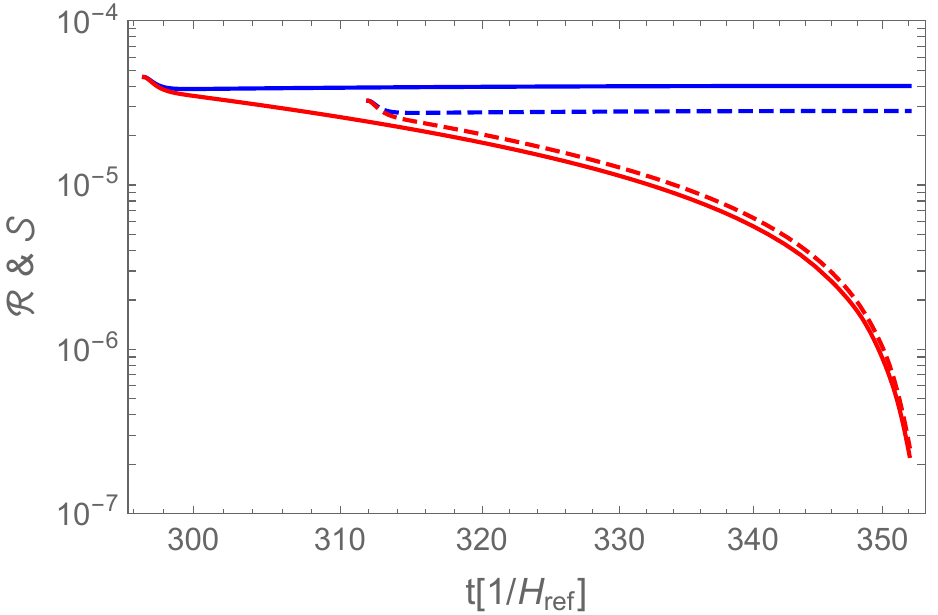}
%\includegraphics[width=0.48\textwidth]{figs/R-for-alpha-10E-6-theta1pi4}
%\hspace{1em}
%\includegraphics[width=0.48\textwidth]{figs/S-for-alpha-10E-6-theta1pi4}
\caption{
Left: the evolution of $Q_N$(the field-perturbation orthogonal to the inflaton direction). 
Right: the evolution of adiabatic($\mathcal{R}$)(blue) and isocurvature($\mathcal{S}$)(red) perturbations.
In both panels, solid and dashed lines correspond to the modes for the comoving scale $k_N = a_N H_I$ associated with the $e$-folds of $N=52$(solid) and $37$(dashed) before the end of inflation.
The parameter set used for the plots is the same as the one in Fig.~\ref{fig:angular-motion}.
The initial condition for the numerical integration of mode equations in \eqs{eq:R-eom}{eq:QN-eom} was set as $Q_T(t_N) = Q_N(t_N) = H(t_N)/2 \pi$ and $\dot{\mathcal{R}}(t_N) = \dot{Q}_N(t_N) = 0$ where $H(t_N)$ is the expansion rate when a comoving scale $k_N$ exit the horizon at $t_N$.
}
\label{fig:perts}
\end{center}
\end{figure}

In Fig.~\ref{fig:perts}, we depict the evolutions of $Q_N, \mathcal{R}$, and $\mathcal{S}$ for comoving scales exiting the horizon at $e$-folds of $N = 37$(dashed) and $52$(solid) before the end of inflation as examples.
%
%%%%%%%%%%%%%%
%%%%%%%%%%%%%%%
From the figure, one can see that, while the adiabatic ones are barely affected, isocurvature perturbations are suppressed mainly because $S = Q_N / \sqrt{2 \epsilon} M_{\rm P} \propto1/\sqrt{2 \epsilon}$.
The uncorrelated isocurvature perturbations is constrained as \cite{Planck:2018jri}
\beq
\beta_{\rm iso} \equiv \frac{P_{\mathcal{S}}}{P_{\mathcal{R}}+P_{\mathcal{S}}} < 3.8 \times 10^{-2}
\eeq 
where $P_{\mathcal{R}}$ and $P_{\mathcal{S}}$ stand for the power spectra of adiabatic and isocurvature perturbations, respectively.
Hence, the potential problem of axion isocurvature perturbation for the relevant cosmological scales of CMB observations is absent in our scenario.

In a multidimensional field-space, if the direction of the motion of the inflaton is changed during inflation, the sound speed of perturbations can be affected due to the mixing of kinetic terms \cite{Tolley:2009fg}.
Specifically, the sound speed $c_s$ is given by \cite{Achucarro:2012sm}
\beq \label{eq:sound-speed}
c_s^{-2}(k) = 1 + \frac{4 \dot{\theta}_T^2}{\l(k/a\r)^2 + m_{\rm eff}^2} \ .
\eeq
However, from Fig.~\ref{fig:m-eff-for-QN} as an illustrative example and subsequent discussion, we see that the change of $c_s$ is negligible, and set $c_s=1$.
%barely changes and confirmed numerically as shown in Fig.~\ref{fig:sound-speed}.
%%
%%%%%%%%%%%%%%%
% \begin{figure}[h] 
%\begin{center}
%\includegraphics[width=0.48\textwidth]{figs/cs-for-alpha0p001-theta1pi4}
%\caption{
%Sound speed $c_s$ during inflation for a comoving scale associated with the horizon of our present universe with $\alpha = 10^{-3}$, $\xi_\phi=1$, $\lambda_\phi=6 \times 10^{-10}$, $\phi_{\rm ini}=20 M_{\rm P}$, and $\theta_{\rm ini}=\pi/4$.
%%The subscript `$_{52}$' stands for quantities associated with $52$ $e$-foldings from the end of inflation.
%}
%\label{fig:sound-speed}
%\end{center}
%\end{figure}
%%%%%%%%%%%%%%%%
%Hence, we ignore such a change and set $c_s=1$.
Then, the amplitude of the power spectrum, the spectral index, and the tensor-to-scalar ratio are given by
\bea \label{eq:PR}
P_{\mathcal{R}} &=& \frac{1}{8 \pi^2 \epsilon} \times \frac{H^2}{M_{\rm P}^2} \ ,
\\
n_s &=& 1 - 4\epsilon + 2 \eta_T \ ,
\\
r_T &=& 16 \epsilon \ .
\eea

There might be a question about the impact of isocurvature perturbations on adiabatic ones as discussed in Refs.~\cite{Cremonini:2010sv,Achucarro:2016fby} for examples.
Including the contribution, the adiabatic perturbation may be written as
\beq
\mathcal{R} = \mathcal{R}_0 + \Delta \mathcal{R} \ ,
\eeq
where $\mathcal{R}_0$ is the pure adiabatic contribution, and $\Delta \mathcal{R}$ is the one from isocurvature perturbations.
In our scenario, ignoring the effect of the effective mass-square ($m_{\rm eff}^2$) of the isocurvature modes, we may approximate $\Delta \mathcal{R}$ as
\beq
\Delta \mathcal{R} \approx \int_0^N dN \l[ - \sqrt{\frac{2}{\epsilon}} \frac{\dot{\theta}_T}{H} \frac{Q_N}{M_{\rm P}} \r] \ ,
\eeq 
where the integration is with respect to the $e$-folding number after the horizon exit of the mode of interest to the end of inflation.
For the uncorrelated isocurvature perturbations, the power spectrum $P_{\mathcal{R}}$ is then given by
\beq
P_{\mathcal{R}} = P_{\mathcal{R}}^0 \times \l[ 1 + \Bigg\la \l( \frac{\Delta \mathcal{R}}{\mathcal{R}_0} \r)^2 \Bigg\ra \r] \ , 
\eeq
where $P_{\mathcal{R}}^0$ is the pure adiabatic contribution to the power spectrum, and the contribution to the bare spectral index $n_s^0$ of the adiabatic modes only is given by
\beq
\Delta n_s \equiv n_s - n_s^0 =  2 \l( \frac{\Delta \mathcal{R}}{\mathcal{R}_0} \r) \l[ \frac{1}{\mathcal{R}_0} \frac{d \Delta \mathcal{R}}{dN} - \l( \frac{\Delta \mathcal{R}}{\mathcal{R}_0} \r) \frac{d \ln \mathcal{R}_0}{dN} \r] \ .
\eeq
Ignoring the minor changes of $Q_N$ and $\dot{\theta}_T/H$ during inflation, one finds
\bea
\frac{\Delta \mathcal{R}}{\mathcal{R}_0} &\approx& - \frac{2 \dot{\theta}_T}{H} \int_0^N dN \sqrt{\frac{\epsilon_*}{\epsilon}} \lesssim \mathcal{O}(10) \times \l( - \frac{2 \dot{\theta}_T}{H} \r) \ , 
\\
\frac{d \Delta \mathcal{R}}{\mathcal{R}_0 dN} &\approx& - \frac{2 \dot{\theta}_T}{H} \lesssim \mathcal{O}(10^{-3})
\eea
with $d \ln \mathcal{R}_0 / dN \approx \l( n_s^0 - 1 \r)/2 = -\mathcal{O}(10^{-2})$.
Hence, it is expected that $\Delta n_s = \mathcal{O}(10^{-5} - 10^{-4})$ which is smaller than the $1-\sigma$ error in the result of the \textit{Planck} satellite experiment \cite{Planck:2018jri} at least by about an order of magnitude, and we ignore the contribution in the subsequent discussion.

For the potential given in \eq{eq:U-approx} with $\alpha \ll 1$, the relations between slow-roll parameters and $e$-folds $N$ in the SM Higgs-inflation \cite{Bezrukov:2007ep} are expected still to hold at least approximately.
The latter is given by \cite{Bezrukov:2007ep}
\beq
N(\phi) \approx \frac{3}{4} \frac{\phi^2 -\phi_e^2}{M_{\rm P}^2 / \xi_\phi}
\eeq
as the $e$-folds from the end of inflation at $\phi_e \approx \l( 4/3 \r)^{1/4} M_{\rm P} / \sqrt{\xi_{\phi,0}}$.
Slow-roll parameters are then given by 
\bea \label{eq:epsilon-Ne}
\epsilon &\simeq& \frac{3}{4} N^{-2} \ ,
\\ \label{eq:eta-Ne}
\eta_T &\simeq& N^{-1} \ , 
\eea
leading to 
\bea \label{eq:PR-Ne}
P_{\mathcal{R}} &\simeq& \frac{\lambda_\phi N^2}{72 \pi^2 \xi_{\phi,0}^2} \ ,
\\ \label{eq:ns-Ne}
n_s &\simeq& 1 - 2N^{-1} \ ,
\\ \label{eq:rT-Ne}
r_T &\simeq& 12 N^{-2} \ .
\eea
%where we used \eqs{eq:H-inflation}{eq:epsilon-Ne} to obtain $P_R$ in \eq{eq:PR-Ne} from \eq{eq:PR}.
Density perturbations of the universe extracted from various observations including CMB observations lead to $P_R \simeq 2.1 (\pm 0.03) \times 10^{-9} \ (1 \sigma \ {\rm CL})$, $n_s = 0.9658 \pm 0.004 \ (1\sigma \ {\rm CL})$, and $r_T < 0.068 \ (2 \sigma \ {\rm CL})$ for the pivot scale, $k_* = 0.05/{\rm Mpc}$ \cite{Planck:2018jri}.
Hence, \eq{eq:PR-Ne} constrains $\lambda_\phi$ and $\xi_{\phi,0}$ to satisfy a relation, 
\beq \label{eq:lambda-cond}
\lambda_\phi \sim 6 \times 10^{-10} \xi_{\phi,0}^2 \l( \frac{52}{N_*} \r)^2 \ .
\eeq
where we used $N_* = 52$ as a representative value of $e$-folds associated with $k_*$.
In Table~\ref{tab:xi-phi-dependence}, we show a result of numerical estimations for some sets of parameters set as 
\beq \label{eq:para-set-for-num}
\lambda_\phi \equiv \lambda_\phi^{\rm ref} \times \xi_{\phi,0}^2, \ \alpha = \frac{10^{-3}}{\xi_{\phi,0}}, \ \phi_{\rm ini} = \frac{20 M_{\rm P}}{\sqrt{\xi_{\phi,0}}}, \ \theta_{\rm ini} = \frac{\pi}{4} \ ,
\eeq
with $\lambda_\phi^{\rm ref} \equiv  6 \times 10^{-10}$ and several different values of $\xi_{\phi,0}$ but $N_*=52$ fixed.
\begin{table}[h]
\begin{center}
\begin{tabular}{|c||c|c|c|}
\hline
$\xi_{\phi,0}$ & $P_R \times 10^9$ & $n_s$ & $r_T \times 10^3$ \\
\hline \hline
1 & 2.067 & 0.9622 & 4.738 \\
\hline 
10 & 2.387 & 0.9620 & 4.128 \\
\hline
$10^2$ & 2.424 & 0.9619 & 4.066 \\
\hline
$10^3$ & 2.428 & 0.9619 & 4.060 \\
\hline
\end{tabular}
\end{center}
\caption{Examples of numerical estimations for inflationary observables with parameters set as \eq{eq:para-set-for-num}.}
\label{tab:xi-phi-dependence}
\end{table}%
The result is in a good agreement with approximations in 
Eq.~(\ref{eq:PR-Ne}) -- (\ref{eq:rT-Ne}), and can match well with observations by adjusting $\lambda_\phi$.
The sizable difference between the cases of $\xi_{\phi,0}=1$ and the other much larger values may be understood from the dependence of $U$ on $\sqrt{\xi_\phi/(1+6 \xi_\phi)}$ in \eq{eq:U-approx}.
For convenience, in the subsequent argument we use $\lambda_\phi$ parametrized as in \eq{eq:para-set-for-num}.

\begin{figure}[t] 
\begin{center}
\includegraphics[width=0.48\textwidth]{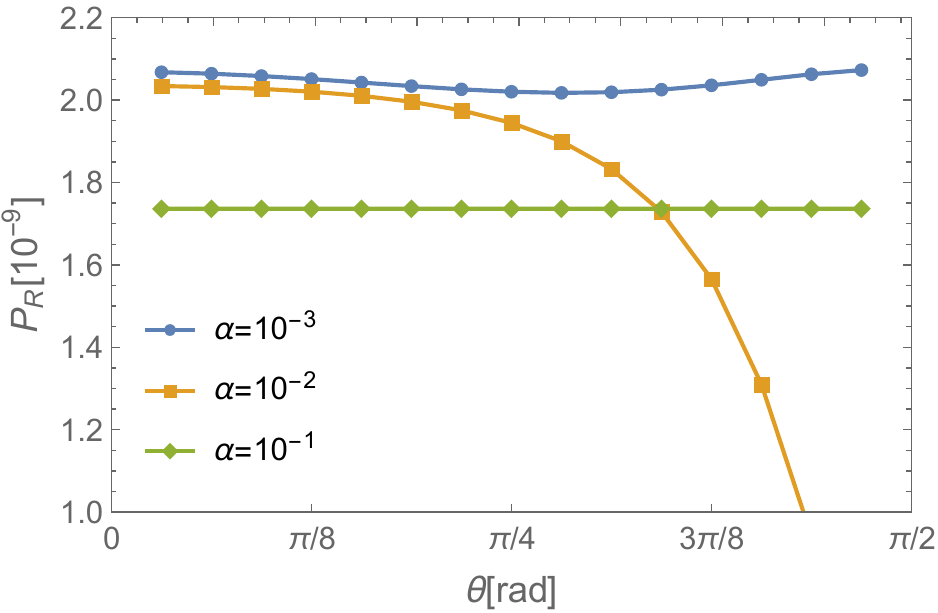}
\hspace{1em}
\includegraphics[width=0.48\textwidth]{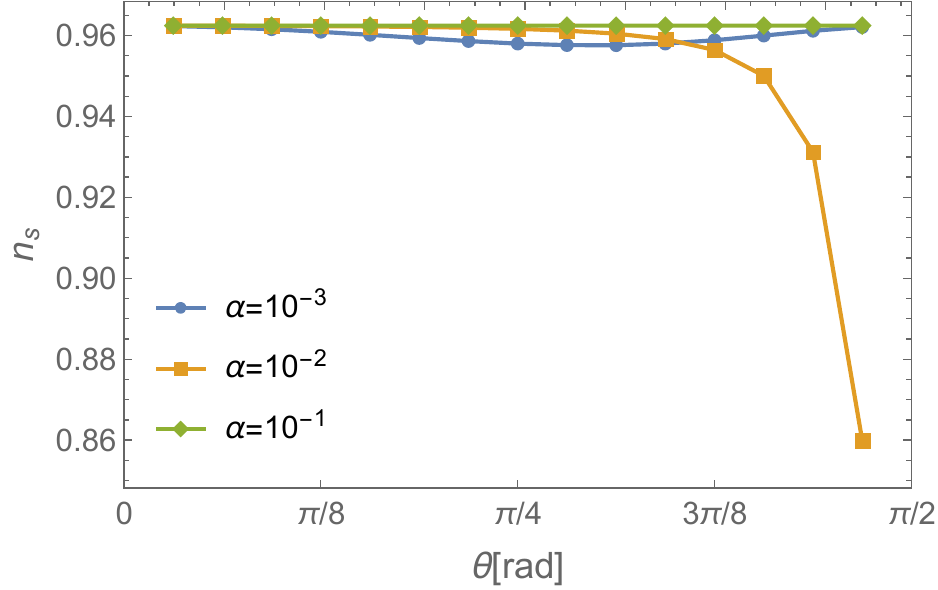}
\includegraphics[width=0.48\textwidth]{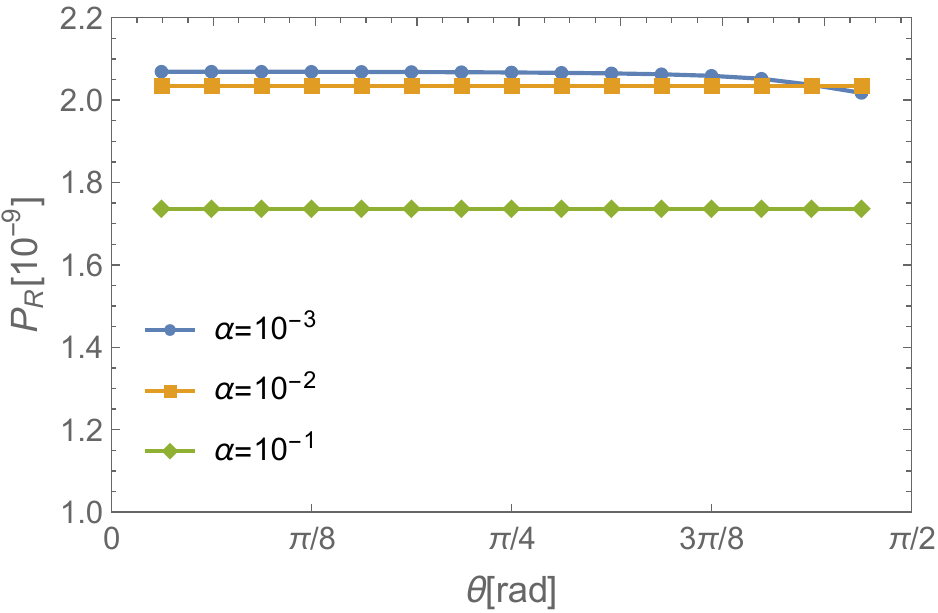}
\hspace{1em}
\includegraphics[width=0.48\textwidth]{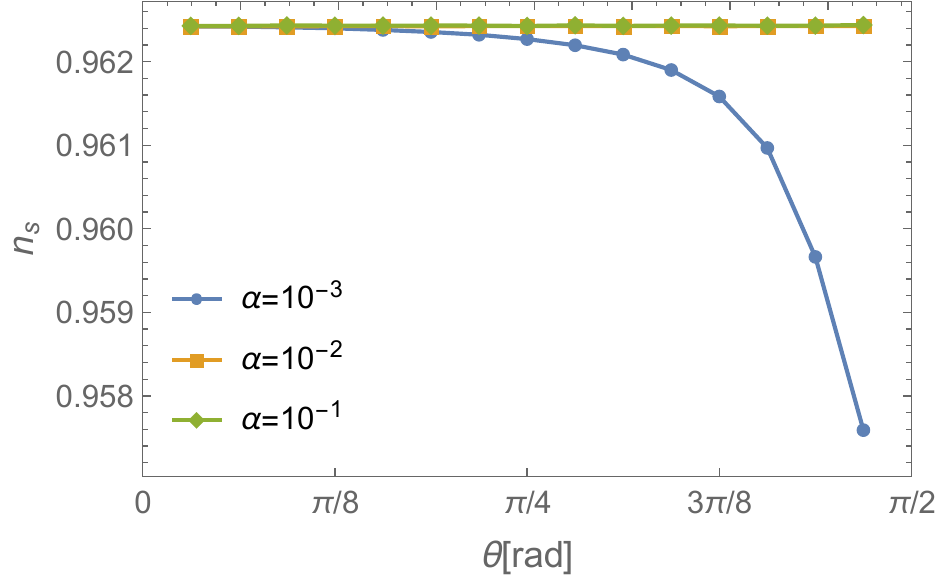}
\caption{
$P_R$(left) and $n_s$(right) as functions of $\theta_{\rm ini}$ for various values of $\alpha$ with $\xi_{\phi,0}=1$, $\lambda_\phi=6 \times 10^{-10}$, $\phi_{\rm ini}/M_{\rm P}=10({\rm top}), 20({\rm bottom})$.
}
\label{fig:para-dependence}
\end{center}
\end{figure}

In our scenario, there are five parameters affecting the observables associated with inflation in the complex field space of PQ-field:
\beq
\xi_{\phi,0}, \lambda_\phi, \alpha, \theta_{\rm ini}, \phi_{\rm ini} \ .
\eeq
%where $\theta_{\rm ini}$ and $\phi_{\rm ini}$ are the initial angle and distance of the field configuration in $(\phi_r, \phi_i)$ space.
Once $\xi_{\phi,0}$ as a free parameter is chosen such that $\xi_{\phi,0} \gtrsim \mathcal{O}(1)$, $\lambda_\phi$ is nearly fixed as long as $\alpha$ is sufficiently smaller than unity.
Dependence of observables on the other parameters for given $\xi_{\phi,0}$ and $\lambda_\phi$ is shown in Fig.~\ref{fig:para-dependence}.
%%%%%%%%%%%%%%
%%%%%%%%%%%%%%%
It turns out that inflationary observables are barely affected by $\theta_{\rm ini}$ for $\sqrt{\xi_{\phi,0}} \phi_{\rm ini} \gtrsim 10 M_{\rm P}$ as long as $\alpha \xi_{\phi,0} \lesssim \mathcal{O}(10^{-2})$ except the case of $\theta_{\rm ini}$ close to $\pi/2$ with $\sqrt{\xi_{\phi,0}}\phi_{\rm ini} \sim 10 M_{\rm P}$ when $\alpha \xi_{\phi,0} \sim 10^{-2}$.
Note that, if angular dynamics is not strong, for a given $e$-folds associated with the pivot scale of observations, the spectral index is nearly fixed irrespective of initial position of the field dynamics.
The amplitude of power spectrum has only minor dependence on the initial position, and the dependence can be compensated by adjusting $\lambda_\phi$ without affecting the spectral index.
This is because, while the former is a function of $e$-folds only, the latter is a function of $e$-folds and energy density of inflation. 
On the other hand, if $\alpha \xi_{\phi,0} \gtrsim \mathcal{O}(10^{-1})$, the observables varies a lot, depending on $\theta_{\rm ini}$, but $\phi_{\rm ini}$-dependence becomes weak if it is sufficiently larger than the minimal value to accommodate about 60 $e$-folds.

\subsection{Reheating after inflation}
\label{subsec:reheating}

\begin{figure}[h] 
\begin{center}
\includegraphics[width=0.46\textwidth]{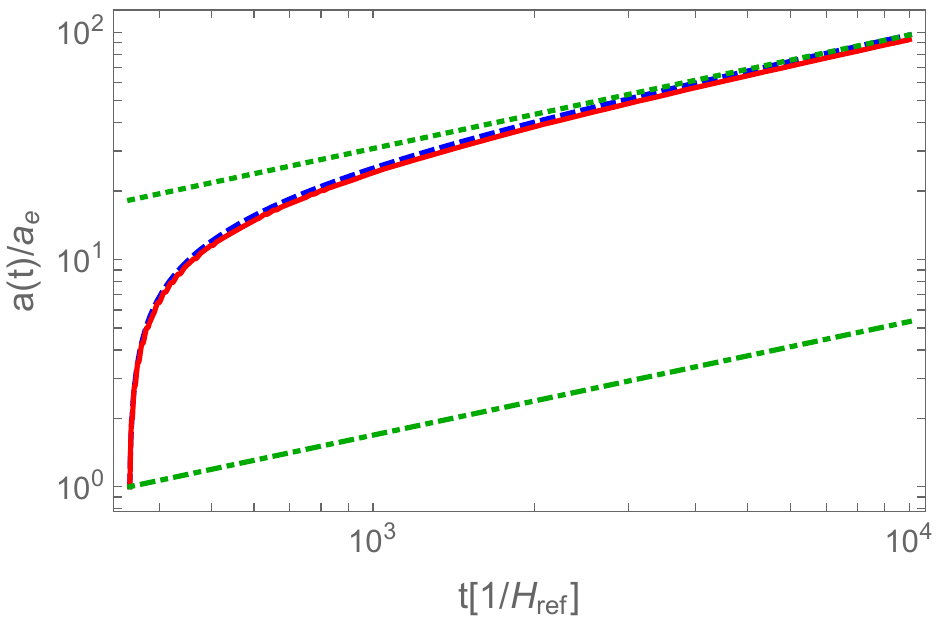}
\includegraphics[width=0.48\textwidth]{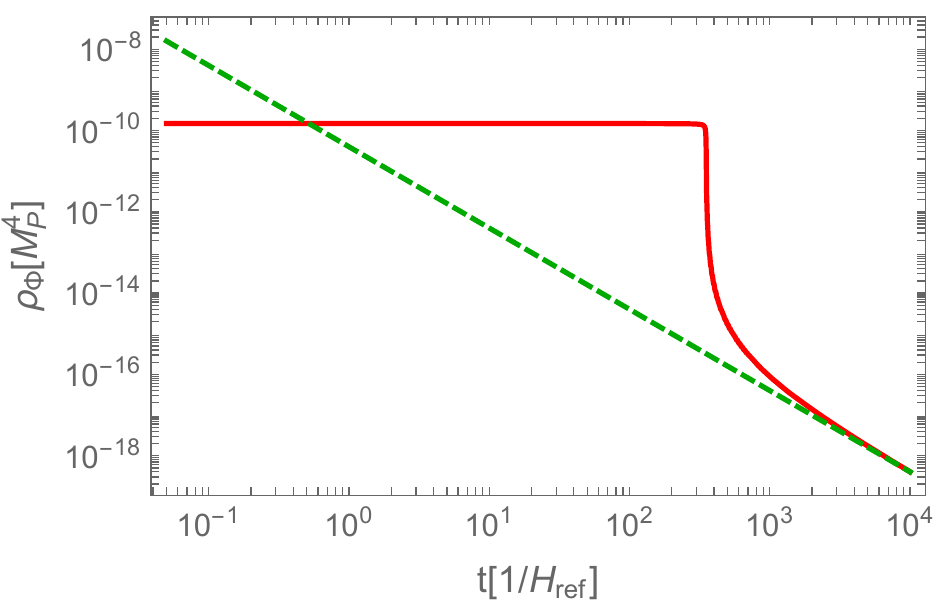}
\caption{
Left: evolutions of the scale factor (dashed blue) and $\sqrt{H_e/H(t)}$ (solid red) as functions of cosmic time $t$ from the end of inflation for the same parameter set as the one used in Fig.~\ref{fig:angular-motion}.
The solid and dashed lines are nearly overlapped, implying $H(t) \propto a(t)^{-2}$.
The dot-dashed green line is of $\l( t/t_e \r)^{1/2}$.
The dotted green line is the same as the dot-dashed green line but just up-shifted for a clear comparison to red or dashed blue line.
Right: evolution of the energy density of $\Phi$ (solid red).
The dashed green line corresponds to $t^{-2}$-behavior shown for a comparison.
}
\label{fig:scale-factor-vs-t}
\end{center}
\end{figure}

Once inflation ends, after a transient period, the energy density of $\phi^4$ model evolves in the same way as radiation \cite{Turner:1983he}.
In the left panel of Fig.~\ref{fig:scale-factor-vs-t}, the evolution of the scale factor and expansion rate are shown.
%%%%%%%%%%%%%%
%%%%%%%%%%%%%%%
One can see that $H(t) \propto a(t)^{-2}$ from the end of inflation.
However, note that these lines do not match to the dot-dashed green line (lower diagonal line) which corresponds to $\l( t/t_e \r)^{1/2}$.
Their time dependence becomes equal only some time after inflation. 
The actual value of the scale factor (dashed blue) is larger than the value of the lower green line by about one order of magnitude.
The energy density of $\Phi$ scales like radiation not right after inflation but some time later after getting through a short suppression period (see the right panel of the figure).

For $\phi$ well below the Planck scale, the scalar potential is approximated as $U \simeq {\tilde V}$.
It can be expressed as a familiar form given in \eq{eq:V-low-energy}.
%Without loss of generality, it can be expressed as a familiar form,
%\beq
%{\tilde V}_S(h, \phi) = V_0 + \frac{\lambda_h}{4} \l( h^2 - h_0^2 \r)^2 + \frac{\lambda_\phi'}{4} \l( \phi^2 - \phi_0^2 \r)^2  
%\eeq
%where 
%\beq
%\lambda_\phi' = \lambda_\phi - \lambda_h \zeta_{h\phi}^2.
%\eeq
For simplicity, we consider the case of $\lambda_\phi' = \lambda_\phi$ (i.e., $\zeta_{h\phi}=0 \ \Leftrightarrow \lambda_{\phi h_i}=0$) and $\lambda_{\phi h}=0$.
Then, the mass of $\phi$-quanta for a given background value is given by
\beq \label{eq:m-phi}
m_\phi = \sqrt{3 \lambda_\phi} \phi \ \xrightarrow[]{\phi \to \phi_0} \ \sqrt{2 \lambda_\phi} \phi_0 \ .
\eeq
For $\lambda_{\phi h_i}=\lambda_{\phi h}=0$, the production of the SM Higgs from either preheating processes or decays of $\phi$ particles is negligible.
For sufficiently small $y_N$ which will be discussed in the subsequent discussion, the production of RHNs via preheating is also not efficient \cite{Peloso:2000hy}.
Hence, initially there can be population of $\phi$-particles and axi-Majorons ($a_\phi$) only \footnote{There are thermal particles with a temperature $T \sim H_{\rm e}/(2\pi)$ at the end of inflation \cite{Gibbons:1977mu}.
We ignore this contribution which does not affect our argument.
}.
As shown in Ref.~\cite{Greene:1997fu}, the self-preheating effect on $\phi$-particles is inefficient.
The population of $a_\phi$ is only from the perturbative decay of $\phi$-particles through the kinetic term of $a_\phi$.
Moreover, if $\Phi$ has nonzero angular momentum (i.e., a phase rotation), the preheating effect will be suppressed since $\phi$ does not pass through the origin during its oscillations and the violation of adiabaticity which is necessary for an efficient preheating becomes weak as a result. 
However, there might be a possibility of efficient energy transfer via axi-Majoron's anomalous couplings to non-Abelian gauge fields due to large velocity of axi-Majoron after inflation \cite{Adshead:2015pva}.
Since the velocity of axi-Majoron is periodically changing as shown in Fig.~\ref{fig:angular-motion}, in regard of the polarization state of gauge field which is amplified the situation is similar to the case of parametric resonance in a conformal scalar theory \cite{Greene:1997fu}.
The details will be investigated in other work. 
In any case, as long as the eventual decay of $\phi$-particle is the dominant source of the energy of the universe and the interaction of $\phi$ to RHNs is small enough to avoid the possibility of thermal-trapping, our argument is barely affected.
Hence, in this work we ignore the effect of preheating.

In our scenario, $U(1)_{\rm PQ}$ symmetry is broken only in the gravity sector, and its effect is suppressed by $\alpha \xi_\phi \phi^2/M_{\rm P}^2$ and turned off as $\phi \ll c_\phi \chi$ [see \eq{eq:xi-alpha}]. 
As a result, $a_\phi$ is lighter than $\phi$ for $\phi \lesssim M_{\rm P}$ with $\alpha \xi_\phi \ll 1$, and $\phi$ can always decay to $a_\phi$s via the kinetic term of $a_\phi$. 
The partial decay rate of the channel is given by
\beq \label{eq:Gamma-to-a}
\Gamma_{a_\phi}^\phi = \frac{m_\phi^3}{32\pi \phi^2} \approx 
\l\{
\begin{array}{ll}
\l(3\sqrt{3}/ 32 \pi \r) \lambda_\phi^{3/2} \phi &: \phi > \phi_\times
\\
\l(2\sqrt{2}/ 32 \pi \r) \lambda_\phi^{3/2} \phi_0 &: \phi < \phi_\times
\end{array}
\r. \ ,
\eeq
where $\phi$ and $\phi_\times(\equiv \sqrt{2} \phi_0)$ are respectively the oscillation amplitude and the point where the potential energy of $\phi$-field is the same as the one at the origin.
The inflaton should be able to decay to RHNs too in order to recover the standard thermal background consisting of mainly the SM particles.
For the mass of a RHN mass-eigenstate $N_i$,  $m_{N_i} = y_{N_i} \phi_0/\sqrt{2}$, the channel is open only if 
\beq \label{eq:cond-for-RHN-ch}
%\sqrt{\lambda_\phi } > y_N = 10^{-6} \l( \frac{\sqrt{2} m_N}{10^6 {\rm GeV}} \r) \l( \frac{10^{12} {\rm GeV}}{\phi_0} \r)
y_{N_i} < \sqrt{\lambda_\phi} \ .
\eeq
From the seesaw formula for the mass parameters of the left-handed neutrinos \cite{Minkowski:1977sc,Yanagida:1979as} \footnote{At low energy below the electroweak scale, $y_\nu$ should be replaced to the one obtained from a biunitarity transformation associated with physical mass-eigenstates.
We ignore such details here, since it does not make changes of the main point of our argument.
In this reason, we assumed $y_\nu$ is flavor-diagonal for simplicity.}, 
\beq \label{eq:seesaw-relation}
m_{\nu_i} \approx \frac{y_{\nu_i}^2 v_u^2}{2 m_{N_i}} \ ,
\eeq
where $v_u \equiv h_{2,0}=v_h \sin \beta$, i.e., the VEV of the up-type Higgs.
Then the constraint in \eq{eq:cond-for-RHN-ch} is translated as a 
constraint on $y_{\nu_i}$, which can be expressed as
\bea
y_{\nu_i}^2 
&\approx& 2\sqrt{b_i \lambda_\phi} \l( \frac{m_{\nu_i} \phi_0}{v_u^2} \r) \ ,
\eea
where we have used $b_i \equiv y_{N_i}^2/ 2 \lambda_\phi$ for later convenience.  
For simplicity, we assume that $m_{N_i}$s and $m_{\nu_i}$s have the same hierarchy.
Note that neutrino oscillation data is consistent with the massless lightest left-handed neutrino, which means that in principle $m_{\nu_1}$(correspondingly $y_{\nu_1}$) associated with the lightest mass eigenstate can be arbitrarily small. 
The decay rate $\phi$ to a RHN mass-eigenstate $N_i$ is given by
\beq \label{eq:Gamma-to-N}
\Gamma_{N_i}^\phi 
%= \frac{y_{N_i}^2}{32 \pi} m_\phi \l( 1 - \frac{4m_{N_i}^2}{m_\phi^2} \r)^{3/2} 
= b_i \l( 1 - 2 b_i \r)^{3/2} \Gamma_{a_\phi}^\phi \ .
\eeq
where we used $m_\phi = \sqrt{2 \lambda_\phi} \phi$ for $\phi \gtrsim \phi_0$.
Only these channels are relevant for the decay of $\phi$.
Hence, the total decay rate of $\phi$ is given by
\beq
\Gamma^\phi = \Gamma_{a_\phi}^\phi + \sum_i \Gamma_{N_i}^\phi \ ,
\eeq
and the branching ratio of $N_i$-channel is given by
\beq \label{eq:Bi}
{\rm B}_i = \frac{b_i}{1+\sum_j b_j} < \frac{1}{3} \ ,
\eeq
where the phase space factor of the RHN-channel was ignored. 

For $\phi \gtrsim \phi_\times$ the universe is driven by a radiationlike energy component.
In this case, the decay can happen effectively once $H \approx \Gamma^\phi/2$ is satisfied.
If the subdominant angular motion is ignored, the expansion rate after inflation can be expressed such as
\beq \label{eq:H-early-RD}
H 
%= \frac{\rho^{1/2}}{\sqrt{3} M_{\rm P}} 
= \frac{\lambda_\phi^{1/2} \phi^2}{2\sqrt{3} M_{\rm P}} \propto a^{-2} \ ,
\eeq
where $\phi \propto a^{-1}$ with $a$ being the scale factor of the expansion of the universe was used for the scaling behavior of $H$.
Then, taking $\Gamma^\phi \approx \Gamma_{a_\phi}^\phi$, one finds that the decay of $\phi$ takes place when $\phi = \phi_*$ with 
\beq \label{eq:phi-star}
\frac{\phi_*}{\phi_\times} \approx \frac{9 \lambda_\phi}{32 \sqrt{2} \pi} \frac{M_{\rm P}}{\phi_0} \simeq \l( \frac{\xi_{\phi,0}}{10^2} \r)^2 \l( \frac{\phi_0^{\rm ref}}{\phi_0} \r)
\ .
\eeq
which is valid only for $\phi_* > \phi_\times$.

If $\phi_* < \phi_\times$, decays start only after the field falls into the vicinity of the low energy true vacuum position, i.e., $\phi_0$. 
In such a case, the field amplitude at the epoch of decays is found as follows.
With a redefined field ${\tilde \phi} \equiv \phi - \phi_0 < \phi_0$, the potential of $\phi$ can be approximated as 
\beq
V(\phi) \approx \lambda_\phi \l[ \phi_0^2 {\tilde \phi}^2 + \phi_0 {\tilde \phi}^3 + \frac{1}{4} {\tilde \phi}^4 \r] \ \xrightarrow[]{{\tilde \phi} \ll \phi_0} \ \frac{1}{2} m_{\phi,0}^2 {\tilde \phi}^2
\eeq 
where $m_{\phi,0} \equiv \sqrt{2 \lambda_\phi} \phi_0$.
The energy density of $\phi$ is then given by
\beq \label{eq:rho-phi-MD}
\rho_\phi \approx \frac{1}{2} m_{\phi,0}^2 {\tilde \phi}^2 \ ,
\eeq
where ${\tilde \phi}$ is understood as the amplitude of oscillation.
%\beq
%m_\phi = \sqrt{2 \lambda_\phi} \phi_0 \simeq 4 \times 10^7 {\rm GeV} \l( \frac{\lambda_\phi}{\lambda_\phi^{\rm ref}} \r)^{1/2} \l( \frac{\phi_0}{\phi_0^{\rm ref}} \r) .
%\eeq
In this regime, the universe is in a matter-domination era driven by nonrelativistic massive $\phi$-particles\footnote{There is no kination era driven by the kinetic anergy caused by angular motion of $\Phi$.
This is because the phase velocity of $\Phi$, i.e., $\dot{\theta}$ scales as $a^{-1}$ for $\phi \gtrsim \phi_\times$ and $a^{-3}$ for 
$\phi \lesssim \phi_\times$ in our scenario.
Its energy contribution is always subdominant relative to the contribution from the radial mode.}.
Hence, requiring $H = (2/3) \Gamma^\phi$, one finds the oscillation amplitude ${\tilde \phi}_*$ at the epoch of the decay of $\phi$ as
\beq
{\tilde \phi}_* \approx \frac{\sqrt{6} \lambda_\phi}{24 \pi} M_{\rm P} \simeq 4.7 \times 10^7 {\rm GeV} \times \xi_{\phi,0}^2 \ ,
\eeq
and the corresponding expansion rate is given by
\beq \label{eq:H-star-eMD}
{\tilde H}_* \approx \frac{\sqrt{2}\lambda_\phi^{3/2}}{24 \pi} \phi_0 \simeq 3.5 \times 10^{-4} {\rm GeV} \times \xi_{\phi,0}^3 \l( \frac{\phi_0}{\phi_0^{\rm ref}} \r) \ .
\eeq

One may worry about the possibility of holding $\phi$ around the origin due to thermal effect before $\phi$ falls below $\phi_\times$.
If it happens, the domain-wall problem can arise although the details may also depend on angular motion of $\Phi$.
In the vicinity of the origin, thermal effect is from the interaction of $\Phi$ to RHNs which had been produced already from the partial decay of $\phi$.
If those RHNs were thermalized with axi-Majorons soon after their production, for $\phi_* > \phi_\times$, the temperature around the epoch of $\phi$-particles' decay is expected to be
\beq
T_{N, *} \approx \rho_{\phi, *}^{1/4} = \frac{\lambda_\phi^{1/4}}{\sqrt{2}} \phi_* \ .
\eeq
Then, the effective mass of $\Phi$ in the vicinity of the origin is given by
\bea \label{eq:msq-phi-eff}
m_{\phi, \rm eff}^2(0) 
&=& - \lambda_\phi \phi_0^2 + \sum_i c_T y_{N_i}^2 T_{N, *}^2
\nonumber \\
&\approx& \lambda_\phi \phi_0^2 \l[ -1 + c_T \lambda_\phi^{5/2} \l( \frac{9}{32 \pi} \frac{M_{\rm P}}{\phi_0} \r)^2 \sum_i b_i \r]
\eea
where $c_T = 1/12$ \cite{Dolan:1973qd}.
Requiring $m_{\phi, \rm eff}^2(0) <0$ to avoid a thermal-trap, we find
\beq \label{eq:non-trap-cond}
\xi_{\phi,0} \lesssim 680 \l( \frac{0.1}{b_N} \r)^{1/5} \l( \frac{\phi_0}{\phi_0^{\rm ref}} \r)^{2/5}
\eeq
where we took $b_N \equiv b_3 \approx b_2 \gg b_1$.
In the case of $\phi_* < \phi_\times$, the constraint on $\xi_{\phi,0}$ is much weaker since the would-be temperature of RHNs around the epoch of $\phi = \phi_\times$ is much lower.
Hence, conservatively speaking, as long as \eq{eq:non-trap-cond} is satisfied, the PQ-field can avoid thermal-trap around the origin, and the PQ-symmetry is never restored during and after inflation.

It should be noted that, even though thermal effect is negligible, and the PQ-symmetry is not restored as a result, domain-wall problem can be avoided only if $\Phi$ has angular motion which is large enough to overcome its quantum fluctuations.
Otherwise, $\Phi$ would spread over the whole $2\pi$ angle due to the random scattering of the oscillating field by the barrier around the origin, resulting in the domain-wall problem. 

The details of the postinflation cosmology depends on whether $\phi_*$ is larger than $\phi_\times$ or not.
So, we discuss those cases separately.

\paragraph{\textit{Case of $\phi_* < \phi_\times$:}}
As shown in \eq{eq:phi-star}, $\phi_* < \phi_\times$ if
\beq \label{eq:xiphi-cond-phis-lower}
\xi_{\phi,0} < 100 \times \l( \frac{\phi_0}{\phi_0^{\rm ref}} \r)^{1/2}
\eeq
Once $\phi$ decays at the epoch of ${\tilde H}_*$, the universe will be filled with axi-Majorons and RHNs as a comparable or subdominant component.  
Depending on interaction rates, they may or may not reach thermal or kinetic equilibrium among themselves before the standard thermal bath is established from decays of RHNs.
Initially, the production of $\phi$ via inverse decay is kinematically forbidden.
Hence, thermalization before the decay of RHNs would be driven by interactions of axi-Majoron kinetic term, $y_{N_i}$, and $y_{\nu_i}$. 
The $\phi$-mediating self-interaction cross-section of axi-Majorons is given by
\beq \label{eq:majoron-s-csection}
\sigma_{a, s} = \frac{7 s^3}{80 \pi m_\phi^4 \phi_0^4} \ ,
\eeq
where $s=4E_a^2$ with $E_a$ being the energy of axi-Majoron in the center of momentum(CM) frame.
If they were not in equilibrium, the number density of axi-Majorons would be 
\beq
n_a(E_a) = n_{a, *} \l( \frac{E_a}{{\tilde E}_{a, *}} \r)^3
\eeq
with 
\beq
n_{a, *} \approx \frac{\rho_{\phi, *}}{{\tilde E}_{a, *}}
\eeq
where $\rho_{\phi, *} \equiv \rho_\phi({\tilde \phi} = {\tilde \phi}_*)$ in \eq{eq:rho-phi-MD}, and ${\tilde E}_{a, *}$ and $E_a$ are the energies of an axi-Majoron at its production and some time later, respectively.
Then, comparing the scattering rate to the expansion rate, one finds that axi-Majorons could reach kinetic equilibrium among themselves when its energy is larger than $E_{a, \rm k}$ satisfying
\beq
\frac{E_{a, \rm k}}{{\tilde E}_{a, *}} 
= \l( 2^5 \times \frac{5 \pi^2}{7} \r)^{1/7} \times \l( \frac{\phi_0}{\lambda_\phi M_{\rm P}} \r)^{2/7} 
\eeq
It implies that $E_{a, \rm k} > {\tilde E}_{a, *}$ if $\xi_{\phi,0} \lesssim 100 \l( \phi_0 / \phi_0^{\rm ref} \r)^{1/2}$.
Hence, for $\xi_{\phi,0}$ satisfying \eq{eq:xiphi-cond-phis-lower}, axi-Majorons are never thermalized among themselves after their production in the decay of $\phi$.

In the case of RHNs, the Yukawa-like interaction $y_N$ in \eq{eq:L-ss} induces axi-Majoron interactions to a Majorana field $N$,
\beq \label{eq:majoron-to-rhns}
\mathcal{L} \supset - i \frac{y_N}{2 \sqrt{2}} a_\phi \ov{N} \gamma_5 N + \frac{y_N}{4 \sqrt{2} \phi_0} a_\phi^2 \ov{N} N + {\rm H.c.} \ ,
\eeq
where $N=\l( \nu_R, \nu_R^c \r)^T$ is a four-component Majorana field with the flavor index suppressed, and only relevant leading order terms are shown.
When the condition in \eq{eq:cond-for-RHN-ch} is satisfied for the RHN-channels of the decay of the inflaton, the cross sections of the various interaction between the axi-Majorons and the RHNs are smaller than the one for axi-Majoron self-interactions at high energy limit($s \approx m_{\phi,0}$) at least by a factor of about $b_i/3$ (see Appendix~\ref{sec:csection}).
Moreover, the number density of RHNs is also suppressed by ${\rm B}_i$.
As a result, for $\xi_{\phi,0}$ satisfying \eq{eq:xiphi-cond-phis-lower}, the associated interaction rates cannot catch up the expansion rate for $H<H_\times \equiv H(\phi_\times)$.

Meanwhile, from the ratio of $y_{\nu_i}$ to $y_{N_i}$, which is given by
\beq \label{eq:ynu-to-yN}
\frac{y_{\nu_i}^4}{y_{N_i}^4} 
= \l[ \frac{m_\nu \phi_0}{v_u^2 \sqrt{b_i \lambda_\phi}} \l( \frac{m_{\nu_i}}{m_\nu} \r) \r]^2
\approx \frac{10^3}{b_i \xi_{\phi,0}^2 \sin^4 \beta} \l[ \frac{\phi_0}{\phi_0^{\rm ref}} \frac{m_{\nu_i}}{m_\nu} \r]^2 \ ,
\eeq
one may think that the interaction $y_{\nu_i}$ can cause much more efficient scatterings and annihilations of $N_i$s to SM particles.
However, the decay rate of a RHN mass eigenstate $N_i$ is given by
\beq \label{eq:GammaN-heavy}
\Gamma_{N_i} 
\simeq \frac{y_{\nu_i}^2 m_{N_i}}{8 \pi} 
= \frac{m_{\nu_i}}{4\pi} \l( \frac{m_{N_i}}{v_u} \r)^2
= \frac{b_i \lambda_\phi m_{\nu_i}}{4\pi} \l( \frac{\phi_0}{v_u} \r)^2 \ ,
\eeq
where RHNs were assumed to be much heavier than the particles of the Higgs doublets.
Also, for the heavy RHN mass-eigenstates which we assume to be $N_{2,3}$ we can take $m_{\nu_i} \approx m_\nu$ while both of the normal or inverted mass hierarchy of the left-handed neutrino are still allowed, depending on the details of $y_\nu$ \cite{Gonzalez-Garcia:2007dlo}.
Then, as long as $\tan \beta \gtrsim 1$ which may have to be the case to match electroweak precision data including the stability of the electroweak vacuum \cite{Eberhardt:2013uba}, for $\xi_{\phi,0}$ in the range we are interested in, with $\phi_0 \sim \phi_0^{\rm ref}$, heavy RHN mass-eigenstates decay well before their self-interactions/annihilations involving either axi-Majorons or SM particles become relevant.
The lightest RHNs can also be decoupled from SM particles before their decay if $m_{\nu_1} \ll m_\nu$ even in the case of $\xi_{\phi,0} \ll \mathcal{O}(10^2)$.
Therefore, once produced, axi-Majorons and RHNs are expected to evolve independently without thermalization at least until the epoch of RHNs decay as long as $m_{\nu_1} \ll m_{\nu_2} \approx m_{\nu_3} \approx m_\nu$.

If it were stable, a RHN mass-eigenstate becomes nonrelativistic when the expansion rate becomes 
\beq
{\tilde H}_{i, \rm non} = {\tilde H}_* \l( \frac{2 m_{N_i}}{m_{\phi, 0}} \r)^2 = 2 b_i {\tilde H}_* \ .
\eeq
It gives 
\beq \label{eq:GammN-non-rel-cond}
\frac{\Gamma_{N_i}}{{\tilde H}_{i, \rm non}} 
= \frac{3}{\sqrt{2 \lambda_\phi}} \frac{m_{\nu_i} \phi_0}{v_u^2} 
\simeq \frac{66}{\xi_{\phi,0} \sin^2 \beta} \l( \frac{m_{\nu_i}}{m_\nu} \r) \l( \frac{\phi_0}{\phi_0^{\rm ref}} \r) \ .
\eeq
Also, if stable enough, it would start dominating the universe when the energy of axi-Majoron becomes $E_a = E_{a, \rm eq}^i$ with
\beq
E_{a, \rm eq}^i = {\rm B}_i m_{N_i} \ .
\eeq
The expansion rate at this epoch is 
\beq \label{eq:Heq-i}
{\tilde H}_{i, \rm eq} 
= \sqrt{2} \l( \frac{E_{a, \rm eq}^i}{{\tilde E}_{a,*}} \r)^2 {\tilde H}_*  
= 2 \sqrt{2} {\rm B}_i^2 b_i {\tilde H}_* \ ,
\eeq
and one finds
\beq \label{eq:cond-for-MD}
\frac{\Gamma_{N_i}}{{\tilde H}_{i, \rm eq}} 
= \frac{3}{2 \sqrt{\lambda_\phi} {\rm B}_i^2} \l( \frac{m_{\nu_i} \phi_0}{v_u^2} \r) 
\approx \frac{51}{{\rm B}_i^2 \xi_{\phi,0} \sin^2 \beta} \l( \frac{m_{\nu_i}}{m_\nu} \r) \l( \frac{\phi_0}{\phi_0^{\rm ref}} \r) \ .
\eeq
If one of RHN mass-eigenstates is responsible for eventual recovering the SM thermal bath, it would have to satisfy $\Gamma_{N_i}/{\tilde H}_{i, \rm eq} < 0.1 \times \l( \Delta N_{\rm eff}^{\rm obs}/0.5 \r)^{3/2}$ with $\Delta N_{\rm eff}^{\rm obs}$ being the observational bound on the extra radiation (see Sec.~\ref{sec:DM-DR}).
Combined with the condition in \eq{eq:xiphi-cond-phis-lower}, the branching ratio ${\rm B}_i$ is then constrained as
\beq \label{eq:Bi-bnd-for-MD}
{\rm B}_i \gtrsim \frac{0.40}{\sin \beta} \times \l( \frac{0.5}{\Delta N_{\rm eff}^{\rm obs}} \r)^{3/4} \l( \frac{m_{\nu_i}}{m_\nu} \r)^{1/2} \l( \frac{\phi_0}{10^9 {\rm GeV}} \r)^{1/4} \ ,
\eeq
but it cannot be larger then $1/3$ as shown in \eq{eq:Bi}.
Hence, the standard thermal bath in this case is to be recovered by the decay of the lightest RHN mass-eigenstate $N_1$ with $m_{\nu_1}$ satisfying the following condition:

\beq \label{eq:m-nu1-DR-bnd}
\frac{m_{\nu_1}}{m_\nu} < 2 \times 10^{-4} \times \sin^2 \beta \l( \frac{\Delta N_{\rm eff}^{\rm obs}}{0.5} \r)^{3/2} \l( \frac{\xi_{\phi,0}}{10} \r) \l( \frac{{\rm B}_1}{0.1} \r)^2 \l( \frac{\phi_0^{\rm ref}}{\phi_0} \r) \ .
\eeq
Then, it is necessary to have $\phi_0 \sim \phi_0^{\rm ref}$ for the observed relic density of dark matter (see Sec.~\ref{sec:DM-DR}), and heavy states are expected to decay around or before the epoch they become nonrelativistic, and never dominate the universe.

If heavy RHN states decay before or around the epoch they become nonrelativistic, their decay rate is suppressed by the boosting effect.
For simplicity, we may take ${\rm B}_N \equiv {\rm B}_3 \approx {\rm B}_2$, $\Gamma_N \equiv \Gamma_{N_3} \approx \Gamma_{N_2}$, $m_{\nu_N} \equiv m_{\nu_3} \approx m_{\nu_2}$, $m_N \equiv m_{N_3} \approx m_{N_2}$.
Then, the expansion rate at the epoch of the decays of heavy RHNs is estimated as
\beq \label{eq:H-N}
H_N = \frac{f_N(\xi_{\phi,0}, \phi_0)}{2} \Gamma_N \ ,
\eeq
where 
\beq \label{eq:f-N}
f_N(\xi_{\phi,0}, \phi_0) 
\equiv \l[\frac{{\tilde H}_*}{\Gamma_N/2} \l( \frac{m_N}{m_{\phi,0}} \r)^2 \r]^{1/3}
%= \l[ \frac{2 \sqrt{2} \lambda_\phi^{1/2}}{3} \l( \frac{v_u^2}{m_{\nu_N} \phi_0} \r) \r]^{1/3}
\simeq 0.3 \xi_{\phi,0}^{1/3} \l( \frac{\phi_0^{\rm ref}}{\phi_0} \r)^{1/3} \ ,
\eeq
with \eqs{eq:H-star-eMD}{eq:GammaN-heavy} used for the last approximation.
Note that \eq{eq:f-N} is valid only for $f_N \leq 1$.
%, the expansion rate at the decay of $N_{2,3}$ is taken to be 
%\beq
%H_N = {\rm Min} \l[ {\tilde H}_N, \Gamma_N/2 \r]
%\eeq
The temperature of the SM thermal bath established due to their decay is then given by
\bea \label{eq:T-N-decay}
T_{\gamma, N} 
&\approx& \l( \frac{3 \times 45 {\rm B}_N}{\pi^2 g_*(T_{\gamma, N})} \r)^{1/4} \sqrt{f_N \Gamma_N M_{\rm P}} 
\nonumber \\
&\simeq& 2.4 \times f_N^{1/2} \l( \frac{{\rm B}_N}{10^{-2}} \r)^{1/4} \l( \frac{m_{\nu_N}}{m_\nu} \r)^{1/2} m_N \ ,
\eea
where we used \eq{eq:seesaw-relation} with $g_*(T_{\gamma, N}) = 106.75$ in the second line.
Hence, unless ${\rm B}_N \ll 10^{-4}$, the heavy states $N_{2,3}$ can be thermalized with the SM thermal bath via inverse decay processes.
Note that $m_{\phi,0}/T_{\gamma,N} \simeq 6.7 \times f_N^{-1/2} \l( 10^{-2}/b_N \r)^{3/4}$ with $b_N \equiv b_3 \approx b_2$.
If $b_N = \mathcal{O}(0.01-0.1)$, the inverse decay, $N+N \to \phi$, can also be efficient enough to thermalize $\phi$-particles.
But the impact of the process is minor in our discussion.
axi-Majorons are not thermalized even in this case, since the interaction is suppressed at least by $\mathcal{O}(T_{\gamma, N}^2/\phi_0^2)$.
There would be thermal contribution to the energy of axi-Majorons from partial decays of thermalized $\phi$-particles, but it is subdominant.

\paragraph{\textit{Case of $\phi_* > \phi_\times$:}}
\label{subsec:phis-large}
If $\phi_* > \phi_\times$, i.e., $\xi_{\phi,0} > 100 \sqrt{\phi_0/\phi_0^{\rm ref}}$ (see \eq{eq:phi-star}), $\Phi$ field decays before falling into the region of the true vacuum position.
Even in this case, $N_i$ can be kept out of kinetic or thermal equilibrium as long as $\xi_{\phi,0} \sim \mathcal{O}(10^2)$ with $\phi_0 \approx \phi_0^{\rm ref}$. 
The expansion rate at the epoch when $N_i$ becomes nonrelativistic is found as
\beq
H_{i, \rm non} = \l( \frac{2 m_{N_i}}{m_{\phi, *}} \r)^2 H_*
\eeq
where $m_{\phi, *} = \sqrt{3 \lambda_\phi} \phi_*$ and $H_* = \lambda_\phi^{1/2} \phi_*^2 / 2 \sqrt{3} M_{\rm P}$ with $\phi_*$ given by \eq{eq:phi-star}. 
The energy density of $N_i$ starts dominating when $H=H_{i, \rm eq}$ with
\beq \label{eq:H-eq-RD}
H_{i, \rm eq} 
= \sqrt{2} {\rm B}_i^2 H_{i, \rm non}
= \sqrt{2} {\rm B}_i^2 \l( \frac{2 m_{N_i}}{m_{\phi, *}} \r)^2 H_* \ .
\eeq
Note that $H_{1, \rm eq}$ is always smaller than ${\tilde H}_{1, \rm eq}$.
%It is instructive to compare ${\tilde H}_{1, \rm eq}$ and $H_{1, \rm eq}$:
%\beq
%\frac{H_{1, \rm eq}}{{\tilde H}_{1, \rm eq}} = \frac{4 \sqrt{2/3} \pi \phi_0}{\lambda_\phi M_{\rm P}}
%\eeq
Then, one finds
\beq
\frac{\Gamma_{N_i}}{H_{i, \rm eq}} 
= \frac{3 \sqrt{3 \lambda_\phi /2}}{8 \pi {\rm B}_i^2} \l( \frac{m_{\nu_i} M_{\rm P}}{v_u^2} \r)
> \frac{0.25}{\sin^2 \beta} \times \l( \frac{1/3}{{\rm B}_i} \r)^2 \l( \frac{\phi_0}{10^9 {\rm GeV}} \r)^{1/2} \l( \frac{m_{\nu_i}}{m_\nu} \r)
\eeq
where the last inequality is from the condition, $\phi_* > \phi_\times$.
Hence, the heavy RHN states cannot be responsible for the late time thermal background consisting of the SM particles in this case, too.

The constraint on $m_{\nu_1}$ to recover the SM thermal bath without too much dark radiation in this case is found as
\beq
\frac{m_{\nu_1}}{m_\nu} \lesssim 10^{-3} \times \sin^2 \beta \l( \frac{\Delta N_{\rm eff}^{\rm obs}}{0.5} \r)^{3/2} \l( \frac{{\rm B}_1}{0.1} \r)^2 \l( \frac{100}{\xi_{\phi,0}} \r) \ .
\eeq
Therefore, irrespective of whether $\phi_*$ is smaller than $\phi_\times$ or not, $m_{\nu_1}$ should be smaller than $m_\nu$ by several orders of magnitude.

\section{Leptogenesis}
\label{sec:lepto}

Thanks to the symmetry-breaking $\xi_a$-term in \eq{eq:S-G-field-redef}, the inflation along $\Phi$-direction can generate angular motion at the end of inflation, as long as $\theta_{\rm ini}$ is not very close to zero and the mass scale along the angular direction is somewhat smaller but not extremely smaller than the expansion rate during inflation \footnote{This is a kind of inflationary Affleck-Dine mechanism, and various different realizations have been considered (see for example Refs.~\cite{Cline:2019fxx,Barrie:2021mwi,Mohapatra:2021ozu,Mohapatra:2022tgb,Barrie:2024yhj}).}.
Since $U(1)_{\rm PQ}$ is conserved in ${\tilde V}$, the angular momentum of $\Phi$ at the end of inflation can be conserved for $\phi \ll M_{\rm P}$, and corresponds to the Peccei-Quinn number asymmetry stored in $\Phi$: 
\beq
\Delta n_\Phi = i \l( \dot{\Phi}^\dag \Phi - \Phi^\dag \dot{\Phi} \r) 
= \dot{\theta} \phi^2
= \dot{\phi}_r \phi_i - \phi_r \dot{\phi}_i \ ,
\eeq
with the PQ-charge of $\Phi$ normalized to be unity.

\begin{figure}[t] 
\begin{center}
\includegraphics[width=0.48\textwidth]{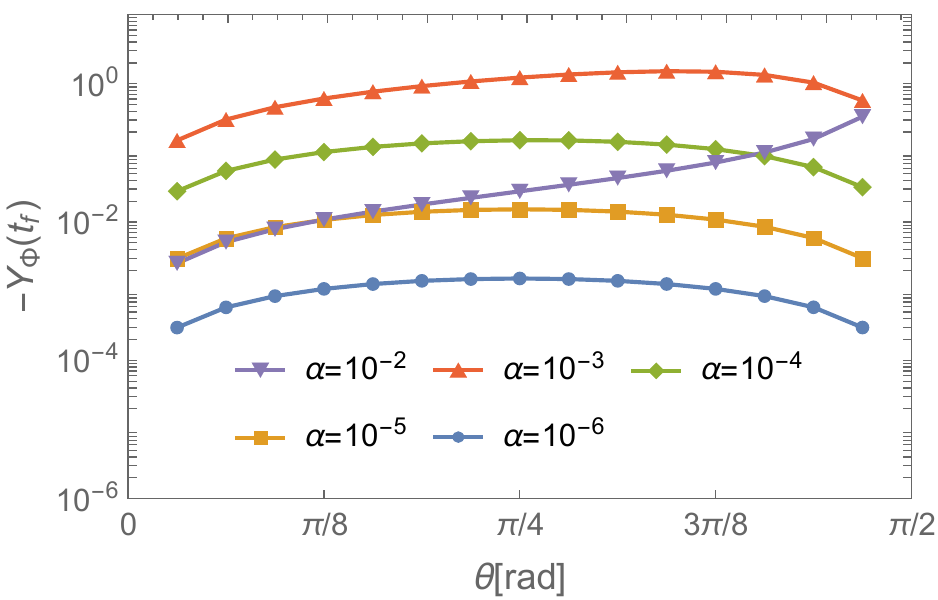}
\hspace{1em}
\includegraphics[width=0.48\textwidth]{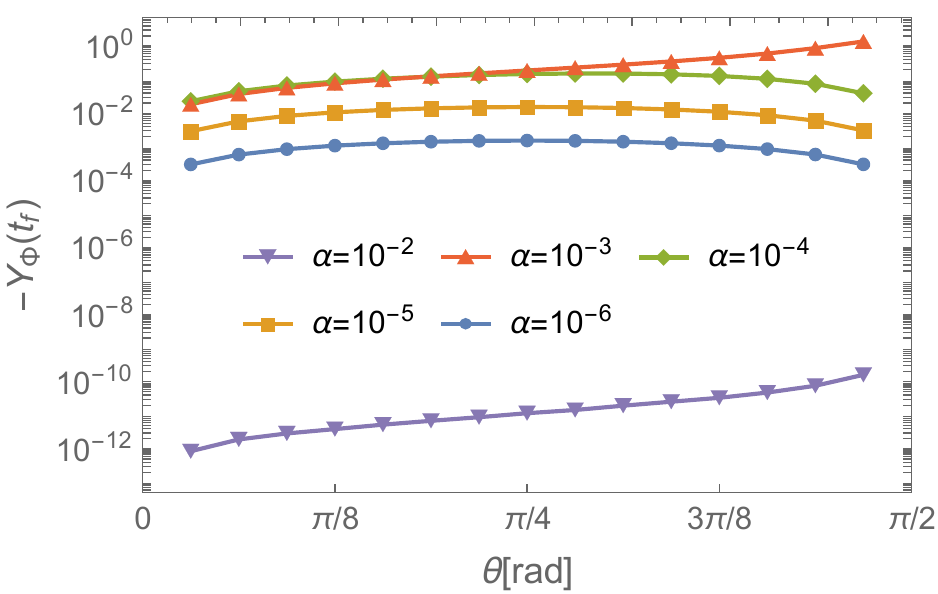}
\caption{
The yield of the PQ-number asymmetry $Y_{\rm PQ}$ at $t_f = 10^4 H_{\rm ref}^{-1}$ as functions of $\theta_{\rm ini}$ for various $\alpha$ with $\phi_{\rm ini}/M_{\rm P}=10$(left), $20$(right).
Inflation ends at $t_e = \mathcal{O}(100) H_{\rm ref}^{-1}$ depending on initial position of the inflaton and $\alpha$.
}
\label{fig:YPQ-e-vs-theta-ini}
\end{center}
\end{figure}

Here is a qualitative idea of how large PQ-number asymmetry can be generated in our model.
At the end of inflation, when $m_{a, e}^2/3 H_e^2 \lesssim \mathcal{O}(10^{-2})$ to avoid too fast angular motion toward the angular minimum position during inflation, it is expected that 
\beq
\dot{\theta}_e \sim \theta_{\rm ini} \l( \frac{m_{a, e}^2}{H_e^2} \r) H_e \simeq \frac{\sqrt{3}}{3} \theta_{\rm ini} \alpha \xi_{\phi,0} \lambda_\phi^{\rm ref} M_{\rm P} 
\ ,
\eeq
where \eq{eq:m-a} and
\beq
H_e \approx \frac{\lambda_\phi^{1/2} \phi_e^2}{2 \sqrt{3} M_{\rm P}} \simeq \frac{\sqrt{\lambda_\phi^{\rm ref}}}{2 \sqrt{3}} M_{\rm P} \ ,
\eeq
with $\phi_e \simeq M_{\rm P}/\sqrt{\xi_{\phi,0}}$ \cite{Bezrukov:2007ep} was used in the last approximation.
Also, for convenience, one may define the effective temperature of the radiationlike energy component at the end of inflation as
\beq \label{eq:T-e}
T_e \equiv \l( \frac{\pi^2}{90} g_*^\Phi \r)^{-1/4} \sqrt{H_e M_{\rm P}} = \l( \frac{2\pi^2 g_*^\Phi}{15} \r)^{-1/4} \lambda_\phi^{1/4} \phi_e \ ,
\eeq
where $g_*^\Phi = 2$ is the degrees of freedom of $\Phi$.
Then, in terms of the yield defined as the ratio of the PQ-number density to the entropy density, the PQ-asymmetry at the end of inflation can be express as  
\bea \label{eq:Y-PQ-e}
Y_{\Phi, e} 
%\equiv \l( \frac{n_{\rm PQ}}{s} \r)_e 
&=& \frac{3}{4} \frac{T_e \Delta n_{\rm PQ, e}}{\rho_{\phi, \rm e}} = \frac{3}{2 \lambda_\phi} \frac{T_e \dot{\theta}_e}{\phi_e^2}
\nonumber \\
&\approx& 
%2.5 \theta_{\rm ini} \times \l( \frac{\alpha}{10^{-2}} \r)
2.5  \l( \frac{\alpha}{10^{-2}} \r) \times \theta_{\rm ini} 
\ .
\eea
It seems that, when $\lambda_\phi$ and $\xi_{\phi,0}$ satisfy the relation in \eq{eq:lambda-cond} to provide inflationary observables matching observations, $Y_{\Phi, e}$ depends only on $\alpha$, modulo the dependence on the initial condition, $\theta_{\rm ini}$.
It should be noted that \eq{eq:Y-PQ-e} provides just a qualitative idea of the maximal possibility of $Y_{\Phi, e}$, though.
The exact parametric dependence of $Y_{\Phi, e}$ is nontrivial, since it involves the motion of $\Phi$ during inflation that is affected by the initial position of $\phi$, i.e., $\phi_{\rm ini}$ in addition to $\theta_{\rm ini}$. 
For later convenience, we may express $\dot{\theta}_e$ in terms of $Y_{\Phi, e}$ as
\beq
\dot{\theta}_e = \frac{2}{3} \l( \frac{2 \pi^2}{15} \r)^{1/4} \lambda_\phi^{3/4} \phi_e Y_{\Phi, e} \ .
\eeq
Figure \ref{fig:YPQ-e-vs-theta-ini} shows numerical estimations of $Y_\Phi$ well after inflation as functions of $\theta_{\rm ini}$ for various values of $\alpha$.
%%%%%%%%%%%%%%
%%%%%%%%%%%%%%%
The very different behaviors of the $\alpha=10^{-2}$ case in the left and right panels of the figure are due the fact that the angular motion during inflation is sizable for a large $\alpha$ so that, if $\phi_{\rm ini}$ is quite large, the field configuration can move the direction of the minimum along angular direction first, and toward the origin, afterward.
From the figure, one can see that, for $\alpha \lesssim 10^{-2}$ it is possible to obtain a quite large amount of asymmetry at least for $\theta_{\rm ini} = \mathcal{O}(0.1 - 1)$ which may be quite natural.

The eventual decays of $\Phi$ field transfer the PQ-asymmetry to the asymmetry of RHN-helicity states via the interaction $y_N$ responsible for generating the majorana mass of RHNs.
One may assign 2-units of lepton number to the Peccei-Quinn field, and one-unit to the right-handed neutrino $\nu_R$.
Then, both interactions $y_N$ and $y_\nu$ respects not only $U(1)_{\rm PQ}$ but also $U(1)_{B-L}$, and the PQ-number asymmetry of $\Phi$ field can be regarded as the lepton number asymmetry as $\Delta n_L^\Phi = 2 \Delta n_\Phi$.
In the broken phase of $U(1)_{\rm PQ}$-symmetry, the nonzero majorana mass-term breaks the lepton number conservation.
However, as eigenstates of the equation of motion, helicity states are not affected by the majorana mass-term and the asymmetry between the helicity states of RHNs is not changed, especially, in the CMB frame of the homogeneous and isotropic universe \cite{Bento:2004xu}.
The Yukawa interaction $y_\nu$ violates the helicity conservation, but the effect depends on the speed of RHNs.
The branching ratio of the channel, $N_i^+ \to \ell + H$, is $\l( 1 + \beta_i \r)/2$ whereas that of the channel, $N_i^+ \to \ov{\ell} + \ov{H}$, is $\l( 1 - \beta_i \r)/2$ with $\beta_i$ being the speed of $N_i$.
Hence, the helicity asymmetry of $N_i$ induced from the decay of $\Phi$-field is only partially transferred to the visible sector by the suppression factor $\beta_i$.
As a result, if $N_i$ is never equilibrated, its decay contributes to the lepton number asymmetry of the visible sector as \cite{Bento:2004xu}
\beq
\Delta n_{L, i} = \Delta n_i \times \beta_i
\eeq
where $\Delta n_i$ denotes the helicity asymmetry between $N_i^+$ and $N_i^-$.

%
%However, the breaking effect depends on the speed of RHNs, and it is still possible to assign a leptonic charge to a RHN (positive and negative) helicity-eigenstate $N_i^\pm$ of the mass-eigenstate $N_i$ such as \cite{Bento:2004xu}
%\beq \label{eq:L-of-RHNs}
%L_ i \equiv \beta_i
%\eeq
%where $\beta_i$ is the speed of $N_i$ in the unit of the speed of light.
%Then, the lepton number asymmetry of a RHN can be expressed as
%\beq
%\Delta n_L^i = \Delta n_{\rm h}^i \times \beta_i
%\eeq
%where $\Delta n_{\rm h}^i$ denotes the asymmetry of between $N_i^+$ and $N_i^-$.
%
%The lepton number asymmetry of $N_i$s can be transferred to that of leptons of the visible sector through the lepton number conserving Yukawa interaction $y_\nu$.
%
The late time asymmetry in the visible sector depends not only on the speeds of RHNs but also on the background temperature of the SM thermal 
bath when RHNs decay.
Depending on parameters causing different situations, there are several possibilities for generating baryon number asymmetry of the universe: Affleck-Dine (AD) baryogenesis \cite{Affleck:1984fy} through the decay chain such as $\Phi \to \nu_R \to \ell$  with RHNs out of equilibrium, 
%\footnote{Recently, singlet Majoron scenario using RHNs in supersymmetric model was discussed in Ref.~\cite{}.}, 
spontaneous leptogenesis thanks to the background motion of the axi-Majoron field with thermal RHNs \cite{Chun:2023eqc}, and resonant leptogenesis \cite{Pilaftsis:2003gt} from decays of nonthermal RHNs.

\subsection{Peccei-Quinn Affleck-Dine leptogenesis}
We consider the case in which the decay of $N_i$ happen in nonrelativistic regime and the background temperature is much lower than $m_{N_i}$ so that washing-out processes are not active.
Looking at \eqss{eq:cond-for-MD}{eq:Bi-bnd-for-MD}{eq:T-N-decay}, we see that this is relevant only for the lightest RHN $N_1$ for which $m_{\nu_1}$ can be much smaller than $m_\nu$, and $N_1$ should be responsible for recovering the late time SM thermal bath.

\begin{figure}[t] 
\begin{center}
\includegraphics[width=0.48\textwidth]{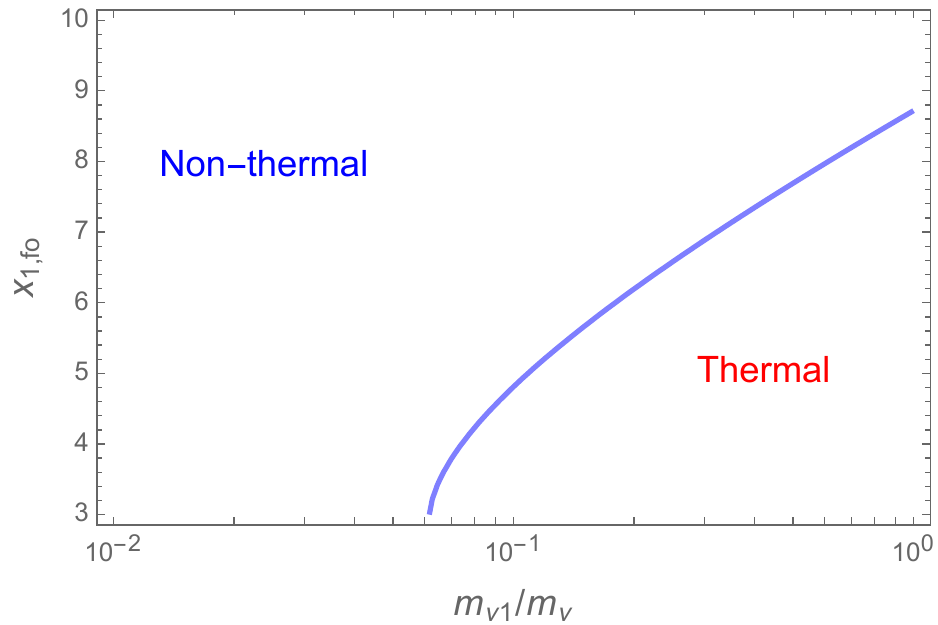}
\caption{
The freeze-out temperature of thermal $N_1$ in terms of $x_{1, \rm fo}$ as a function of $m_{\nu_1}$.
}
\label{fig:N1-freeze-out}
\end{center}
\end{figure}

A part of the PQ-number asymmetry can be transferred to the visible sector through the helicity states of $N_1$, while the contributions from heavier states are washed to a much smaller value.
It is possible for the heavy states to decay when they just become nonrelativistic while producing SM particles with a temperature $T \gtrsim m_N \gtrsim m_{N_1}$.
%For a diagonal mass matrix of RHNs, which we are assuming in this work, the scattering and inverse process to thermalize $N_1$ is efficient only when  $T \sim m_{N_1}$ with $T$ being the temperature in the standard radiation-dominated universe.
In such a thermal bath, the scattering rate of $N_1$ to SM particles when $T \gtrsim m_{N_1}$ is expected to be $\Gamma_{1, s} = c_{1,s} y_{\nu_1}^2 T$ with $c_{1, s}=\mathcal{O}(10^{-2})$ \cite{Bento:2004xu} being a numerical factor including the effect of SM couplings and the number of allowed channels. 
The expansion rate at $T=m_{N_1}$ in our scenario is given by 
\beq
H(T=m_{N_1})=H_{\rm SM}(T=m_{N_1})/\sqrt{2 {\rm B}_N}
\eeq
with $H_{\rm SM}(T)$ being the expansion rate in the standard cosmology in 
a radiation-dominated universe with temperature $T$.
It gives
\bea
\frac{\Gamma_{1,s}}{H(T=m_{N_1})}
%&=& 2 \l( \frac{90}{\pi^2 g_*(T=m_{N_1})} \r)^{1/2} c_{1, s} \sqrt{2 {\rm B}_N} \frac{m_{\nu_1} M_{\rm P}}{v_u^2}
%\nonumber \\
&\approx& 5.2 \times \l( \frac{c_{1, s}}{10^{-2}} \r) \l( \frac{{\rm B}_N}{0.1} \r)^{1/2} \l( \frac{m_{\nu_1}}{m_\nu} \r) \ ,
\eea
where $g_*(T=m_{N_1})=106.75$ was used for the numerical approximation.
Meanwhile, the freeze-out temperature($T_{1, \rm fo}$) of the dominant inverse processes washing-out the $N_1$'s contribution to the visible sector lepton number asymmetry can be somewhat lower than $m_{N_1}$. 
It can be found by equating the expansion rate to the rate of the processes ($\gamma_{1, \rm ID}$) \cite{Chun:2023eqc}:
\beq
H(T_{1, \rm fo}) = \gamma_{1, \rm ID} \equiv \frac{n_{N_1}^{({\rm eq})}}{n_\ell^{({\rm eq})}} \frac{K_1(x_{1, \rm fo})}{K_2(x_{1, \rm fo})} \Gamma_{N_1}
\eeq 
where $x_{1, \rm fo} \equiv m_{N_1}/T_{1, \rm fo}$, $n^{({\rm eq})}$ is the equilibrium number density, and $K_{1,2}$ are the modified Bessel function.
It turns out that $x_{1, \rm fo}$ depends only on $m_{\nu_1}$, and the dependence is shown in Fig.~\ref{fig:N1-freeze-out}.
%%%%%%%%%%%%%%
From the figure, one sees that $N_1$ would be never in equilibrium if $m_{\nu_1} / m_\nu < 6 \times 10^{-2}$.
Note that the upper limit is already well above the bound in \eq{eq:m-nu1-DR-bnd} for the late time recovering of the SM thermal bath without too much dark raditaion. 
Hence, as long as \eq{eq:m-nu1-DR-bnd} is satisfied, even though there would be thermal contribution to the abundance of $N_1^\pm$s as a subdominant portion, $N_1$ produced in the decay of the inflaton can be never thermalized with the SM particles.

The details of the transfer processes of the PQ-number asymmetry to lepton-number asymmetry depend on whether $\phi_*$ is larger or smaller than $\phi_\times$.
If $\phi_* < \phi_\times$, contrary to the conventional scenarios of AD-mechanism in which a AD-field carrying the asymmetry decay always in a symmetric phase, $\Phi$ falls into the vicinity of the true vacuum position before it finally decays.
At the broken phase, $\Phi$ is decomposed into radial and angular (axi-Majoron) modes.
The radial mode does not carry any asymmetry, but it will heat up the universe by its decay.
Hence, AD-mechanism is expected to be relevant only when the oscillation amplitude of $\Phi$ is larger than $\phi_\times$ at which the energy of field is the same as the one at the origin.
The precise criterion to distinguish those two regime is not clear.
For simplicity, we assume that it happens at the epoch of $H_\times$ and include the uncertainty as a suppression parameter $c_{\rm AD}$.
Then, the helicity asymmetry associated with $N_1$ from the partial decay of $\Phi$ at the epoch is given by
\beq
\Delta n_{1, \times} \approx \frac{4c_{\rm AD} {\rm B}_1}{5} \frac{\Gamma^\phi}{H_\times} \Delta n_{\Phi, \times} \ .
\eeq 
It has a suppression factor 
\beq
\Delta_{\rm AD} \equiv c_{\rm AD} \times \frac{2\Gamma^\phi}{5H_\times} \ ,
\eeq 
relative to the expectation in conventional scenarios.
For the decoupled $N_1$, its speed $\beta_{1, \rm d}^{<}$ when it decays is given by
\bea
\beta_{1, \rm d}^{<} 
&=& \frac{m_{\phi, \times}}{2 m_{N_1}}  \l( \frac{{\tilde H}_*}{H_\times} \r)^{2/3} \l( \frac{{\tilde H}_{1, \rm eq}}{{\tilde H}_*} \r)^{1/2} \l( \frac{H_{\rm d}}{{\tilde H}_{1, \rm eq}} \r)^{2/3}
%\nonumber \\
%&=& \frac{1}{4} \l( \frac{\sqrt{3} \lambda_\phi}{2 \pi^2 {\rm B}_1} \r)^{1/3} \l( \frac{m_{\nu_1} M_{\rm P}}{v_u^2} \r)^{2/3}
\nonumber \\
&\simeq& 10^{-3} \times \l( \frac{\xi_{\phi,0}}{10} \r)^{2/3} \l( \frac{0.1}{{\rm B}_1} \r)^{1/3} \l( \frac{10^4 m_{\nu_1}}{m_\nu} \r)^{2/3}  \ ,
\eea
where $m_{\phi, \times} \equiv \sqrt{3 \lambda_\phi} \phi_\times$.
Also, the yield of the helicity asymmetry of $N_1$ at the epoch is found as (see Appendix~\ref{app:Y1d})
\beq \label{eq:Y1d-MD}
Y_{1, \rm d}^{<} 
= 4 \Delta_{\rm AD} \l( \frac{{\tilde H}_*}{{\tilde H}_{1, \rm eq}} \r)^{1/2} \frac{T_{\rm d}}{T_{\Phi, \times}} Y_{\Phi, e} \ .
\eeq
Then, the late time asymmetry transferred through the decays of $N_1$s is found to be \cite{Davidson:2008bu} 
\bea \label{eq:YB-AD-N1-MD}
Y_{B, {\rm AD}}^{<} 
&=& \frac{12}{37} \times \beta_{1, \rm d}^{<} Y_{1, \rm d}^{<}
\nonumber \\
%&=& -\Delta_{\rm AD} \times \beta_{1, \rm d} \times \frac{12}{37} \l( \frac{\sqrt{3}}{8 \pi} \r)^{1/2} \l( \frac{g_*^\Phi}{g_*(T_{\rm d})} \r)^{1/4} \lambda_\phi^{1/4} \l( \frac{m_{\nu_1} M_{\rm P}}{v_u^2} \r)^{1/2} Y_{\Phi, e}
%\nonumber \\
&\simeq& 8.4 \times 10^{-11} \times \l( \frac{0.1}{{\rm B}_1} \r)^{4/3} \l( \frac{\xi_{\phi,0}}{10} \r)^{19/6} \l( \frac{10^4 m_{\nu_1}}{m_\nu} \r)^{7/6} \l( \frac{\phi_0^{\rm ref}}{\phi_0} \r) \l( \frac{Y_{\Phi, e}}{10^{-2}} \r) \ ,
\eea
where we used $c_{\rm AD}=0.1, \ \Gamma^\phi \approx \Gamma_a^\phi$, and $g_{*, \rm d}=106.75$ in the second line.
Since axi-Majorons produced by misalignment are the only contribution to dark matter and kinetic misalignment does not work in our scenario (see Sec.~\ref{sec:DM-DR}), it is required to have $\phi_0 \sim \phi_0^{\rm ref}$.
Also, the validity of  \eq{eq:YB-AD-N1-MD} is limited to the case of $\phi_* < \phi_\times$ which puts a limit $\xi_{\phi,0} \lesssim 100$ (see \eq{eq:xiphi-cond-phis-lower}).
Moreover, the constraint of $\Gamma_{N_1} \ll H_{1, \rm eq}$ requires ${\rm B}_1 \gtrsim \l( m_{\nu_1}/m_\nu \r)^{1/2} / \sin \beta$ (see \eq{eq:cond-for-MD}).
Therefore, if $c_{\rm AD}$ is not much smaller than unity, $Y_{\Phi, e}$ should be smaller than unity by a couple of orders of magnitude in order to match observation.

If $\phi_* > \phi_\times$, the speed of $N_1$ at its decay is then given by
\beq
\beta_{1, \rm d}^{>}
= \frac{m_{\phi, *}}{2 m_{N_1}}  \l( \frac{H_{1, \rm eq}}{H_*} \r)^{1/2} \l( \frac{H_{\rm d}}{H_{1, \rm eq}} \r)^{2/3}
%= \l( \frac{3}{16} \r)^{1/6} \beta_{1, \rm d}^{<}
\eeq
with $H_*$ and $H_{1, \rm eq}$ given in Sec.~\ref{subsec:phis-large}.
The yield of the helicity asymmetry of $N_1$ at the epoch in this case is given by
\beq \label{eq:Y1d-RD}
Y_{1, \rm d}^{>} 
= \l( \frac{H_*}{H_{1, \rm eq}} \r)^{1/2} \frac{T_{\rm d}}{T_{\Phi, *}} Y_{\Phi, e} \ .
\eeq
Note that, compared to \eq{eq:Y1d-MD}, \eq{eq:Y1d-RD} does not have the suppression factor $\Delta_{\rm AD}$.
As a result, the late time baryon number asymmetry is given by
\bea
\label{eq:YB-AD-N1-RD}
Y_{B, \rm AD}^{>} 
&=& \frac{12}{37} \times \beta_{1, \rm d}^{>} Y_{1, \rm d}^{>} 
%\nonumber \\
%&=& \frac{5 \times 2^{7/12} \pi}{3^{4/3} c_{\rm AD}} \l( \frac{\phi_0}{\lambda_\phi M_{\rm P}} \r) Y_{B, \rm AD}^{<}
\nonumber \\
&\simeq& 6.2 \times 10^{-11} \times \l( \frac{0.1}{{\rm B}_1} \r)^{4/3} \l( \frac{\xi_{\phi,0}}{200} \r)^{7/6} \l( \frac{10^4 m_{\nu_1}}{m_\nu} \r)^{7/6} \l( \frac{Y_{\Phi, e}}{10^{-6}} \r) \ .
\eea
Matching observation in this case requires a much smaller $Y_{\Phi, e}$ relative to the case of $\phi_* < \phi_\times$.
It is notable that \eq{eq:YB-AD-N1-RD} does not depend on $\phi_0$ although the relic density of dark matter requires $\phi_0 \sim \phi_0^{\rm ref}$.

%%%%%%%%%%%%%%
 \begin{figure}[t] 
\begin{center}
\includegraphics[width=0.48\textwidth]{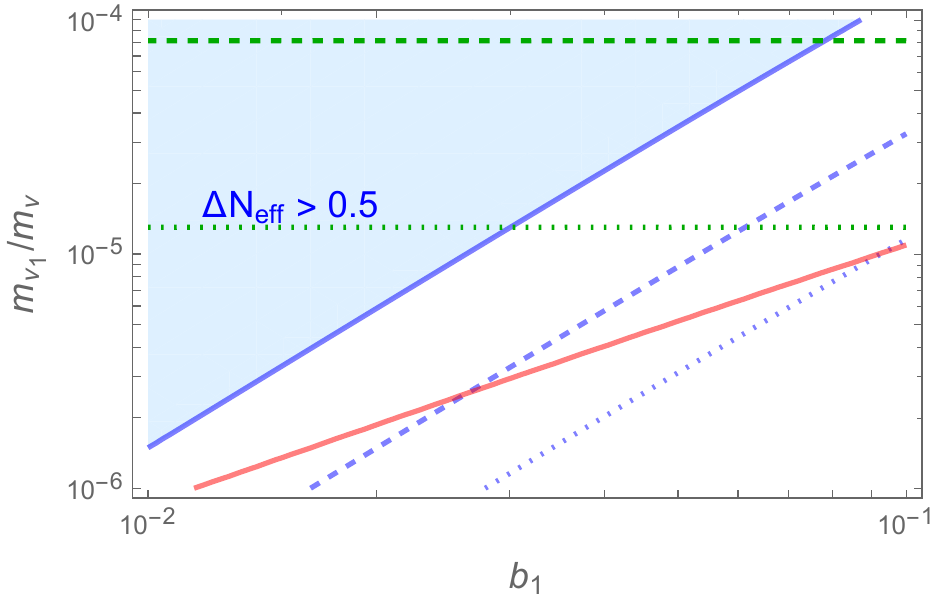}
\hspace{0.1cm}
\includegraphics[width=0.48\textwidth]{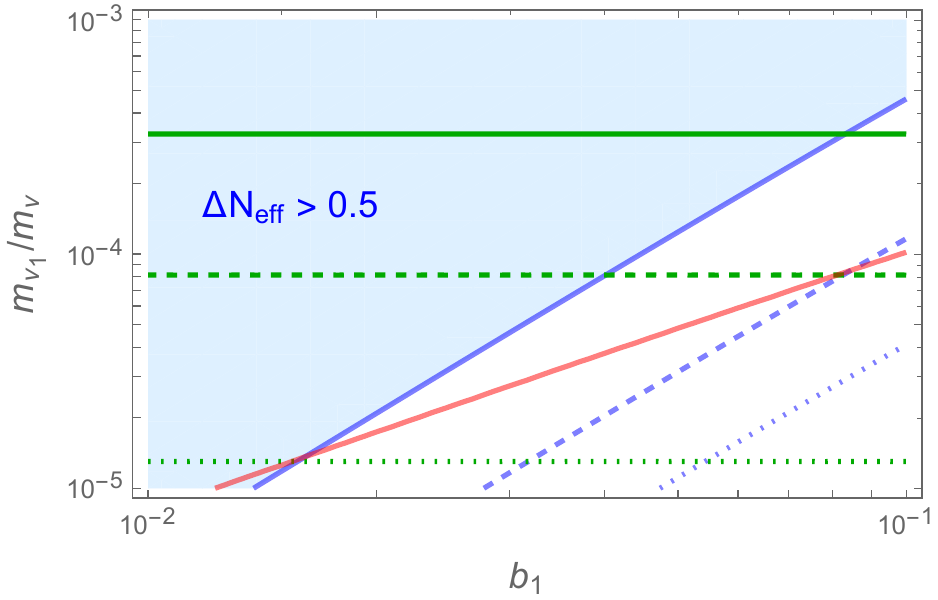}
\caption{
Parameter for a baryon number asymmetry (red solid line), matching observations $Y_B^{\rm obs} \simeq 9 \times 10^{-11}$ \cite{ParticleDataGroup:2022pth}, from the contribution of the lightest RHN via AD-mechanism as a function of $m_{\nu_1}$ and $b_1$ for $b_N=0.1$, and $\phi_0=\phi_0^{\rm ref}$.
Left: $Y_{B, \rm AD}^{<}$ with  $Y_{\Phi, e}=10^{-1}$ in \eq{eq:YB-AD-N1-MD} for $\xi_{\phi,0}=10$ leading to $\phi_*/\phi_\times = 9.1 \times 10^{-3}$.
Right: $Y_{B, \rm AD}^{>}$ with $Y_{\Phi, e}=10^{-6}$ in \eq{eq:YB-AD-N1-RD} for $\xi_{\phi,0}=200$ leading to $\phi_*/\phi_\times = 3.6$.
For a smaller $Y_{\Phi, e}$, the red line is pushed up.
The blue shaded region may be excluded due to too much dark radiation.
Blue lines corresponds to $\Delta N_{\rm eff} = 0.1$(dotted), $0.2$(dashed), and $0.5$(solid).
Green lines are for $T_{\rm d}/m_{N_1} = 0.02$(dotted), $0.05$(dashed), and $0.1$(solid).
}
\label{fig:para-for-AD-N1}
\end{center}
\end{figure}
Figure \ref{fig:para-for-AD-N1} shows the parameter space for $Y_{B, \rm AD}$ matching observations.
As shown in the figure, we can obtain the right amount of baryon asymmetry if $m_{\nu_1}$ is smaller than $m_\nu$ by several orders of magnitude while a sizable extra-radiation can also be obtained from hot axi-Majorons (see Sec.~\ref{sec:DM-DR}).

A remark is in order before closing this subsection.
The initial asymmetry stored in $\Phi$ is only partially transferred to the visible sector with the help of RHNs.
The remaining part of the asymmetry is kept as the background motion of axi-Majoron.
Since $U(1)_{\rm PQ}$ symmetry is broken only spontaneously in the matter sector, the amount of the net PQ-number asymmetry in a comoving volume after inflation should be conserved.
This means that the loss of asymmetry in the visible sector should appear as the motion of axi-Majoron, but has minor impact on the overall homogeneous motion.
Meanwhile, if the background axi-Majoron motion is large enough, it might become a large source of baryon number asymmetry via spontaneous leptogenesis as discussed in Refs.~\cite{Co:2019wyp,Chao:2023ojl,Chun:2023eqc} for example.
We discuss this possibility in the next subsection.

\subsection{Spontaneous leptogenesis via axi-Majoron}
If RHNs are in thermal bath while they are coupled to the derivative of the axi-Majoron field, the idea of spontaneous baryogenesis \cite{Cohen:1987vi} can be applied. 
In Ref.~\cite{Chun:2023eqc}, it was shown that in the presence of a nonzero $\dot{\theta}$, the derivative couplings of the Majoron(the axi-Majoron in our scenario) to fermion currents can act as a source term for generating nonzero lepton number asymmetry via inverse decay processes ($\ell + H \to N_i^\pm$) in equilibrium through the Yukawa interaction $y_\nu$.
In our scenario, due to its nature of sharing the properties of axion and Majoron, axi-Majoron contributes to QCD sphaleron \cite{Mohapatra:1991bz} as well as electroweak sphaleron processes, and Boltzmann equations are to be modified correspondingly.
However, once $L$-number violation is frozen when the inverse decay to $N_i^\pm$ is closed, $B-L$ is also frozen.
Also, $B+L$ is eventually driven to cancel axi-Majoron contribution in the electroweak sphaleron contribution in Boltzmann equations, and $\dot{\theta}/T$ at the freeze-out epoch of the sphalerons is negligible in our scenario as will be clear in the subsequent discussion.
Hence, the eventual baryon and lepton number asymmetries are determined by the nonzero $B-L$ number which is determined in a way similar to Ref.~\cite{Chun:2023eqc}.
The effect of QCD/electroweak sphalerons on the results is minor, and we will ignore such a difference since it is easily compensated by other parameters.

As discussed in the previous section, $N_1$ decays out of equilibrium and is never thermalized.
Hence, only heavy RHNs are relevant for spontaneous leptogenesis in our scenario.
The temperature of the SM thermal bath established by their decays is given 
in \eq{eq:T-N-decay}, and one finds
\beq
H(T_\gamma = m_{N_i}) = \frac{\Gamma_{N_i}}{2} \l( \frac{m_{N_i}}{T_{\gamma, N}} \r)^2
\ , 
\eeq
leading to
\bea \label{eq:dilution-factor}
\eta_i &\equiv& \frac{\Gamma_{N_i}}{H(T_\gamma = m_{N_i})} 
\simeq 66 \times {\rm B}_N^{1/2} \l( \frac{m_{\nu_i}}{m_\nu} \r) \ .
\eea
Therefore, the wash-in effect is strong enough as long as ${\rm B}_N \gtrsim \mathcal{O}(10^{-4})$.
The processes will be shut down as the background temperature becomes sufficiently smaller than the mass of relevant RHN mass-eigenstates.
The expansion rate at that epoch is found to be \cite{Chun:2023eqc}
\beq \label{eq:H-fo}
{\tilde H}_{\rm fo} = c_{\rm inv} \Gamma_N \ ,
\eeq
with $c_{\rm inv} \simeq 2.7 \times 10^{-3}$.
In the standard thermal bath, the freeze-out temperature is then given by
\beq \label{eq:Tfo}
T_{\gamma, {\rm fo}} = \frac{m_N}{x_{\rm fo}} =  \frac{\sqrt{b_N \lambda_\phi}}{x_{\rm fo}} \phi_0
\simeq 2.6 \times 10^6 {\rm GeV} \times b_N^{1/2} \xi_{\phi,0} \l( \frac{\phi_0}{\phi_0^{\rm ref}} \r) \ , 
\eeq
with $x_{\rm fo} \approx 10$.
In our scenario, ${\tilde H}_{\rm fo}$ is larger than the one related to $T_{\gamma, \rm fo}$ in the standard cosmology, leading to a smaller $x_{\rm fo}$.
However, such a dependence is mainly logarithmic on ${\rm B}_N$, and minor as long ${\rm B}_N =\mathcal{O}(0.1)$ which we assume in this work.

The baryon and lepton number asymmetry at the freeze-out in the singlet Majoron scenario is found as \cite{Chun:2023eqc}
\beq \label{eq:n-BL-spon}
\l( n_{B,L} \r)_{\rm fo} = \frac{c_{B,L}}{6} \l( \dot{\theta} T_{\gamma}^2 \r)_{\rm fo}
\ , 
\eeq
where $c_B$ and $c_L$ are model-dependent numerical coefficients of order unity.
Angular velocity $\dot{\theta}$ after inflation scales as
\beq
\dot{\theta} \propto \l\{
\begin{array}{ll}
a^{-1} &: \ \phi \gtrsim \phi_\times \ ,
\\
a^{-3} &: \ \phi \lesssim \phi_\times \ .
\end{array}
\r. 
\eeq
Hence, $\dot{\theta} T_\gamma^2 \propto a^{-5}$ for ${\tilde \phi}_* \ll \phi_0$.
The standard thermal bath is expected to be established by the decay of the lightest RHN $N_1$ well before the EWPT. 
Since $N_1$s should dominate the universe for a while as nonrelativistic particles, there would be a dilution of preexisting baryon and lepton number asymmetries.
The late time baryon number asymmetry through the spontaneous baryogenesis mechanism is given by
\beq
Y_{B, \rm sb} = \frac{12}{37} Y_{B-L}^{\rm d}
\eeq
where $Y_{B-L}^{\rm d}$ is the yield of $B-L$ charge at the epoch of $N_1$'s decay.
It depends on whether the freeze-out of inverse processes producing heavy RHNs takes place before or after $N_1$ starts dominating the universe (i.e., whether ${\tilde H}_{\rm fo} \geq {\tilde H}_{1, \rm eq}$ or not).

For $\phi_* < \phi_\times$, one finds
\bea \label{eq:Hfo-to-Heq}
\frac{{\tilde H}_{\rm fo}}{{\tilde H}_{1, \rm eq}}
%&=& c_{\rm inv} \l( \frac{b_N m_{\nu_N}}{b_1 m_{\nu_1}} \r) \l( \frac{\Gamma_{N_1}}{{\tilde H}_{1, \rm eq}} \r)
%\nonumber \\
&=& \frac{3 c_{\rm inv}}{2} \frac{b_N}{b_1 {\rm B}_1^2 \lambda_\phi^{1/2}} \l( \frac{m_{\nu_N} \phi_0}{v_u^2} \r)
\nonumber \\
&\simeq& \frac{1.4}{\sin^2 \beta} \times \l( \frac{b_N}{b_1} \r) \l( \frac{0.1}{{\rm B}_1} \r)^2 \l( \frac{10}{\xi_{\phi,0}} \r) \l( \frac{\phi_0}{\phi_0^{\rm ref}} \r) \ .
\eea
%%%%%%%%%%%%%%
 \begin{figure}[t] 
\begin{center}
\includegraphics[width=0.48\textwidth]{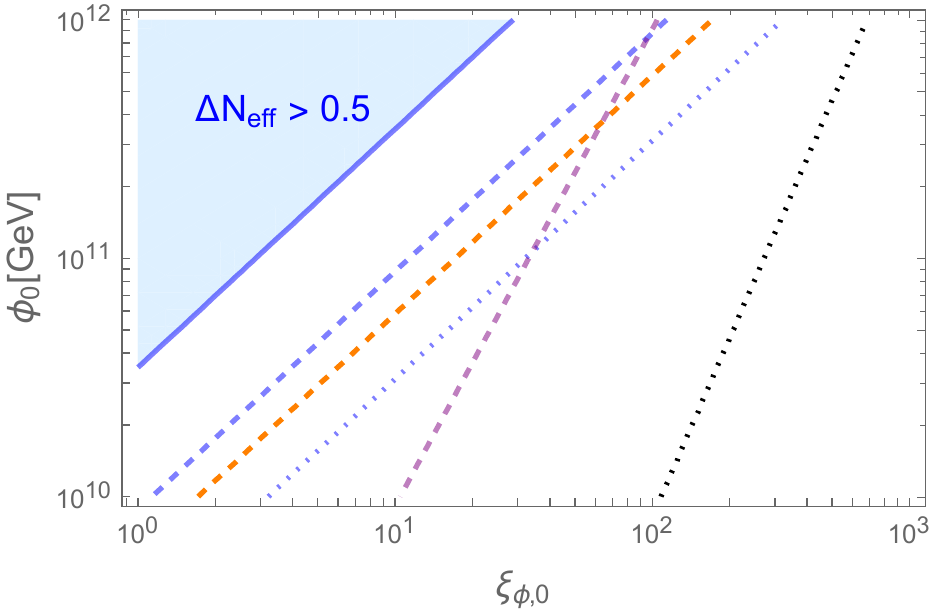}
\caption{
Parameter space of ${\tilde H}_{\rm fo}/{\tilde H}_{1, \rm eq}=1$(orange dashed line), $\Delta N_{\rm eff} = 0.5$(blue solid line), $0.2$(blue dashed line), $0.1$(blue dotted line), and $\phi_*/\phi_\times=1$(purple dashed line) for $b_1=0.05, \ b_N=0.1$, $m_{\nu_N}=m_\nu = 10^4 m_{\nu_1}$, and $\sin \beta = 1$ as an example.
The black dotted line is the upper bound of $\xi_{\phi,0}$ given in \eq{eq:non-trap-cond}.
The position of purple line does not depend on $b_i$s and $m_{\nu_i}$s.
A larger/smaller $b_1$ with other parameters fixed shifts the orange dashed line to the left/right (see \eq{eq:Hfo-to-Heq}).
A smaller $\sin \beta$ shifts the line to the right.
The blue shaded regions and blue lines are move in the same way, but a smaller $m_{\nu_1}$ pushes them to the left.
}
\label{fig:para-space-for-Hfo}
\end{center}
\end{figure}
%%%%%%%%%%%%%%%
Figure \ref{fig:para-space-for-Hfo} shows the parameter spaces for ${\tilde H}_{\rm fo}/{\tilde H}_{1, \rm eq}$.
In the region of ${\tilde H}_{\rm fo} \geq {\tilde H}_{1, \rm eq}$, one finds 
\bea \label{eq:YBL-d}
Y_{B-L}^{\rm d} &=& \l( \frac{s_{\rm fo}}{s_{\rm d}} \r) \l( \frac{a_{\rm fo}}{a_{\rm d}} \r)^3 Y_{B-L}^{\rm fo}
\nonumber \\
&=& 4 {\rm B}_N x_{\rm fo} \l( \frac{T_{\rm d}}{m_N} \r) \l( \frac{{\tilde H}_{\rm fo}}{{\tilde H}_{1, \rm eq}} \r)^{1/2} Y_{B-L}^{\rm fo} \ .
\eea
The $B-L$ asymmetry at freeze-out is given by
\beq \label{eq:YBL-fo}
Y_{B-L}^{\rm fo} = \frac{45 c_{B-L}}{12 \pi^2 g_{*S}(T_{\gamma, \rm fo})} \l( \frac{\dot{\theta}}{T} \r)_{\rm fo} \ ,
\eeq
where $c_{B-L} = \mathcal{O}(1)$ \cite{Chun:2023eqc}.
The velocity of axi-Majoron at the epoch of ${\tilde H}_{\rm fo}$ is given by
\beq \label{eq:dth-fo}
\dot{\theta}_{\rm fo} \approx \dot{\theta}_e \l( \frac{\phi_\times}{\phi_e} \r) \l( \frac{{\tilde H}_*}{H_\times} \r)^2 \l( \frac{{\tilde H}_{\rm fo}}{{\tilde H}_*} \r)^{3/2} \ .
\eeq
%Also, the decay temperature of $N_1$ is found as  
%\beq
%\frac{T_{\rm d}}{m_N} = \l( \frac{5}{2 \pi^4 g_*(T_{\rm d})} \r)^{1/4} \l( \frac{m_{\nu_1} M_{\rm P}}{v_u^2} \r)^{1/2} \l( \frac{b_1}{b_N} \r)^{1/2}, 
%\eeq
%and
%\bea \label{eq:Hfo-to-Heq}
%\l( \frac{H_{\rm fo}}{H_{1, \rm eq}} \r)^{1/2}
%&=& c_{\rm inv}^{1/2} \l( \frac{b_N m_{\nu_N}}{b_1 m_{\nu_1}} \r)^{1/2} \l( \frac{\Gamma_{N_1}}{H_{\rm eq}} \r)^{1/2} 
%\nonumber \\
%&=& \l[ \frac{3 c_{\rm inv}}{4\sqrt{2}} \frac{b_N}{b_1 {\rm B}_1^2 \lambda_\phi^{1/2}} \l( \frac{m_{\nu_N} \phi_0}{v_u^2} \r) \r]^{1/2}  \ .
%\eea
Combining all, we find the late-time baryon number asymmetry for $\phi_* < \phi_\times$ as
\bea \label{eq:YB-SB-MD}
Y_{B, \rm sb}^{<} 
&\approx& \frac{3.3 c_{B-L} c_{\rm inv}^2 x_{\rm fo}^2}{\pi^{9/2} g_{*, \rm d}^{1/4} g_{*, \rm fo}} \frac{b_N^2 \lambda_\phi^{5/4}}{b_1} \l( \frac{m_{\nu_1} M_{\rm P}}{v_u^2} \r)^{1/2} \l( \frac{m_{\nu_N} M_{\rm P}}{v_u^2} \r)^2 Y_{\Phi, e}
\nonumber \\
&\simeq& \frac{1.5 \times 10^{-10}}{\sin^5 \beta} \l( \frac{0.05}{b_1} \r) \l( \frac{b_N}{0.1} \r)^2 \l( \frac{\xi_{\phi,0}}{50} \r)^{5/2} \l( \frac{10^4 m_{\nu_1}}{m_\nu} \r)^{1/2} \l( \frac{Y_{\Phi, e}}{0.1} \r) \ ,
\eea
where we used $c_{B-L}=2, \ g_{*, \rm d} = g_{*, \rm fo} \equiv g_*(T_{\rm fo}) = 106.75$, and $m_{\nu_N} = m_\nu$ in the second line.
Note that, $Y_B^{<}$ does not depend on the symmetry-breaking scale.

Comparing \eq{eq:YB-SB-MD} to \eq{eq:YB-AD-N1-MD}, one finds
\beq
\frac{Y_{B, \rm sb}^{<}}{Y_{B, \rm AD}^{<}} \simeq \frac{55}{\sin^5 \beta} \times {\rm B}_1^{1/3} b_3 {\rm B}_3 \l( \frac{m_{\nu}}{10^8 m_{\nu_1}} \r)^{2/3} \ .
\eeq
It implies that, as long as $\beta \gtrsim 1$, in order for the contribution from spontaneous leptogenesis to become relevant, $m_{\nu_1}$ should be smaller than $m_{\nu}$ by many orders of magnitude.
Meanwhile, $Y_{\Phi, e} \lesssim \mathcal{O}(1)$ as shown in \eq{eq:Y-PQ-e} and Fig.~\ref{fig:YPQ-e-vs-theta-ini}.
Hence, $Y_{B, \rm sb}^{<}$ is always subdominant, relative to $Y_{B, \rm AD}^{<}$ for a given set of parameters. 

Depending on $b_1$, the freeze-out of the inverse processes could take place when $H={\tilde H}_{\rm fo}' < {\tilde H}_{1, \rm eq}$ while $\phi_* < \phi_\times$ is maintained.
From the scaling behavior of $Y_{B-L}$ after the freeze-out epoch, one can see that, relative to the case of  ${\tilde H}_{\rm fo} > {\tilde H}_{1, \rm eq}$,  $Y_{B-L}^{\rm d}$ in this case is suppressed by a factor 
\beq
\l( \frac{{\tilde H}_{1, \rm eq}} {{\tilde H}_{\rm fo}}\r) \l( \frac{{\tilde H}_{\rm fo}'}{{\tilde H}_{1, \rm eq}} \r)^{1/2} < 1 \ .
\eeq
That is, the contribution to the baryon number asymmetry in the visible sector is always subdominant than the one in \eq{eq:YB-SB-MD}. 
Hence, the contribution from AD-mechanism is expected to be always dominant in this case, too, and we do not consider the details of this case further.

As the last possibility arising in spontaneous leptogenesis in our scenario, as shown in Fig.~\ref{fig:para-space-for-Hfo}, and inferred from \eq{eq:Hfo-to-Heq}, if $\xi_{\phi,0} \gtrsim \mathcal{O}(10^2)$ with $b_1$ not much smaller than $b_N$, the case of $H_{\rm fo} < H_{1, \rm eq}$ can arise with $\phi_* > \phi_\times$, too.
The $B-L$ asymmetry at the epoch of $N_1$'s decay in this case is given by
\bea \label{eq:YBL-d-RD}
Y_{B-L, \rm d}^{>} &=& \l( \frac{s_{\rm fo}}{s_{\rm d}} \r) \l( \frac{a_{\rm fo}}{a_{\rm d}} \r)^3 Y_{B-L}^{\rm fo}
\nonumber \\
&=& 4 {\rm B}_N x_{\rm fo} \l( \frac{T_{\rm d}}{m_N} \r) \l( \frac{H_{\rm fo}}{H_{1, \rm eq}} \r)^{2/3} Y_{B-L}^{\rm fo} \ ,
\eea
where $H_{1, \rm eq}$ is to be taken from \eq{eq:H-eq-RD}.
Since the freezing-out of the inverse processes in this case takes place during matter-domination era, $H_{\rm fo}$ in \eq{eq:YBL-d-RD} is larger than the one in radiation-domination era.
If $H_{1, \rm eq} > H_{\rm fo} \gg H_{\rm d}$, ignoring the contribution of the partial decay of $N_1$ to the SM thermal bath, one can express $H_{\rm fo}$ as
\beq
H_{\rm fo} = H_{1, \rm eq} \l( \frac{x_{1, \rm eq}}{x_{\rm fo}} \r)^{3/2}  \ ,
\eeq
where $x_{1, \rm eq} \equiv m_N/T_{1, \rm eq}$ with $T_{1, \rm eq}$ being the temperature when $N_1$ starts dominating the universe.
In this case, $x_{1, \rm eq}$ is found as
\beq
x_{1, \rm eq} 
= \l( \frac{3 \pi^2}{80} g_{*, \rm eq} \r)^{1/4} \frac{b_N^{1/2} \lambda_\phi^{1/4}}{b_1^{1/2} {\rm B}_1 {\rm B}_N^{1/4}} \ ,
\eeq
where $g_{*, \rm eq} \equiv g_*(T_{1, \rm eq})$.
Equating the expansion rate to the interaction rate of the inverse processes, one finds that $x_{\rm fo} \sim 10$ whereas $x_{1, \rm eq} \approx 0.3 - 9$ for $b_1 = 0.01 - 1$ with $b_N=0.1$ as an example. 
Hence, considering the case as an illustrative set of parameters, we may take $x_{\rm fo}=10$ for simplicity in the subsequent discussion.
The velocity of axi-Majoron at the epoch of $H_*$ is given by
\beq
\dot{\theta}_* = \dot{\theta}_e \l( \frac{\phi_0}{\phi_e} \r) \l( \frac{\phi_*}{\phi_0} \r)^3 \ , 
\eeq
where the change of the background field value from $\phi_* \to \phi_0$ was taken into account under the conservation of angular momentum(charge-conservation).
The freeze-out velocity of axi-Majoron is given by
\beq \label{eq:dth-fo}
\dot{\theta}_{\rm fo} \approx \dot{\theta}_*\l( \frac{H_{1, \rm eq}}{H_*} \r)^{3/2} \l( \frac{H_{\rm fo}}{H_{1, \rm eq}} \r)^2 \ .
\eeq
Combining all, in this case we find
\bea \label{eq:YB-SB-RD}
Y_{B, \rm sb}^{>} 
&\approx& \frac{0.54 \times c_{B-L} b_N \lambda_\phi^{5/4}}{\pi^{1/2} x_{\rm fo}^2 g_{*, \rm d}^{1/4} {\rm B}_1} \l( \frac{m_{\nu_1} M_{\rm P}}{v_u^2} \r)^{1/2} Y_{\Phi, e}
\nonumber \\
&\simeq& \frac{1.8 \times 10^{-10}}{\sin \beta} \l( \frac{b_N}{{\rm B}_1} \r) \l( \frac{\xi_{\phi,0}}{200} \r)^{5/4} \l( \frac{10^4 m_{\nu_1}}{m_\nu} \r)^{1/2} \l( \frac{Y_{\Phi, e}}{0.1} \r) \ ,
\eea
where we have used $x_{\rm fo}=10$ and $c_{B-L}=2$ in the second line.
The observed baryon asymmetry can be obtained by interplay between $b_i, \ \xi_{\phi,0}, \ m_{\nu_1}$, and $Y_{\Phi, e}$.
Comparing \eq{eq:YB-SB-RD} to \eq{eq:YB-AD-N1-RD}, we see that the contribution via spontaneous leptogenesis mechanism cannot catch up the one from AD-mechanism in this case, too.

\subsection{Resonant leptogenesis}
If the inflaton does not carry enough asymmetry, the baryon number asymmetry of the present universe should be obtained by decays of RHNs, following the usual way of out-of-equilibrium decay.
Because of the tiny $y_\nu$ in our scenario, the so-called ``resonant leptogenesis'' \cite{Pilaftsis:2003gt} may be the only possibility in this case.
The essential point of the mechanism is that the \textit{CP} violation in the decay of $N_i$ mass-eigenstate can be given by
\beq
\delta_i 
\approx {\tilde \delta}_{ij} \times \frac{\l( m_i^2 - m_j^2 \r) m_i \Gamma_j}{\l( m_i^2 - m_j^2 \r)^2 + m_i^2 \Gamma_j^2}
\ \xrightarrow[]{\l( \Delta m \r)_{ij} \ll m_i} \
{\tilde \delta}_{ij} \times \frac{2 \l( \Delta m \r)_{ij} m_i^2 \Gamma_j}{4 m_i^2 \l( \Delta m \r)_{ij}^2 + m_i^2 \Gamma_j^2} \ ,
\eeq
where 
\beq
{\tilde \delta}_{ij} \equiv \l[ \frac{{\rm Im} \l( y_\nu^\dag y_\nu \r)_{ij}^2}{\l( y_\nu^\dag y_\nu \r)_{ii} \l( y_\nu^\dag y_\nu \r)_{jj}} \r]
\eeq
and $\l(\Delta m \r)_{ij} \equiv |m_i - m_j|$.
Hence, if $\l(\Delta m \r)_{ij} \sim \Gamma_i \simeq \Gamma_j$, it is possible to have $\delta_i \sim \mathcal{O}(1)$.
Since the lightest RHN with $\l| y_{\nu_1} \r| \ll \l| y_{\nu_{2,3}} \r|$ is to recover the standard thermal bath via its delayed decay (see Sec.~\ref{subsec:reheating}), two heavier RHN mass-eigenstates, $N_{2,3}$, are responsible for the resonant leptogenesis.
Lepton number asymmetry is produced well before the electroweak phase transition.
It is diluted by the later entropy injection from the decay of the long-lived lightest RHN, $N_1$.
The present baryon number asymmetry is then expected to be \cite{Pilaftsis:2003gt}
\beq \label{eq:Y-B-0}
Y_{B, \rm res} \sim \sum_{i = 2,3} \frac{\delta_i}{1400 \eta_i} \l( \frac{\Gamma_1}{{\tilde H}_{1, \rm eq}} \r)^{1/2} 
\eeq
where the dilution factor $\eta_i$ is given in \eq{eq:dilution-factor}, and as an example, we used the ${\tilde H}_{1, \rm eq}$ as the expansion rate at the epoch when $N_1$ starts dominating the universe. 
The use of $H_{1, \rm eq}$ instead of ${\tilde H}_{1, \rm eq}$ would require a smaller $\delta_{2,3}$ to match observations for a given set of parameters.

\begin{figure}[t] 
\begin{center}
\includegraphics[width=0.48\textwidth]{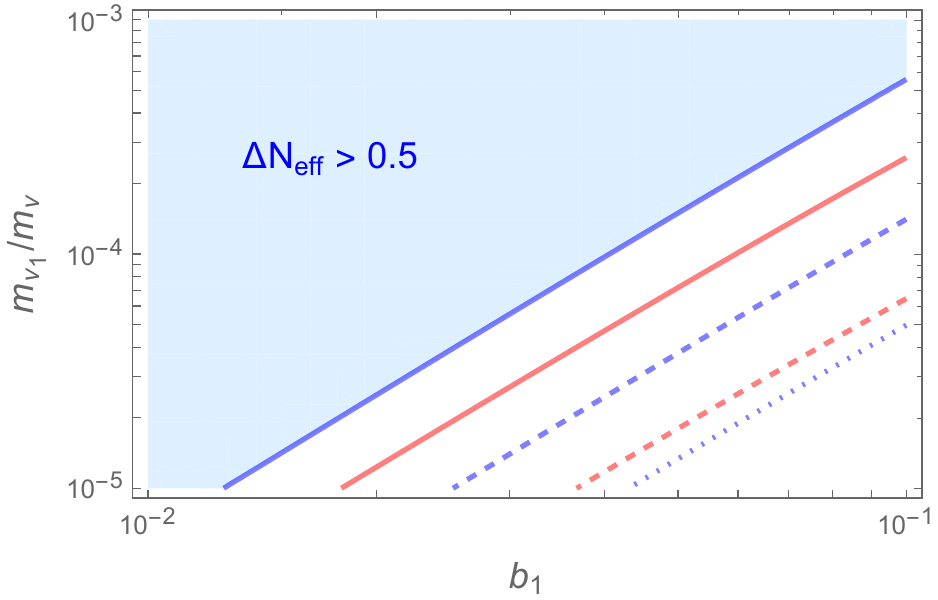}
\caption{
The case of resonant leptogenesis for the late time baryon number asymmetry, $Y_{B, \rm res}$ in \eq{eq:YB-res} matching observations as a function of $m_{\nu_1}$ and ${\rm b_1}$ for ${\tilde \delta}_{23}= 2 \times 10^{-5}$(solid red), $4 \times 10^{-5}$(dashed red) wtih $c_m=1, \ \sin \beta=1, \ b_N=0.15$, $\xi_{\phi,0}=50$, and $\phi_0=\phi_0^{\rm ref}$.
The scheme of the blue region and the blue lines is the same as the one in Fig.~\ref{fig:para-for-AD-N1}.
%Blue lines and shaded region are the same as Fig.~\ref{fig:YB}.
}
\label{fig:YB-res}
\end{center}
\end{figure}

The mass splitting between $N_2$ and $N_3$ before their decay is mainly from 1-loop corrections \footnote{The SM thermal bath is initially established from the decay of those heavy RHNs.
Thermal correction to the mass splitting of $N_2$ and $N_3$ before their decay is negligible.}.
It is given by \cite{Pilaftsis:1997jf}
\beq
\l( \Delta m \r) = c_m \frac{y_\nu^2}{16 \pi} m_N = \frac{c_m}{2} \Gamma_N
\eeq
where $c_m = \mathcal{O}(1)$ is a numerical coefficient.
%%
%\beq \label{eq:eta}
%\eta = \frac{\sqrt{{\rm B}_N}}{4 \pi} \l( \frac{90}{\pi^2 g_*} \r)^{1/2} \l( \frac{m_\nu M_{\rm P}}{v_u^2} \r)
%\simeq 47 \times \sqrt{{\rm B}_N} \l( \frac{m_\nu}{0.05 {\rm eV}} \r)
%\eeq
%In \eq{eq:eta}, $\eta$ is proportional to $\sqrt{{\rm B}_N}$, but the equation is valid only for $\eta \geq 1$.
%%
Combining all of these with \eq{eq:cond-for-MD} for $\Gamma_{N_1}/{\tilde H}_{1, \rm eq}$, the present value of $Y_{B, \rm res}$ in \eq{eq:Y-B-0} is given by 
\bea \label{eq:YB-res}
Y_{B, \rm res} &\sim& \frac{\pi^2 {\tilde \delta}_{23} c_m^2}{700 \l( c_m^2+1 \r) \lambda_\phi^{1/4} {\rm B}_1 \sqrt{{\rm B}_N}} \l( \frac{g_*(T_{\gamma, N})}{15 \sqrt{2}} \r)^{1/2} \frac{\sqrt{m_{\nu_1} \phi_0} v_u}{m_\nu M_{\rm P}}
\nonumber \\
&\simeq&  \frac{3.0 \times 10^{-6} \times {\tilde \delta}_{23} c_m^2 \sin \beta}{\l( c_m^2 + 1 \r) {\rm B}_1 \sqrt{{\rm B}_N \xi_{\phi,0}}} \l( \frac{10^3 m_{\nu_1}}{m_\nu} \r)^{1/2} \l( \frac{\phi_0}{\phi_0^{\rm ref}} \r)^{1/2} \ .
\eea
Therefore, as shown in Fig.~\ref{fig:YB-res}, even for ${\tilde \delta}_{23} \ll 1$ we can obtain a right amount of baryon asymmetry consistent with observations for $\phi_0 \sim \phi_0^{\rm ref}$ which saturates the relic density of dark matter consisting of axi-Majorons.
%%%%%%%%%%%%%%
%%%%%%%%%%%%%%%

\section{Dark matter and dark radiation}
\label{sec:DM-DR}

In our scenario, two types of axi-Majorons appear: hot and cold axi-Majorons.
Hot axi-Majorons are produced in decays of the Peccei-Quinn inflaton as discussed in Sec.~\ref{sec:inf}.
At the epoch of the late time decay of the lightest RHN mass-eigenstate, the energy density of axi-Majorons to the radiation density of SM particles is given by
\bea
\frac{\rho_a}{\rho_r} &=& \l( \frac{H_{\rm d}}{H_{1, \rm eq}} \r)^{2/3} \ .
\eea
where $H_{1, \rm eq}$ is either from \eq{eq:Heq-i} or \eq{eq:H-eq-RD}.
Depending on the ratio, $\Gamma_{N_1} / H_{1, \rm eq}$, which determines the duration of an early matter-domination era, the contribution of hot axi-Majorons to the total energy density of the universe can be still sizable and contributes to the effective number of extra neutrino species, $\Delta N_{\rm eff}$, as a component of dark radiation.
In the present universe, it is expected to be
\bea \label{eq:Delta-N-eff}
\Delta N_{\rm eff} 
&=& \l( \frac{\rho_a}{\rho_\nu} \r)_0
\nonumber \\
&=& \frac{g_*(T_{\gamma, \rm d})}{\l(7/8\r) g_\nu} \l( \frac{T_{\gamma, 0}}{T_{\nu, 0}} \r)^4 \l( \frac{g_{*S}(T_{\gamma, 0})}{g_{*S}(T_{\gamma, \rm d})} \r)^{4/3} \l( \frac{2}{3} \frac{\Gamma_{N_1}}{H_{1, \rm eq}} \r)^{2/3}
\nonumber \\
&\simeq& 0.47 \times \l( \frac{10 \Gamma_{N_1}}{H_{1, \rm eq}} \r)^{2/3} \ .
\eea
where the subscript $_0$ stands for quantities at the present universe, $\rho_\nu$ is the energy density of a single neutrino species, $T_{\gamma, \rm d}$ and $T_{\gamma, 0}$ are respectively the photon temperatures at the epoch of $N_i$'s decay and the present time, and we used $g_*(T_{\gamma, \rm d})=g_{*S}(T_{\gamma, \rm d})=106.75$ and $g_{*S}(T_{\gamma, 0})=3.9$ in the last line.
It may help to ameliorate the tension in observations for the expansion rate of the present universe \cite{DiValentino:2021izs}.

Cold axi-Majorons in our scenario are produced only via the misalignment mechanism, since $U(1)_{\rm PQ}$ is already broken during inflation and never restored.
It is the only component of cold dark matter in our scenario.
In the minimal DFSZ axion scenario we are considering here, the axion decay constant is given by $f_a = \phi_0/c_a$ with $c_a=6$ being the QCD-anomaly coefficient of $U(1)_{\rm PQ}$ symmetry with the charge of PQ-field $\Phi$ being normalized to be unity.
In conventional scenarios where the initial kinetic energy of the axion is ignored, for $\theta_{\rm mis} = \mathcal{O}(1)$ as the misalignment angle, a right amount of dark matter is obtained for $f_a \approx 3 \times 10^{11} {\rm GeV}$ \cite{Marsh:2015xka}. 
If axi-Majoron has a sufficiently large kinetic energy to delay the onset of its low energy coherent oscillation due to angular misalignment, there would be a chance of the kinetic misalignment \cite{Co:2019jts} even for $\phi_0$ much smaller than $\phi_0^{\rm ref}$.
It requires $Y_\Phi = \mathcal{O}(10) \times \l( \phi_0 / 10^9 {\rm GeV} \r)$ as the ratio of the PQ-number density to the entropy density of the SM particles around the epoch of axi-Majoron oscillations in conventional scenarios.
However, in our scenario, as shown in \eq{eq:Y-PQ-e}, the angular motion is limited as $Y_{\Phi, e} \lesssim \mathcal{O}(1)$ with $\alpha \lesssim \mathcal{O}(10^{-2})$ which is necessary to obtain sufficiently large motion of axi-Majoron after inflation [see \eq{eq:Y-PQ-e}].
Hence, the kinetic misalignment does not work, and the symmetry-breaking scale is constrained as $\phi_0 \sim \phi_0^{\rm ref}$ for $\mathcal{O}(1)$ misalignment angle.  

The amount of dark radiation is determined by the duration of the matter-domination (MD) era driven by the lightest RHN, as shown in \eq{eq:Delta-N-eff}.
The presence of an early MD-era reduces the $e$-folds associated with cosmic scales relative to the case of the standard cosmology.
As a result, the spectral index in our scenario is expected to be smaller than in the standard case.
The difference relative to the expectation in the standard cosmology is small, but this implies that, if some amount of dark radiation is detected in future experiments, in principle, it can be used for a consistency check of our scenario with the spectral index of density perturbations, although it will strongly depend on the sensitivity of the experiments.
Also, such an early MD-era causes dilution of preexisting particles due to the entropy release at the end of the era.
Specially, the nearly scale-invariant inflationary gravitational waves would be diluted for cosmic scales which were inside the horizon at the epoch of the decay of the dominating particle.
If detected, firstly, the inflationary gravitational waves provide information of the energy scale during inflation, i.e., $H_I$ in \eq{eq:H-inflation}.
When combined with inflationary observables, $\lambda_\phi$ and $\xi_{\phi,0}$ can be nearly fixed, modulo the minor variation due to the possible angular motion during inflation when $\alpha$ is somewhat sizable.  
Additionally, the spectra of the primordial gravitational waves at the present universe will have spectral changes at two frequencies:
The characteristic frequencies at the present universe are given by
\bea \label{eq:feq0}
f_{{\rm eq}, 0} &=& f_{N_1, 0} \l( \frac{H_{1, \rm eq}}{H_{\rm d}} \r)^{2/3},
\\
\label{eq:f10}
f_{N_1,0} 
&=& f_{N_1, \rm d} \l( \frac{g_{*S}(T_0)}{g_{*S}(T_{\rm d})} \r)^{1/3} \frac{T_0}{T_{\rm d}}
%\nonumber \\
%&=& f_{{\rm ew},0} \l( \frac{2}{3} \frac{\Gamma_{N_1}}{H_{\rm ew}} \r) \l( \frac{g_{*S}(T_{\gamma, \rm ew})}{g_{*S}(T_{\gamma, \rm d})} \r)^{1/3} \frac{T_{\gamma, \rm ew}}{T_{\gamma, \rm d}}
\nonumber \\
&\simeq& 2.7 \times 10^{-6} {\rm Hz} \times \l(\frac{H_{\rm d}}{H_{\rm ew}} \r)^{1/2} 
\ .
\eea
where $T_0$ is the background temperature of the present universe.
In the second line of \eq{eq:f10}, we used $f_{N_1, \rm d} = H_{\rm d}/2 \pi$ with $H_{\rm d}= (2/3)\Gamma_{N_1}$, and $H_{\rm ew}$ is the expansion rate of the universe at the epoch of the EWPT with photon temperature $T_{\rm ew} = 100 {\rm GeV}$.
These frequencies are associated with the beginning [\eq{eq:feq0}] and the end [\eq{eq:f10}] of the matter-domination era.
All modes of $f > f_{{\rm eq},0}$ are suppressed by the factor $\l( H_{\rm d}/H_{\rm eq} \r)^{2/3}$ while modes of $f<f_{N_1,0}$ are not affected.
Once such a feature of GWs is detected in some future experiments sensitive to the energy scale of inflation in our scenario(corresponding to $\Omega_{\rm GW}h^2 \sim \mathcal{O}(10^{-17})$ as the fractional energy of GWs at the present universe), it would provide information of $H_{1, \rm eq}$ and $\Gamma_{N_1}$ separately \cite{Bernal:2019lpc}, as long as $\Gamma_{N_1}/H_{1, \rm eq}$ is not much smaller than unity.
Even in the case only $f_{N_1,0}$ is detected, if available, $\Delta N_{\rm eff}$ can be used to determine $H_{\rm eq}$.
Then, since dark matter relic density would nearly fix $\phi_0$, it may be possible to extract information of $m_{\nu_1}$ and $m_{N_1}$ if $\tan \beta$ is tightly constrained by collider experiments so as to lead to $\sin \beta \approx 1$ \cite{Su:2019dsf}.
To our knowledge, this is a unique way of probing those mass parameters in the seesaw sector.

\section{Conclusions}
\label{sec:conc}
In this work, we showed the details of `\textit{the minimal cosmological standard model (MCSM)}' which we proposed recently in Ref.~\cite{Barenboim:2024akt}.
The model is a minimal scenario beyond the standard model, which includes the seesaw sector too in the form of the singlet Majoron scenario. 
Scale-invariance was imposed for both of the matter and the gravity sectors. 
the Planck scale emerges dynamically from the spontaneous breaking of 
scale-invariance.
The global Peccei-Quinn symmetry, $U(1)_{\rm PQ}$, was imposed on the matter sector, but it was assumed to be broken in the gravity sector in a scale-invariant manner.
The model is a combination of DFSZ axion model and the singlet Majoron scenario, in which Peccei-Quinn field not only provides the axion solution but also is responsible for generation of the masses of right-handed neutrinos for the seesaw mechanism.

We showed that our scenario provides a simple unified framework simultaneously addressing the following phenomenological/cosmological/theoretical puzzles:  (i) the origin of scales, (ii) a period of primordial inflation, (iii) baryogenesis, (iv) tiny neutrino masses, (v) dark matter, and (vi) strong \textit{CP}-problem.
Primordial inflation in our scenario is realized along the Peccei-Qinn field direction thanks to a nonminimal gravitational interaction and scale-invariance.
The presence of the symmetry-breaking nonminimal coupling causes in general angular motion of the inflaton, and can generate a large PQ-number asymmetry at the end of inflation.
It is nothing but the way of generating particle asymmetry in Affleck-Dine mechanism of baryogenesis, which is naturally realized in our setup.
Depending on parameters, the asymmetry can be transferred to the visible sector through right-handed neutrinos in various ways.
It can be through a nonthermal decay of the lightest RHNs which have a helicity-asymmetry induced by the asymmetry of PQ-number.
This is a kind of usual Affleck-Dine scenario, although details are somewhat different in our scenario.
Additionally, the asymmetry in the visible sector can be obtained via thermal processes, as in the case of spontaneous baryogenesis mechanism. 
In this case, the remnant angular motion of axi-Majoron after the decay of the Peccei-Quinn inflaton provides the effective chemical potential to the fields charged under $U(1)_{\rm PQ}$-symmetry.
Compared to the contribution by AD-mechanism, this contribution to the baryon number asymmetry is always subdominant, though.
The third possibility is the resonant leptogenesis from the decay of nearly degenerate heavy RHN mass-eigenstates.
This case is relevant when the initial PQ-number asymmetry after inflation is not enough for neither AD-mechanism nor spontaneous baryogenesis.

Cold dark matter is purely axi-Majorons produced only through the misalignment mechanism.
Kinetic misalignment mechanism does not work, since the kick of the axi-Majoron at the end of inflation is limited by axi-Majoron's initial position and velocity during inflation.
Hence, the symmetry-breaking scale of $U(1)_{\rm PQ}$, i.e., $\phi_0$ is constrained to be of $\mathcal{O}(10^{12}) {\rm GeV}$.
A sizable amount of dark radiation can be obtained naturally, and may be helpful for ameliorating the tension between observations for the expansion rate of the present universe.

In regard to axion solution, the symmetry-breaking term in the gravity sector does not invalidate the solution, since it is effectively turned off at low energy due to its field-dependent nature which we assumed.
The axion-quality problem may also be absent in the presence of Gauss-Bonnet term.
Although our scenario is a DSFZ axion model having more than one domain, it does not have a domain wall problem thanks to the sizable nonzero angular motion and weak interactions of Peccei-Quinn field to RHNs that make the symmetry never restored during the whole history of the universe.
Isocurvature perturbations of the axi-Majoron are also suppressed because of the following reasons:
(i) the magnitude of Peccei-Quinn field during inflation is comparable to or larger than the Planck scale,  (ii) the effective curvature along the angular direction is positive and much smaller than the expansion rate when the relevant cosmological scales exit the horizon (i.e., the perturbations of the field orthogonal to the inflaton direction is nearly frozen or suppressed.).

Our scenario is a very simple minimal extension of the SM, but it provides a unified picture of the unknown history of the Universe from inflation to the BBN epoch, thanks to a scale-invariant $U(1)_{\rm PQ}$-breaking nonminimal gravitational interaction.
This model predicts a presence of a short early matter-domination era.
It determines the amount of axi-Majoron dark radiation, and leaves a characteristic fingerprint on inflationary primordial gravitational waves.
If such a feature is detected in future experiments, and once combined with inflationary observables, the amount of dark radiation and the spectra of gravitational waves can provide information of the mass parameters of the lightest states of the left-handed and the right-handed neutrinos. 
Although complicated, it may end up being the unique way of probing those physical quantities.

\medskip

% \noindent {\bf \emph{Note Added. - }} 

\medskip

%\noindent {\bf \emph{Acknowledgement}} \\
%This work is supported by the National Research Foundation of Korea grants %2017R1D1A1B06035959 (WIP) and 2022R 1A4A5030362 (WIP), .... 
%, and by Research Base Construction Fund Support Program funded by Jeonbuk National University in 2022 (WIP).   

\section*{Acknowledgments}

\noindent
This work was supported by the National Research Foundation of Korea grants by the Korea government: No. 2017R1D1A1B06035959 (W.I.P.), No.  2022R1A4A5030362 (W.I.P.), No. NRF-2019R1A2C3005009 (P.K.) and KIAS Individual Grants under Grant No. PG021403 (P.K.).
It was also supported by the Spanish grants  No. CIPROM/2021/054 (Generalitat Valenciana), No. PID2020-113775GB-I00 (AEI/10.13039/501100011033) (G.B. and W.I.P.) and  by the European ITN project HIDDeN (No. H2020-MSCA-ITN-2019/860881-HIDDeN) (G.B.).

%This work was supported by the National Research Foundation of Korea grants by the Korea %government: 2017R1D1A1B06035959 (W.I.P.),  2022R1A4A5030362 (W.I.P.), and ....
%It was also supported by the Spanish grants PID2020-113775GB-I00 
%(AEI/10.13039/501100011033) and CIPROM/2021/054 (Generalitat Valenciana) (W.I.P.).

\appendix

\section{Metric component in various bases}
\label{appen:metric-compo}
For a fixed Planck scale with the SM Higgs field ignored, we define the frame function as
\beq
\Omega 
\equiv 1 + \frac{\xi_\phi \phi^2}{M_{\rm P}^2} \l[ 1  + \alpha \cos \l( 2 \theta_\Phi \r) \r]
= 1 + \frac{\xi_r \phi_r^2}{M_{\rm P}^2} + \frac{\xi_i \phi_i^2}{M_{\rm P}^2} \ ,
\eeq
where $\alpha \equiv \xi_a/\xi_\phi$, and the condition  
\beq
\xi_r > \xi_i \quad \Leftrightarrow \quad \xi_\phi > \xi_a
\eeq
is required for $\Omega$ to be positive definite even at the large field region of $\xi_\phi^{1/2} \phi > M_{\rm P}$ which is of our interest.
The symmetric field-space metric $\mathcal{K}_{ab}$ is given by
\beq
\mathcal{K}_{ab} = \Omega^{-2} \l[ \Omega \delta_{ab} + \frac{3M_{\rm P}^2}{2} \frac{\partial \Omega}{\partial \phi_a} \frac{\partial \Omega}{\partial \phi_b} \r] \equiv \Omega^{-2} K_{ab} \ .
\eeq
In the basis of ($\phi_r, \phi_i$), 
\bea
\frac{\partial \Omega}{\partial \phi_r} &=& \frac{2 \xi_r}{M_{\rm P}^2} \phi_r \ ,
\\
\frac{\partial \Omega}{\partial \phi_i} &=& \frac{2 \xi_i}{M_{\rm P}^2} \phi_i \ ,
\eea
and one finds
\bea
K_{11} &=& 1 + \frac{\xi_r \l( 1 + 6 \xi_r \r) \phi_r^2}{M_{\rm P}^2} + \frac{\xi_i \phi_i^2}{M_{\rm P^2}} \ ,
\\
K_{22} &=& 1 + \frac{\xi_r \phi_r^2}{M_{\rm P}^2} + \frac{\xi_i \l( 1 + 6 \xi_i \r) \phi_i^2}{M_{\rm P^2}}  \ ,
\\
K_{12} &=& \frac{6 \xi_r \xi_i \phi_r \phi_i}{M_{\rm P}^2} \ .
\eea
In the ($\phi, \theta$)-basis,
\bea
\frac{\partial \Omega}{\partial \phi} &=& \frac{2 \xi_\phi \phi}{M_{\rm P}^2} \l[ 1  + \alpha \cos \l( 2 \theta_\Phi \r) \r] \ ,
\\
\frac{\partial \Omega}{\partial \theta} &=& -\frac{2 \xi_\phi \phi^2}{M_{\rm P}^2} \alpha \sin \l( 2 \theta_\Phi \r)  \ ,
\eea
and 
\bea
K_{11} &=& \Omega + 6 \l( \frac{\xi_\phi \phi}{M_{\rm P}} \r)^2 \l[ 1 + \alpha \cos \l( 2 \theta_\Phi \r) \r]^2 \ ,
\\
K_{22} &=& \Omega \phi^2 + 6 \l( \frac{\xi_\phi \phi}{M_{\rm P}} \r)^2 \phi^2 \l[ \alpha \sin \l( 2 \theta_\Phi\r) \r]^2 \ ,
\\
K_{12} &=& -6 \l( \frac{\xi_\phi \phi}{M_{\rm P}} \r)^2 \phi  \l[ 1 + \alpha \cos \l( 2 \theta_\Phi\r) \r] \l[ \alpha \sin \l( 2 \theta_\Phi \r) \r] \ .
\eea

\section{Derivatives of the Einstein frame potential with respect to the canonical fields}
\label{appen:U-deriv}
In the large field region of $\xi_\phi^{1/2} \phi \gg M_{\rm P}$, in terms of the approximate canonical fields of the polar field-space coordinates, the potential is read as
\bea
U(\varphi_1,\varphi_2) 
&=& \frac{\lambda_\Phi M_{\rm P}^4}{4 \xi_\phi^2}  \l[ 1  +  \alpha \cos \l( \frac{2 \sqrt{\xi_\phi} \varphi_2}{M_{\rm P}} \r) + \exp\l[ - 2 \sqrt{\frac{\xi_\phi}{1+6\xi_\phi}} \frac{\varphi_1}{M_{\rm P}} \r] \r]^{-2} \ .
\eea
Let us define $A_1 \equiv 2 \sqrt{\xi_\phi/(1+6 \xi_\phi)} $ and $A_2 \equiv 2 \sqrt{\xi_\phi}$ for notational convenience.
Then, derivatives of the potential up to the second order are found to be
\bea
\frac{\partial U}{\partial \varphi_1} &=& \frac{2 A_1U/M_{\rm P}}{\l[ 1 + \l( 1  +  \alpha \cos \l( \frac{A_2 \varphi_2}{M_{\rm P}} \r) \r) \exp\l[ \frac{A_1 \varphi_1}{M_{\rm P}} \r] \r]} \ ,
\\
\frac{\partial^2 U}{\partial \varphi_1^2} &=& \frac{2 A_1^2 \l[ 2 - \l( 1  +  \alpha \cos \l( \frac{A_2 \varphi_2}{M_{\rm P}} \r)  \r) \exp\l[ \frac{A_1 \varphi_1}{M_{\rm P}} \r] \r] U/M_{\rm P}^2}{\l[ 1 + \l( 1  +  \alpha \cos \l( \frac{A_2 \varphi_2}{M_{\rm P}} \r) \r) \exp\l[ \frac{A_1 \varphi_1}{M_{\rm P}} \r] \r]^2} \ ,
%\\
%\frac{\partial^3 U}{\partial \varphi_1^3} &=& \frac{2 A_1^3 \l[ 4 - 7 \exp\l[ \frac{A_1 \varphi_1}{M_{\rm P}} \r] \l( 1  +  \alpha \cos \l( \frac{A_2 \varphi_2}{M_{\rm P}} \r) \r) + \exp\l[ \frac{2 A_1 \varphi_1}{M_{\rm P}} \r]  \l( 1  +  \alpha \cos \l( \frac{A_2 \varphi_2}{M_{\rm P}} \r) \r)^2 \r]}{ \l( U/M_{\rm P}^3 \r)^{-1} \l[ 1 + \l( 1  +  \alpha \cos \l( \frac{A_2 \varphi_2}{M_{\rm P}} \r) \r) \exp\l[ \frac{A_1 \varphi_1}{M_{\rm P}} \r] \r]^3}
\\
\frac{\partial U}{\partial \varphi_2} &=& \frac{2 A_2 \alpha \sin \l( A_2 \varphi_2/M_{\rm P} \r) U/M_{\rm P}}{\l[ 1  +  \alpha \cos \l( \frac{A_2 \varphi_2}{M_{\rm P}} \r) + \exp\l[ - \frac{A_1 \varphi_1}{M_{\rm P}} \r] \r]} \ , 
\\
\frac{\partial^2 U}{\partial \varphi_2^2} &=& \frac{2 A_2^2 \alpha \exp\l[ \frac{A_1 \varphi_1}{M_{\rm P}} \r] \l[ \l( 1 + \l( 1 - \alpha \r) \exp\l[ \frac{A_1 \varphi_1}{M_{\rm P}} \r]  \r) \cos \l( \frac{A_2 \varphi_2}{M_{\rm P}} \r) + 2 \alpha \exp\l[ \frac{A_1 \varphi_1}{M_{\rm P}} \r] \r] }{ \l( U/M_{\rm P}^2 \r)^{-1} \l[ 1 + \l( 1  +  \alpha \cos \l( \frac{A_2 \varphi_2}{M_{\rm P}} \r) \r) \exp\l[ \frac{A_1 \varphi_1}{M_{\rm P}} \r] \r]^2} \ ,
%\\
%\frac{\partial^3 U}{\partial \varphi_2^3} &=& \frac{2 A_2^3 \alpha \exp\l[ \frac{A_1 \varphi_1}{M_{\rm P}} \r] f(\varphi_1, \varphi_2) \sin \l( \frac{A_2 \varphi_2}{M_{\rm P}} \r) }{ \l( U/M_{\rm P}^3 \r)^{-1} \l[ 1 + \l( 1  +  \alpha \cos \l( \frac{A_2 \varphi_2}{M_{\rm P}} \r) \r) \exp\l[ \frac{A_1 \varphi_1}{M_{\rm P}} \r] \r]^3}
\\
\frac{\partial^2 U}{\partial \varphi_1\partial \varphi_2} &=& \frac{6 A_1 A_2 \alpha \exp\l[ \frac{A_1 \varphi_1}{M_{\rm P}} \r] \sin \l( \frac{A_2 \varphi_2}{M_{\rm P}} \r) U/M_{\rm P}^2 }{\l[ 1 + \l( 1  +  \alpha \cos \l( \frac{A_2 \varphi_2}{M_{\rm P}} \r) \r) \exp\l[ \frac{A_1 \varphi_1}{M_{\rm P}} \r] \r]^2} \ .
\eea
%where
%\bea
%f(\varphi_1, \varphi_2) 
%&\equiv& -1 + \exp\l[ \frac{A_1 \varphi_1}{M_{\rm P}} \r]  \l[ -2 + 7 \alpha \cos \l( \frac{A_2 \varphi_2}{M_{\rm P}} \r) \r]
%\nonumber \\
%&& + \exp\l[ \frac{2 A_1 \varphi_1}{M_{\rm P}} \r] \l[ -1 + 10 \alpha^2 + 7 \alpha \cos \l( \frac{A_2 \varphi_2}{M_{\rm P}} \r)  - 2 \alpha^2 \cos \l( \frac{A_2 \varphi_2}{M_{\rm P}} \r) \r] \ .
%\eea
For $\alpha \lll 1$, they are approximated as
\bea
\frac{\partial U}{\partial \varphi_1} &=& 2 A_1\exp\l[ -\frac{A_1 \varphi_1}{M_{\rm P}} \r] \frac{U}{M_{\rm P}} \ ,
\\
\frac{\partial^2 U}{\partial \varphi_1^2} &=& - 2 A_1^2 \exp\l[ - \frac{A_1 \varphi_1}{M_{\rm P}} \r] \frac{U}{M_{\rm P}^2} \ ,
%\\
%\frac{\partial^3 U}{\partial \varphi_1^3} &=& 2 A_1^3 \l( -7 + \exp\l[ \frac{A_1 \varphi_1}{M_{\rm P}} \r] \r) \exp\l[ - 2 \frac{A_1 \varphi_1}{M_{\rm P}} \r] \frac{U}{M_{\rm P}^3} 
\\
\frac{\partial U}{\partial \varphi_2} &=&  2 A_2 \alpha \sin \l( \frac{A_2 \varphi_2}{M_{\rm P}} \r) \frac{U}{M_{\rm P}} \ ,
\\
\frac{\partial^2 U}{\partial \varphi_2^2} &=& 2 A_2^2 \alpha \cos \l( \frac{A_2 \varphi_2}{M_{\rm P}} \r) \frac{U}{M_{\rm P}^2} \ ,
%\\
%\frac{\partial^3 U}{\partial \varphi_2^3} &=& -2 A_2^3 \alpha \sin \l( \frac{A_2 \varphi_2}{M_{\rm P}} \r) \frac{U}{M_{\rm P}^3} \xrightarrow[]{\theta_\Phi \ll 1} -4 \alpha \theta_\Phi \frac{\lambda_\Phi M_{\rm P}}{\sqrt{\xi_\phi}} = - 8 \sqrt{3/\lambda_\Phi} \alpha \theta_\Phi H_I \ .
\\
\frac{\partial^2 U}{\partial \varphi_1\partial \varphi_2} &=& 6 A_1 A_2 \alpha \exp\l[ - \frac{A_1 \varphi_1}{M_{\rm P}} \r] \sin \l( \frac{A_2 \varphi_2}{M_{\rm P}} \r) \frac{U}{M_{\rm P}^2} \ .
\eea

%Inflation like Higgs-inflation is expected along $\varphi_1$.
%The slow-roll parameters are given by
%\bea
%\epsilon_T &\equiv& \frac{1}{2} \l( \frac{M_{\rm P}}{U} \frac{\partial U}{\partial \varphi_1} \r)^2 \approx \l( \frac{8 \xi_\phi}{1+6 \xi_\phi} \r) \frac{M_{\rm P}^4}{\xi_\phi^2 \phi^4}
%\\
%\eta_T &\equiv& \frac{M_{\rm P}^2}{U} \frac{\partial^2 U}{\partial \varphi_1^2}  \approx - \l( \frac{8 \xi_\phi}{1+6 \xi_\phi} \r) \frac{M_{\rm P}^2}{\xi_\phi \phi^2}
%\\
%\zeta_T &\equiv& \frac{M_{\rm P}^4}{U^2} \frac{\partial^3 U}{\partial \varphi_1^3} \frac{\partial U}{\partial \varphi_1}  \approx \l( \frac{8 \xi_\phi}{1+6 \xi_\phi} \r)^2 \frac{M_{\rm P}^4}{\xi_\phi^2 \phi^4}
%\eea 
%The inflationary observables can be reproduced by taking $\lambda_\Phi/\xi_\phi^2 \approx 7 \times 10^{-10}$ in the region of $\xi_\phi^{1/2} \phi \gg M_{\rm P}$ \cite{}.

\section{Scattering/annihilation cross sections}
\label{sec:csection}
The kinetic term of Peccei-Quinn field gives the following interaction of axi-Majorons $a_\phi(x) \equiv \theta(x)/\phi_0$,
\beq
\mathcal{L} 
\supset \frac{1}{2} \l( \partial \delta \phi \r)^2 - \frac{1}{2} m_\phi^2 \l( \delta \phi \r)^2 + \frac{1}{2} \l( 1 + \frac{\delta \phi}{\phi_0} \r)^2 \l( \partial a_\phi \r)^2 \supset \frac{\delta \phi}{\phi_0} \l( \partial a_\phi \r)^2 
%\longrightarrow
%\frac{1}{2 m_\phi^2 \phi_0^2} \l( \partial_\mu a \partial^\mu a \r)^2
\eeq
where $m_\phi^2 = 2 \lambda_\phi \phi_0^2$.
When the scalar field $\delta \phi$ is integrated out with its EOM given by
\beq
\partial^2 \delta \phi + m_\phi^2 \delta \phi = - \frac{1}{\phi_0} \l( \partial a_\phi \r)^2
\longrightarrow 
\delta \phi \simeq  \frac{1}{m_\phi^2 \phi_0} \l( \partial a_\phi \r)^2 \ , 
\eeq
one obtains an effective self-interaction of axi-Majorons as
\beq
\frac{\delta \phi}{\phi_0} \l( \partial a_\phi \r)^2 
\longrightarrow
\frac{1}{m_\phi^2 \phi_0^2} \l( \partial a_\phi \r)^4 \ .
\eeq

Also, once $U(1)_{\rm PQ}$ is spontaneously broken, Peccei-Quinn field $\Phi$ can be express as $\Phi = \phi_0 e^{ia_\phi}/\sqrt{2}$, and the mass term of right-handed neutrinos leads to the interactions of axi-Majoron and RHN neutrinos as follows.
\bea 
\mathcal{L} 
%&\supset& - \frac{y_N}{2} \Phi^* \ov{\nu_R^c} \nu_R + {\rm h.c.}
%\nonumber \\
&\supset& - \frac{m_N}{2} \ov{N} N + i \frac{y_N}{2 \sqrt{2}} a_\phi \ov{N} \gamma_5 N + \frac{y_N}{4 \sqrt{2} \phi_0} a_\phi^2 \ov{N} N
\\
& = &  - \frac{m_N}{2} \overline{N} N + i \frac{y_N}{2 \sqrt{2}} a_\phi \overline{N} 
\gamma_5 N + \frac{y_N^2}{8 m_N} a_\phi^2 \overline{N} N \ ,
\eea
where we have used $m_N = y_N \phi_0 / \sqrt{2}$ in the last term of the second line, and $N \equiv \l( \nu_R, \nu_R^c \r)^T$ is the four-component Majorana field.
In the following expressions for averaged amplitude-squared and cross sections, the mass of axi-Majoron is set to be zero.

\subsection{$a_\phi + a_\phi \rightarrow a_\phi + a_\phi$}

The amplitude for the elastic scattering $a_\phi (p_1) + a_\phi (p_2) \rightarrow a_\phi (p_3) + a_\phi (p_4)$ 
is given by 
\begin{equation}
{\cal M}  =  \frac{1}{\phi_0^2}  \left[ \frac{s^2}{s - m_\phi^2} +   \frac{t^2}{t - m_\phi^2} + \frac{u^2}{u - m_\phi^2} \right]  \ ,
\end{equation}
where the Mandelstamd variables $s,t,u$ defined as
\[
s \equiv ( p_1 + p_2 )^2 \ , \ t \equiv ( p_1 - p_3 )^2 \ , \ u \equiv ( p_1 - p_4 )^2 \ ,
\]
were used with $s+t+u = 4 m_a^2 \simeq 0$, assuming $m_a^2 = 0$.   
Then the total cross section for $a_\phi + a_\phi \rightarrow a_\phi + a_\phi$ can be obtained as follows, after integrating over the angular variables:
\begin{equation}
\sigma (a_\phi a_\phi \rightarrow a_\phi  a_\phi) = 
\frac{7 s^3}{80 \pi m_\phi^4 \phi_0^4} = \frac{7 \lambda_\phi^2}{20 \pi m_\phi^2} \l( \frac{s}{m_\phi^2} \r)^3
\end{equation}

\subsection{$N + a_\phi \rightarrow N + a_\phi $}
The amplitude for the elastic scattering  $N (p_1) + a_\phi (p_2)   \rightarrow  N (p_3) + a_\phi (p_4) $ 
is given by 
\begin{equation}
{\cal M} =  \frac{y_N^2}{2}  \overline{u_3} \left[ \frac{1}{m_N} +  \gamma_5 \left( \frac{\slashed{p_1} 
+ \slashed{p_2}
+ m_N}{s - m_N^2} \right) \gamma_5  + \gamma_5 \left( \frac{ \slashed{p_1} - \slashed{p_4} + 
m_N }{u - m_N^2} \right) \gamma_5 \right] u_1 
\end{equation}
where 
%\begin{equation}
%s \equiv (p_1+ p_2)^2 \ , \ t \equiv ( p_1 - p_3 )^2 \ , \ u \equiv ( p_1 - p_4 )^2 \ ,
%\label{eq:stu}
%\end{equation} 
%with 
$s + t + u = 2 m_N^2$.  We have abbreviated $u (p_1,s_1) \equiv u_1$, etc..
After squaring the amplitude and averaging and summing over the initial and the final spins 
of the RH neutrinos, one gets 
%\begin{equation}
%\overline{| {\cal M} |^2} = 
%\end{equation}
%Finally after integrating over the angular variables, one gets 
\begin{equation}
\sigma ( a_\phi + N \rightarrow a_\phi + N ) = \frac{y_N^4}{64 \pi m_N^2 } \left[ 
\frac{ s^2 - m_N^4 + 2 s m_N^2 \ln ( m_N^2 /s) }{s (s - m_N^2) }  \right] \ .
\end{equation}
At high energy limit, $s \gg m_N^2$, one has 
\begin{equation}
\sigma ( a_\phi + N \rightarrow a_\phi + N ) \simeq  \frac{y_N^4}{64 \pi m_N^2 }  \  .
\end{equation}
Near the low energy threshold ($s\rightarrow m_N^2$), one has 
\begin{equation}
\sigma ( a_\phi + N \rightarrow a_\phi + N ) = \frac{y_N^4}{192\pi m_N^2 } \l( 1 - \frac{m_N^2}{s} \r)^2 \  .
\end{equation}

\subsection{$N + N \rightarrow a_\phi + a_\phi$}
The amplitude for this process $N (p_1) + N (p_2)   \rightarrow  a_\phi (p_3) + a_\phi (p_4) $ can be 
obtained from the results obtained on $N (p_1) + a_\phi (p_2)   \rightarrow  N (p_3) + a_\phi (p_4) $
 in the previous subsection  by crossing $s \leftrightarrow t$, {\it i.e.} $p_2 \leftrightarrow -p_3$.  Then 
\begin{equation}
{\cal M} =  \frac{y_N^2}{2}  \overline{v_2} \left[ \frac{1}{m_N} 
+  \gamma_5 \left( \frac{\slashed{p_1} - \slashed{p_3}+ m_N}{t - m_N^2} \right) \gamma_5  
+ \gamma_5 \left( \frac{\slashed{p_1} - \slashed{p_4} + m_N}{u - m_N^2} \right) \gamma_5 \right] u_1 
\end{equation}
After squaring the amplitude and averaging and summing over the initial and the final spins 
of the RH neutrinos, one gets 
\begin{equation}
\sigma  = \frac{y_N^4 s}{64 \pi m_N^2 \l( s - 4 m_N^2 \r)} \l\{ \l( 1 - \frac{4 m_N^2}{s} \r)^{1/2} + \frac{2 m_N^2}{s} \ln \l[ \frac{1 - \l( 1 - \frac{4 m_N^2}{s} \r)^{1/2} }{1 + \l( 1 - \frac{4 m_N^2}{s} \r)^{1/2} } \r]  \r\} \ . 
\end{equation}
We have ignored the axi-Majoron mass to simplify the expression. 
Expanding near the threshold with $v_1 - v_2 = v$, one has $s \simeq 4 m_N^2 ( 1 + v^2 / 4 )$. 
Then one gets 
\begin{equation}
 \sigma  \xrightarrow[]{v \ll 1}  \frac{y_N^4 v}{192 \pi m_N^2}   \ . 
\end{equation}
%Note that, because of Majorana fermion nature of $N$, there is a $S-$wave annihilation which is independent of $v$.  

\subsection{$N + N \rightarrow N + N$}
Finally let us consider $N (p_1) + N (p_2) \rightarrow N (p_3) + N (p_4)$, which occurs 
through the axi-Majoron exchanges in the $s,t,$ and $u-$channels. 
%The Mandelstamd variables $s,t,u$ are defined as before, Eq. (\ref{eq:stu}), and we have $s + t + u = 4 m_N^2$ for this process. 
We have $s + t + u = 4 m_N^2$ for this process. 
Then the amplitude for this process is given by 
\begin{equation}
{\cal M} = \left( \frac{y_N}{\sqrt{2}} \right)^2 \left[ \frac{(\overline{v_2} \gamma_5 u_1 )( \overline{u_3} \gamma_5 v_4)}{s} +  \frac{(\overline{u_3} \gamma_5 u_1 )(\overline{u_4} \gamma_5 u_2)}{t} - \frac{(\overline{u_4} \gamma_5 u_1 )(\overline{u_3} \gamma_5 u_2)}{u}  \right] \ ,
\end{equation}
neglecting the $a_\phi$ mass in the $a_\phi$-propagators.
Note the relative minus sign in the $u$-channel contribution compared to the $t$ and $s$-channel contributions.
After summing over the spins and taking average over the initial spins yield 
\begin{eqnarray}
\overline{| {\cal M} |^2} & = &  \left( \frac{y_N}{\sqrt{2}} \right)^4 \times  \frac{24}{2 \times 2} = \frac{3}{2} y_N^4 \ .  
\end{eqnarray}
Integrating over the angular variables (or, equivalently, over $t$) and dividing by a factor of 2 for the two identical 
particles in the final state, we get 
\begin{equation}
\sigma (N + N \rightarrow N + N) = \frac{3 y_N^4}{64 \pi s} \ . 
\end{equation}

\section{The helicity asymmetry of $N_1$s at the epoch of their decays}
\label{app:Y1d}
In our scenario, the standard thermal bath is to be recovered from the decay of the lightest right-handed neutrinos ($N_1$s).
At the epoch of their decays, the yield of the helicity asymmetry of $N_1$s can be defined as
\beq
Y_{1, \rm d} \equiv \frac{\Delta n_{1, \rm d}}{s_{\rm d}}
\eeq
where $\Delta n_{1, \rm d}$ is the helicity asymmetry of $N_1$s just before their decay, and $s_{\rm d}$ is the entropy density established by the decay of $N_1$s in the sudden decay approximation.

\subsection{Case of $\phi_* < \phi_\times$}
In this case, one finds
\beq
\Delta n_{1, \rm d} = \Delta n_{1, \times} \l( \frac{a_\times}{a_{\rm d}} \r)^3
\eeq
where $\Delta n_{1, \times} = {\rm B}_1 \Delta_{\rm AD} \Delta n_{\Phi, \times}$ with $\Delta n_{\Phi, \times}$ being the asymmetry of $\Phi$ particles at the epoch of $\phi = \phi_\times$. 
The entropy density $s_{\rm d}$ is given by 
\beq
s_{\rm d} = \frac{4}{3} \frac{\rho_{1, \rm d}}{T_{\rm d}}
\eeq
where $\rho_{1, \rm d}$ is the energy density of $N_1$ at its decay.
One finds
\beq
\rho_{1, \rm d} 
= {\rm B}_1 \rho_{\Phi, \times} \l( \frac{a_*}{a_{1, \rm eq}} \r) \l( \frac{a_\times}{a_{\rm d}} \r)^3
\eeq
where the energy density of $\Phi$ when $\phi=\phi_\times$ is given by
\beq
\rho_{\Phi, \times} 
= \frac{\lambda_\phi}{4} \phi_0^4 
= \frac{1}{4} \times \frac{\lambda_\phi}{4} \phi_\times^4
= \frac{1}{4} \times \frac{3}{4} T_{\Phi, \times} s_{\Phi, \times}
\eeq
with $T_{\Phi, \times}$ defined as
\beq \label{eq:T-phix}
T_{\Phi, \times} = \l( \frac{\pi^2}{30} g_*^\Phi \r)^{-1/4} \l( \frac{\lambda_\phi}{4} \phi_\times^4 \r)^{1/4} \ .
\eeq
As a result, one finds
\beq
Y_{1, \rm d} = 4 \Delta_{\rm AD} \l( \frac{{\tilde H}_*}{{\tilde H}_{1, \rm eq}} \r)^{1/2} \frac{T_{\rm d}}{T_{\Phi, \times}} Y_{\Phi, e} \ .
\eeq

\subsection{Case of $\phi_* > \phi_\times$}
In this case, one finds
\beq
\Delta n_{1, \rm d} = \Delta n_{1, *} \l( \frac{a_*}{a_{\rm d}} \r)^3
\eeq
where $\Delta n_{1, *} = {\rm B}_1 \Delta n_{\Phi, *}$ with $\Delta n_{\Phi, *}$ being the asymmetry of $\Phi$ particles at the epoch of $\phi = \phi_*$. 
Also, $\rho_{1, \rm d}$ is given by
\beq
\rho_{1, \rm d} 
= {\rm B}_1 \rho_{\Phi, *} \l( \frac{a_*}{a_{1, \rm eq}} \r) \l( \frac{a_*}{a_{\rm d}} \r)^3
\eeq
where
\beq
\rho_{\Phi, *} 
= \frac{\lambda_\phi}{4} \phi_*^4 
= \frac{3}{4} T_{\Phi, *} s_{\Phi, *}
\eeq
with $T_{\Phi, *}$ defined in the same way as \eq{eq:T-phix} but $\phi_\times$ replaced to $\phi_*$.
As a result, one finds
\beq
Y_{1, \rm d} = \l( \frac{H_*}{H_{1, \rm eq}} \r)^{1/2} \frac{T_{\rm d}}{T_{\Phi, *}} Y_{\Phi, e} \ .
\eeq

\bibliographystyle{JHEP}
\bibliography{mcsm-ref}

\providecommand{\href}[2]{#2}\begingroup\raggedright\begin{thebibliography}{10}

\bibitem{Glashow:1961tr}
S.L.~Glashow, \emph{{Partial Symmetries of Weak Interactions}},
  \href{https://doi.org/10.1016/0029-5582(61)90469-2}{\emph{Nucl. Phys.}
  {\bfseries 22} (1961) 579}.

\bibitem{Weinberg:1967tq}
S.~Weinberg, \emph{{A Model of Leptons}},
  \href{https://doi.org/10.1103/PhysRevLett.19.1264}{\emph{Phys. Rev. Lett.}
  {\bfseries 19} (1967) 1264}.

\bibitem{Wetterich:2020cxq}
C.~Wetterich, \emph{{Fundamental scale invariance}},
  \href{https://doi.org/10.1016/j.nuclphysb.2021.115326}{\emph{Nucl. Phys. B}
  {\bfseries 964} (2021) 115326}
  [\href{https://arxiv.org/abs/2007.08805}{{\ttfamily 2007.08805}}].

\bibitem{Garcia-Bellido:2011kqb}
J.~Garcia-Bellido, J.~Rubio, M.~Shaposhnikov and D.~Zenhausern,
  \emph{{Higgs-Dilaton Cosmology: From the Early to the Late Universe}},
  \href{https://doi.org/10.1103/PhysRevD.84.123504}{\emph{Phys. Rev. D}
  {\bfseries 84} (2011) 123504}
  [\href{https://arxiv.org/abs/1107.2163}{{\ttfamily 1107.2163}}].

\bibitem{Bezrukov:2012hx}
F.~Bezrukov, G.K.~Karananas, J.~Rubio and M.~Shaposhnikov, \emph{{Higgs-Dilaton
  Cosmology: an effective field theory approach}},
  \href{https://doi.org/10.1103/PhysRevD.87.096001}{\emph{Phys. Rev. D}
  {\bfseries 87} (2013) 096001}
  [\href{https://arxiv.org/abs/1212.4148}{{\ttfamily 1212.4148}}].

\bibitem{Salvio:2014soa}
A.~Salvio and A.~Strumia, \emph{{Agravity}},
  \href{https://doi.org/10.1007/JHEP06(2014)080}{\emph{JHEP} {\bfseries 06}
  (2014) 080} [\href{https://arxiv.org/abs/1403.4226}{{\ttfamily 1403.4226}}].

\bibitem{Ferreira:2016vsc}
P.G.~Ferreira, C.T.~Hill and G.G.~Ross, \emph{{Scale-Independent Inflation and
  Hierarchy Generation}},
  \href{https://doi.org/10.1016/j.physletb.2016.10.036}{\emph{Phys. Lett. B}
  {\bfseries 763} (2016) 174}
  [\href{https://arxiv.org/abs/1603.05983}{{\ttfamily 1603.05983}}].

\bibitem{Ferreira:2016wem}
P.G.~Ferreira, C.T.~Hill and G.G.~Ross, \emph{{Weyl Current, Scale-Invariant
  Inflation and Planck Scale Generation}},
  \href{https://doi.org/10.1103/PhysRevD.95.043507}{\emph{Phys. Rev. D}
  {\bfseries 95} (2017) 043507}
  [\href{https://arxiv.org/abs/1610.09243}{{\ttfamily 1610.09243}}].

\bibitem{Coleman:1973jx}
S.R.~Coleman and E.J.~Weinberg, \emph{{Radiative Corrections as the Origin of
  Spontaneous Symmetry Breaking}},
  \href{https://doi.org/10.1103/PhysRevD.7.1888}{\emph{Phys. Rev. D} {\bfseries
  7} (1973) 1888}.

\bibitem{Blas:2011ac}
D.~Blas, M.~Shaposhnikov and D.~Zenhausern, \emph{{Scale-invariant alternatives
  to general relativity}},
  \href{https://doi.org/10.1103/PhysRevD.84.044001}{\emph{Phys. Rev. D}
  {\bfseries 84} (2011) 044001}
  [\href{https://arxiv.org/abs/1104.1392}{{\ttfamily 1104.1392}}].

\bibitem{Ferreira:2016kxi}
P.G.~Ferreira, C.T.~Hill and G.G.~Ross, \emph{{No fifth force in a scale
  invariant universe}},
  \href{https://doi.org/10.1103/PhysRevD.95.064038}{\emph{Phys. Rev. D}
  {\bfseries 95} (2017) 064038}
  [\href{https://arxiv.org/abs/1612.03157}{{\ttfamily 1612.03157}}].

\bibitem{Kannike:2015apa}
K.~Kannike, G.~H\"utsi, L.~Pizza, A.~Racioppi, M.~Raidal, A.~Salvio et~al.,
  \emph{{Dynamically Induced Planck Scale and Inflation}},
  \href{https://doi.org/10.1007/JHEP05(2015)065}{\emph{JHEP} {\bfseries 05}
  (2015) 065} [\href{https://arxiv.org/abs/1502.01334}{{\ttfamily
  1502.01334}}].

\bibitem{Ferreira:2018qss}
P.G.~Ferreira, C.T.~Hill, J.~Noller and G.G.~Ross, \emph{{Inflation in a scale
  invariant universe}},
  \href{https://doi.org/10.1103/PhysRevD.97.123516}{\emph{Phys. Rev. D}
  {\bfseries 97} (2018) 123516}
  [\href{https://arxiv.org/abs/1802.06069}{{\ttfamily 1802.06069}}].

\bibitem{Gonzalez-Garcia:2007dlo}
M.C.~Gonzalez-Garcia and M.~Maltoni, \emph{{Phenomenology with Massive
  Neutrinos}}, \href{https://doi.org/10.1016/j.physrep.2007.12.004}{\emph{Phys.
  Rept.} {\bfseries 460} (2008) 1}
  [\href{https://arxiv.org/abs/0704.1800}{{\ttfamily 0704.1800}}].

\bibitem{Kim:1979if}
J.E.~Kim, \emph{{Weak Interaction Singlet and Strong CP Invariance}},
  \href{https://doi.org/10.1103/PhysRevLett.43.103}{\emph{Phys. Rev. Lett.}
  {\bfseries 43} (1979) 103}.

\bibitem{Shifman:1979if}
M.A.~Shifman, A.I.~Vainshtein and V.I.~Zakharov, \emph{{Can Confinement Ensure
  Natural CP Invariance of Strong Interactions?}},
  \href{https://doi.org/10.1016/0550-3213(80)90209-6}{\emph{Nucl. Phys. B}
  {\bfseries 166} (1980) 493}.

\bibitem{Dine:1981rt}
M.~Dine, W.~Fischler and M.~Srednicki, \emph{{A Simple Solution to the Strong
  CP Problem with a Harmless Axion}},
  \href{https://doi.org/10.1016/0370-2693(81)90590-6}{\emph{Phys. Lett. B}
  {\bfseries 104} (1981) 199}.

\bibitem{Zhitnitsky:1980tq}
A.R.~Zhitnitsky, \emph{{On Possible Suppression of the Axion Hadron
  Interactions. (In Russian)}}, {\emph{Sov. J. Nucl. Phys.} {\bfseries 31}
  (1980) 260}.

\bibitem{Shin:1987xc}
M.~Shin, \emph{{Light Neutrino Masses and Strong {CP} Problem}},
  \href{https://doi.org/10.1103/PhysRevLett.59.2515}{\emph{Phys. Rev. Lett.}
  {\bfseries 59} (1987) 2515}.

\bibitem{Barenboim:2024akt}
G.~Barenboim, P.~Ko and W.-i.~Park, \emph{{The minimal cosmological standard
  model}},  \href{https://arxiv.org/abs/2403.05390}{{\ttfamily 2403.05390}}.

\bibitem{Peccei:1977hh}
R.D.~Peccei and H.R.~Quinn, \emph{{CP Conservation in the Presence of
  Instantons}}, \href{https://doi.org/10.1103/PhysRevLett.38.1440}{\emph{Phys.
  Rev. Lett.} {\bfseries 38} (1977) 1440}.

\bibitem{Peccei:1977ur}
R.D.~Peccei and H.R.~Quinn, \emph{{Constraints Imposed by CP Conservation in
  the Presence of Instantons}},
  \href{https://doi.org/10.1103/PhysRevD.16.1791}{\emph{Phys. Rev. D}
  {\bfseries 16} (1977) 1791}.

\bibitem{Volkas:1988cm}
R.R.~Volkas, A.J.~Davies and G.C.~Joshi, \emph{{NATURALNESS OF THE INVISIBLE
  AXION MODEL}},
  \href{https://doi.org/10.1016/0370-2693(88)91084-2}{\emph{Phys. Lett. B}
  {\bfseries 215} (1988) 133}.

\bibitem{Clarke:2015bea}
J.D.~Clarke and R.R.~Volkas, \emph{{Technically natural nonsupersymmetric model
  of neutrino masses, baryogenesis, the strong CP problem, and dark matter}},
  \href{https://doi.org/10.1103/PhysRevD.93.035001}{\emph{Phys. Rev. D}
  {\bfseries 93} (2016) 035001}
  [\href{https://arxiv.org/abs/1509.07243}{{\ttfamily 1509.07243}}].

\bibitem{Ballesteros:2016xej}
G.~Ballesteros, J.~Redondo, A.~Ringwald and C.~Tamarit, \emph{{Standard
  Model\textemdash{}axion\textemdash{}seesaw\textemdash{}Higgs portal
  inflation. Five problems of particle physics and cosmology solved in one
  stroke}}, \href{https://doi.org/10.1088/1475-7516/2017/08/001}{\emph{JCAP}
  {\bfseries 08} (2017) 001}
  [\href{https://arxiv.org/abs/1610.01639}{{\ttfamily 1610.01639}}].

\bibitem{Sopov:2022bog}
A.H.~Sopov and R.R.~Volkas, \emph{{VISH\ensuremath{\nu}: solving five Standard
  Model shortcomings with a Poincar\'e-protected electroweak scale}},
  \href{https://doi.org/10.1016/j.dark.2023.101381}{\emph{Phys. Dark Univ.}
  {\bfseries 42} (2023) 101381}
  [\href{https://arxiv.org/abs/2206.11598}{{\ttfamily 2206.11598}}].

\bibitem{Takahashi:2015waa}
F.~Takahashi and M.~Yamada, \emph{{Strongly broken Peccei-Quinn symmetry in the
  early Universe}},
  \href{https://doi.org/10.1088/1475-7516/2015/10/010}{\emph{JCAP} {\bfseries
  10} (2015) 010} [\href{https://arxiv.org/abs/1507.06387}{{\ttfamily
  1507.06387}}].

\bibitem{Hashimoto:2021xgu}
T.~Hashimoto, N.S.~Risdianto and D.~Suematsu, \emph{{Inflation connected to the
  origin of CP violation}},
  \href{https://doi.org/10.1103/PhysRevD.104.075034}{\emph{Phys. Rev. D}
  {\bfseries 104} (2021) 075034}
  [\href{https://arxiv.org/abs/2105.06089}{{\ttfamily 2105.06089}}].

\bibitem{Affleck:1984fy}
I.~Affleck and M.~Dine, \emph{{A New Mechanism for Baryogenesis}},
  \href{https://doi.org/10.1016/0550-3213(85)90021-5}{\emph{Nucl. Phys. B}
  {\bfseries 249} (1985) 361}.

\bibitem{Pilaftsis:2003gt}
A.~Pilaftsis and T.E.J.~Underwood, \emph{{Resonant leptogenesis}},
  \href{https://doi.org/10.1016/j.nuclphysb.2004.05.029}{\emph{Nucl. Phys. B}
  {\bfseries 692} (2004) 303}
  [\href{https://arxiv.org/abs/hep-ph/0309342}{{\ttfamily hep-ph/0309342}}].

\bibitem{Kallosh:1995hi}
R.~Kallosh, A.D.~Linde, D.A.~Linde and L.~Susskind, \emph{{Gravity and global
  symmetries}}, \href{https://doi.org/10.1103/PhysRevD.52.912}{\emph{Phys. Rev.
  D} {\bfseries 52} (1995) 912}
  [\href{https://arxiv.org/abs/hep-th/9502069}{{\ttfamily hep-th/9502069}}].

\bibitem{DiValentino:2021izs}
E.~Di~Valentino, O.~Mena, S.~Pan, L.~Visinelli, W.~Yang, A.~Melchiorri et~al.,
  \emph{{In the realm of the Hubble tension\textemdash{}a review of
  solutions}}, \href{https://doi.org/10.1088/1361-6382/ac086d}{\emph{Class.
  Quant. Grav.} {\bfseries 38} (2021) 153001}
  [\href{https://arxiv.org/abs/2103.01183}{{\ttfamily 2103.01183}}].

\bibitem{Vagnozzi:2019ezj}
S.~Vagnozzi, \emph{{New physics in light of the $H_0$ tension: An alternative
  view}}, \href{https://doi.org/10.1103/PhysRevD.102.023518}{\emph{Phys. Rev.
  D} {\bfseries 102} (2020) 023518}
  [\href{https://arxiv.org/abs/1907.07569}{{\ttfamily 1907.07569}}].

\bibitem{Branco:2011iw}
G.C.~Branco, P.M.~Ferreira, L.~Lavoura, M.N.~Rebelo, M.~Sher and J.P.~Silva,
  \emph{{Theory and phenomenology of two-Higgs-doublet models}},
  \href{https://doi.org/10.1016/j.physrep.2012.02.002}{\emph{Phys. Rept.}
  {\bfseries 516} (2012) 1} [\href{https://arxiv.org/abs/1106.0034}{{\ttfamily
  1106.0034}}].

\bibitem{Chikashige:1980ui}
Y.~Chikashige, R.N.~Mohapatra and R.D.~Peccei, \emph{{Are There Real Goldstone
  Bosons Associated with Broken Lepton Number?}},
  \href{https://doi.org/10.1016/0370-2693(81)90011-3}{\emph{Phys. Lett. B}
  {\bfseries 98} (1981) 265}.

\bibitem{Ferreira:2018itt}
P.G.~Ferreira, C.T.~Hill and G.G.~Ross, \emph{{Inertial Spontaneous Symmetry
  Breaking and Quantum Scale Invariance}},
  \href{https://doi.org/10.1103/PhysRevD.98.116012}{\emph{Phys. Rev. D}
  {\bfseries 98} (2018) 116012}
  [\href{https://arxiv.org/abs/1801.07676}{{\ttfamily 1801.07676}}].

\bibitem{Wald:1984rg}
R.M.~Wald, \emph{{General Relativity}}, Chicago Univ. Pr., Chicago, USA (1984),
  \href{https://doi.org/10.7208/chicago/9780226870373.001.0001}{10.7208/chicago/9780226870373.001.0001}.

\bibitem{Hill:2020oaj}
C.T.~Hill and G.G.~Ross, \emph{{Gravitational Contact Interactions and the
  Physical Equivalence of Weyl Transformations in Effective Field Theory}},
  \href{https://doi.org/10.1103/PhysRevD.102.125014}{\emph{Phys. Rev. D}
  {\bfseries 102} (2020) 125014}
  [\href{https://arxiv.org/abs/2009.14782}{{\ttfamily 2009.14782}}].

\bibitem{Burgess:2009ea}
C.P.~Burgess, H.M.~Lee and M.~Trott, \emph{{Power-counting and the Validity of
  the Classical Approximation During Inflation}},
  \href{https://doi.org/10.1088/1126-6708/2009/09/103}{\emph{JHEP} {\bfseries
  09} (2009) 103} [\href{https://arxiv.org/abs/0902.4465}{{\ttfamily
  0902.4465}}].

\bibitem{Burgess:2010zq}
C.P.~Burgess, H.M.~Lee and M.~Trott, \emph{{Comment on Higgs Inflation and
  Naturalness}}, \href{https://doi.org/10.1007/JHEP07(2010)007}{\emph{JHEP}
  {\bfseries 07} (2010) 007} [\href{https://arxiv.org/abs/1002.2730}{{\ttfamily
  1002.2730}}].

\bibitem{Barbon:2009ya}
J.L.F.~Barbon and J.R.~Espinosa, \emph{{On the Naturalness of Higgs
  Inflation}}, \href{https://doi.org/10.1103/PhysRevD.79.081302}{\emph{Phys.
  Rev. D} {\bfseries 79} (2009) 081302}
  [\href{https://arxiv.org/abs/0903.0355}{{\ttfamily 0903.0355}}].

\bibitem{Sasaki:1995aw}
M.~Sasaki and E.D.~Stewart, \emph{{A General analytic formula for the spectral
  index of the density perturbations produced during inflation}},
  \href{https://doi.org/10.1143/PTP.95.71}{\emph{Prog. Theor. Phys.} {\bfseries
  95} (1996) 71} [\href{https://arxiv.org/abs/astro-ph/9507001}{{\ttfamily
  astro-ph/9507001}}].

\bibitem{GrootNibbelink:2001qt}
S.~Groot~Nibbelink and B.J.W.~van Tent, \emph{{Scalar perturbations during
  multiple field slow-roll inflation}},
  \href{https://doi.org/10.1088/0264-9381/19/4/302}{\emph{Class. Quant. Grav.}
  {\bfseries 19} (2002) 613}
  [\href{https://arxiv.org/abs/hep-ph/0107272}{{\ttfamily hep-ph/0107272}}].

\bibitem{Gordon:2000hv}
C.~Gordon, D.~Wands, B.A.~Bassett and R.~Maartens, \emph{{Adiabatic and entropy
  perturbations from inflation}},
  \href{https://doi.org/10.1103/PhysRevD.63.023506}{\emph{Phys. Rev. D}
  {\bfseries 63} (2000) 023506}
  [\href{https://arxiv.org/abs/astro-ph/0009131}{{\ttfamily
  astro-ph/0009131}}].

\bibitem{Bezrukov:2007ep}
F.L.~Bezrukov and M.~Shaposhnikov, \emph{{The Standard Model Higgs boson as the
  inflaton}}, \href{https://doi.org/10.1016/j.physletb.2007.11.072}{\emph{Phys.
  Lett. B} {\bfseries 659} (2008) 703}
  [\href{https://arxiv.org/abs/0710.3755}{{\ttfamily 0710.3755}}].

\bibitem{Co:2019wyp}
R.T.~Co and K.~Harigaya, \emph{{Axiogenesis}},
  \href{https://doi.org/10.1103/PhysRevLett.124.111602}{\emph{Phys. Rev. Lett.}
  {\bfseries 124} (2020) 111602}
  [\href{https://arxiv.org/abs/1910.02080}{{\ttfamily 1910.02080}}].

\bibitem{Cohen:1987vi}
A.G.~Cohen and D.B.~Kaplan, \emph{{Thermodynamic Generation of the Baryon
  Asymmetry}}, \href{https://doi.org/10.1016/0370-2693(87)91369-4}{\emph{Phys.
  Lett. B} {\bfseries 199} (1987) 251}.

\bibitem{Peterson:2010np}
C.M.~Peterson and M.~Tegmark, \emph{{Testing Two-Field Inflation}},
  \href{https://doi.org/10.1103/PhysRevD.83.023522}{\emph{Phys. Rev. D}
  {\bfseries 83} (2011) 023522}
  [\href{https://arxiv.org/abs/1005.4056}{{\ttfamily 1005.4056}}].

\bibitem{Sasaki:1986hm}
M.~Sasaki, \emph{{Large Scale Quantum Fluctuations in the Inflationary
  Universe}}, \href{https://doi.org/10.1143/PTP.76.1036}{\emph{Prog. Theor.
  Phys.} {\bfseries 76} (1986) 1036}.

\bibitem{Mukhanov:1988jd}
V.F.~Mukhanov, \emph{{Quantum Theory of Gauge Invariant Cosmological
  Perturbations}}, {\emph{Sov. Phys. JETP} {\bfseries 67} (1988) 1297}.

\bibitem{GrootNibbelink:2000vx}
S.~Groot~Nibbelink and B.J.W.~van Tent, \emph{{Density perturbations arising
  from multiple field slow roll inflation}},
  \href{https://arxiv.org/abs/hep-ph/0011325}{{\ttfamily hep-ph/0011325}}.

\bibitem{Achucarro:2012yr}
A.~Achucarro, V.~Atal, S.~Cespedes, J.-O.~Gong, G.A.~Palma and S.P.~Patil,
  \emph{{Heavy fields, reduced speeds of sound and decoupling during
  inflation}}, \href{https://doi.org/10.1103/PhysRevD.86.121301}{\emph{Phys.
  Rev. D} {\bfseries 86} (2012) 121301}
  [\href{https://arxiv.org/abs/1205.0710}{{\ttfamily 1205.0710}}].

\bibitem{Peterson:2010mv}
C.M.~Peterson and M.~Tegmark, \emph{{Non-Gaussianity in Two-Field Inflation}},
  \href{https://doi.org/10.1103/PhysRevD.84.023520}{\emph{Phys. Rev. D}
  {\bfseries 84} (2011) 023520}
  [\href{https://arxiv.org/abs/1011.6675}{{\ttfamily 1011.6675}}].

\bibitem{Planck:2018jri}
{\scshape Planck} collaboration, \emph{{Planck 2018 results. X. Constraints on
  inflation}}, \href{https://doi.org/10.1051/0004-6361/201833887}{\emph{Astron.
  Astrophys.} {\bfseries 641} (2020) A10}
  [\href{https://arxiv.org/abs/1807.06211}{{\ttfamily 1807.06211}}].

\bibitem{Tolley:2009fg}
A.J.~Tolley and M.~Wyman, \emph{{The Gelaton Scenario: Equilateral
  non-Gaussianity from multi-field dynamics}},
  \href{https://doi.org/10.1103/PhysRevD.81.043502}{\emph{Phys. Rev. D}
  {\bfseries 81} (2010) 043502}
  [\href{https://arxiv.org/abs/0910.1853}{{\ttfamily 0910.1853}}].

\bibitem{Achucarro:2012sm}
A.~Achucarro, J.-O.~Gong, S.~Hardeman, G.A.~Palma and S.P.~Patil,
  \emph{{Effective theories of single field inflation when heavy fields
  matter}}, \href{https://doi.org/10.1007/JHEP05(2012)066}{\emph{JHEP}
  {\bfseries 05} (2012) 066} [\href{https://arxiv.org/abs/1201.6342}{{\ttfamily
  1201.6342}}].

\bibitem{Cremonini:2010sv}
S.~Cremonini, Z.~Lalak and K.~Turzynski, \emph{{On Non-Canonical Kinetic Terms
  and the Tilt of the Power Spectrum}},
  \href{https://doi.org/10.1103/PhysRevD.82.047301}{\emph{Phys. Rev. D}
  {\bfseries 82} (2010) 047301}
  [\href{https://arxiv.org/abs/1005.4347}{{\ttfamily 1005.4347}}].

\bibitem{Achucarro:2016fby}
A.~Ach\'ucarro, V.~Atal, C.~Germani and G.A.~Palma, \emph{{Cumulative effects
  in inflation with ultra-light entropy modes}},
  \href{https://doi.org/10.1088/1475-7516/2017/02/013}{\emph{JCAP} {\bfseries
  02} (2017) 013} [\href{https://arxiv.org/abs/1607.08609}{{\ttfamily
  1607.08609}}].

\bibitem{Turner:1983he}
M.S.~Turner, \emph{{Coherent Scalar Field Oscillations in an Expanding
  Universe}}, \href{https://doi.org/10.1103/PhysRevD.28.1243}{\emph{Phys. Rev.
  D} {\bfseries 28} (1983) 1243}.

\bibitem{Peloso:2000hy}
M.~Peloso and L.~Sorbo, \emph{{Preheating of massive fermions after inflation:
  Analytical results}},
  \href{https://doi.org/10.1088/1126-6708/2000/05/016}{\emph{JHEP} {\bfseries
  05} (2000) 016} [\href{https://arxiv.org/abs/hep-ph/0003045}{{\ttfamily
  hep-ph/0003045}}].

\bibitem{Gibbons:1977mu}
G.W.~Gibbons and S.W.~Hawking, \emph{{Cosmological Event Horizons,
  Thermodynamics, and Particle Creation}},
  \href{https://doi.org/10.1103/PhysRevD.15.2738}{\emph{Phys. Rev. D}
  {\bfseries 15} (1977) 2738}.

\bibitem{Greene:1997fu}
P.B.~Greene, L.~Kofman, A.D.~Linde and A.A.~Starobinsky, \emph{{Structure of
  resonance in preheating after inflation}},
  \href{https://doi.org/10.1103/PhysRevD.56.6175}{\emph{Phys. Rev. D}
  {\bfseries 56} (1997) 6175}
  [\href{https://arxiv.org/abs/hep-ph/9705347}{{\ttfamily hep-ph/9705347}}].

\bibitem{Adshead:2015pva}
P.~Adshead, J.T.~Giblin, T.R.~Scully and E.I.~Sfakianakis,
  \emph{{Gauge-preheating and the end of axion inflation}},
  \href{https://doi.org/10.1088/1475-7516/2015/12/034}{\emph{JCAP} {\bfseries
  12} (2015) 034} [\href{https://arxiv.org/abs/1502.06506}{{\ttfamily
  1502.06506}}].

\bibitem{Minkowski:1977sc}
P.~Minkowski, \emph{{$\mu \to e\gamma$ at a Rate of One Out of $10^{9}$ Muon
  Decays?}}, \href{https://doi.org/10.1016/0370-2693(77)90435-X}{\emph{Phys.
  Lett. B} {\bfseries 67} (1977) 421}.

\bibitem{Yanagida:1979as}
T.~Yanagida, \emph{{Horizontal gauge symmetry and masses of neutrinos}},
  {\emph{Conf. Proc. C} {\bfseries 7902131} (1979) 95}.

\bibitem{Dolan:1973qd}
L.~Dolan and R.~Jackiw, \emph{{Symmetry Behavior at Finite Temperature}},
  \href{https://doi.org/10.1103/PhysRevD.9.3320}{\emph{Phys. Rev. D} {\bfseries
  9} (1974) 3320}.

\bibitem{Eberhardt:2013uba}
O.~Eberhardt, U.~Nierste and M.~Wiebusch, \emph{{Status of the
  two-Higgs-doublet model of type II}},
  \href{https://doi.org/10.1007/JHEP07(2013)118}{\emph{JHEP} {\bfseries 07}
  (2013) 118} [\href{https://arxiv.org/abs/1305.1649}{{\ttfamily 1305.1649}}].

\bibitem{Cline:2019fxx}
J.M.~Cline, M.~Puel and T.~Toma, \emph{{Affleck-Dine inflation}},
  \href{https://doi.org/10.1103/PhysRevD.101.043014}{\emph{Phys. Rev. D}
  {\bfseries 101} (2020) 043014}
  [\href{https://arxiv.org/abs/1909.12300}{{\ttfamily 1909.12300}}].

\bibitem{Barrie:2021mwi}
N.D.~Barrie, C.~Han and H.~Murayama, \emph{{Affleck-Dine Leptogenesis from
  Higgs Inflation}},
  \href{https://doi.org/10.1103/PhysRevLett.128.141801}{\emph{Phys. Rev. Lett.}
  {\bfseries 128} (2022) 141801}
  [\href{https://arxiv.org/abs/2106.03381}{{\ttfamily 2106.03381}}].

\bibitem{Mohapatra:2021ozu}
R.N.~Mohapatra and N.~Okada, \emph{{Unified model for inflation,
  pseudo-Goldstone dark matter, neutrino mass, and baryogenesis}},
  \href{https://doi.org/10.1103/PhysRevD.105.035024}{\emph{Phys. Rev. D}
  {\bfseries 105} (2022) 035024}
  [\href{https://arxiv.org/abs/2112.02069}{{\ttfamily 2112.02069}}].

\bibitem{Mohapatra:2022tgb}
R.N.~Mohapatra and N.~Okada, \emph{{Affleck-Dine leptogenesis with one loop
  neutrino masses and a solution to the strong CP problem}},
  \href{https://doi.org/10.1103/PhysRevD.106.115014}{\emph{Phys. Rev. D}
  {\bfseries 106} (2022) 115014}
  [\href{https://arxiv.org/abs/2207.10619}{{\ttfamily 2207.10619}}].

\bibitem{Barrie:2024yhj}
N.D.~Barrie and C.~Han, \emph{{Affleck-Dine Dirac Leptogenesis}},
  \href{https://arxiv.org/abs/2402.15245}{{\ttfamily 2402.15245}}.

\bibitem{Bento:2004xu}
L.~Bento and F.C.~Santos, \emph{{Neutrino helicity asymmetries in
  leptogenesis}}, \href{https://doi.org/10.1103/PhysRevD.71.096001}{\emph{Phys.
  Rev. D} {\bfseries 71} (2005) 096001}
  [\href{https://arxiv.org/abs/hep-ph/0411023}{{\ttfamily hep-ph/0411023}}].

\bibitem{Chun:2023eqc}
E.J.~Chun and T.H.~Jung, \emph{{Leptogenesis driven by a Majoron}},
  \href{https://doi.org/10.1103/PhysRevD.109.095004}{\emph{Phys. Rev. D}
  {\bfseries 109} (2024) 095004}
  [\href{https://arxiv.org/abs/2311.09005}{{\ttfamily 2311.09005}}].

\bibitem{Davidson:2008bu}
S.~Davidson, E.~Nardi and Y.~Nir, \emph{{Leptogenesis}},
  \href{https://doi.org/10.1016/j.physrep.2008.06.002}{\emph{Phys. Rept.}
  {\bfseries 466} (2008) 105}
  [\href{https://arxiv.org/abs/0802.2962}{{\ttfamily 0802.2962}}].

\bibitem{ParticleDataGroup:2022pth}
{\scshape Particle Data Group} collaboration, \emph{{Review of Particle
  Physics}}, \href{https://doi.org/10.1093/ptep/ptac097}{\emph{PTEP} {\bfseries
  2022} (2022) 083C01}.

\bibitem{Chao:2023ojl}
W.~Chao and Y.-Q.~Peng, \emph{{Majorana Majoron and the Baryon Asymmetry of the
  Universe}},  \href{https://arxiv.org/abs/2311.06469}{{\ttfamily 2311.06469}}.

\bibitem{Mohapatra:1991bz}
R.N.~Mohapatra and X.-m.~Zhang, \emph{{QCD sphalerons at high temperature and
  baryogenesis at electroweak scale}},
  \href{https://doi.org/10.1103/PhysRevD.45.2699}{\emph{Phys. Rev. D}
  {\bfseries 45} (1992) 2699}.

\bibitem{Pilaftsis:1997jf}
A.~Pilaftsis, \emph{{CP violation and baryogenesis due to heavy Majorana
  neutrinos}}, \href{https://doi.org/10.1103/PhysRevD.56.5431}{\emph{Phys. Rev.
  D} {\bfseries 56} (1997) 5431}
  [\href{https://arxiv.org/abs/hep-ph/9707235}{{\ttfamily hep-ph/9707235}}].

\bibitem{Marsh:2015xka}
D.J.E.~Marsh, \emph{{Axion Cosmology}},
  \href{https://doi.org/10.1016/j.physrep.2016.06.005}{\emph{Phys. Rept.}
  {\bfseries 643} (2016) 1} [\href{https://arxiv.org/abs/1510.07633}{{\ttfamily
  1510.07633}}].

\bibitem{Co:2019jts}
R.T.~Co, L.J.~Hall and K.~Harigaya, \emph{{Axion Kinetic Misalignment
  Mechanism}},
  \href{https://doi.org/10.1103/PhysRevLett.124.251802}{\emph{Phys. Rev. Lett.}
  {\bfseries 124} (2020) 251802}
  [\href{https://arxiv.org/abs/1910.14152}{{\ttfamily 1910.14152}}].

\bibitem{Bernal:2019lpc}
N.~Bernal and F.~Hajkarim, \emph{{Primordial Gravitational Waves in Nonstandard
  Cosmologies}}, \href{https://doi.org/10.1103/PhysRevD.100.063502}{\emph{Phys.
  Rev. D} {\bfseries 100} (2019) 063502}
  [\href{https://arxiv.org/abs/1905.10410}{{\ttfamily 1905.10410}}].

\bibitem{Su:2019dsf}
W.~Su, M.~White, A.G.~Williams and Y.~Wu, \emph{{Exploring the low $\tan \beta
  $ region of two Higgs doublet models at the LHC}},
  \href{https://doi.org/10.1140/epjc/s10052-021-09609-4}{\emph{Eur. Phys. J. C}
  {\bfseries 81} (2021) 810}
  [\href{https://arxiv.org/abs/1909.09035}{{\ttfamily 1909.09035}}].

\end{thebibliography}\endgroup

\end{document}